\documentclass[11pt,a4paper]{article}
\usepackage{cite}
\usepackage{graphicx}
\usepackage{amssymb}
\usepackage{amsmath}
\usepackage{amsfonts}
\usepackage{dsfont}
\usepackage{mathtools}
\usepackage{rotating}
\usepackage{bbold,amsfonts}

\usepackage[utf8]{inputenc}
\usepackage{bm}
\usepackage{xcolor}
\usepackage{braket}

\usepackage[height=8.8in,width=6.45in]{geometry}
\usepackage[font=small,labelfont=bf]{caption}
\usepackage[hidelinks]{hyperref}
\usepackage{booktabs,float,slashed}
\numberwithin{equation}{section}

\newcommand{\beq}{\begin{equation}}
\newcommand{\eeq}{\end{equation}}

\newcommand{\mc}{\mathcal}

\newcommand{\overbar}[1]{\mkern 1.5mu\overline{\mkern-1.5mu#1\mkern-1.5mu}\mkern 1.5mu}

\newcommand{\z}{\zeta}

\newcommand{\n}{\nu}

\newcommand{\p}{\pi}

\DeclareMathOperator{\Tr}{Tr}
\DeclareMathOperator{\tr}{tr}

\newcommand{\trp}{\Tr_{\cR}^\prime}

\newcommand{\ii}{\mathrm{i}}
\newcommand{\bI}{\,\mathbf{I}}

\makeatletter
\newcommand*{\letterdef@}{}
\newcommand*{\letterdef}[3]{%
	\def\letterdef@##1{\expandafter\newcommand\csname #1\endcsname{#2{##1}}}%
	\@tfor\@tempa :=#3\do{\expandafter\letterdef@\expandafter{\@tempa}}}
\makeatother
\letterdef{c#1} {\mathcal}{ABCDEFGHIJKLMNOPQRSTUVWXYZ} 
\letterdef{rm#1}{\mathrm} {dDeimM} 
\newcommand{\tmb}[1]{{\mbox{\tiny{#1}}}}

\newdimen\tableauside\tableauside=1.0ex
\newdimen\tableaurule\tableaurule=0.4pt
\newdimen\tableaustep
\def\phantomhrule#1{\hbox{\vbox to0pt{\hrule height\tableaurule
			width#1\vss}}}
\def\phantomvrule#1{\vbox{\hbox to0pt{\vrule width\tableaurule
			height#1\hss}}}
\def\sqr{\vbox{%
		\phantomhrule\tableaustep
		\hbox{\phantomvrule\tableaustep\kern\tableaustep\phantomvrule\tableaustep}%
		\hbox{\vbox{\phantomhrule\tableauside}\kern-\tableaurule}}}
\def\squares#1{\hbox{\count0=#1\noindent\loop\sqr
		\advance\count0 by-1 \ifnum\count0>0\repeat}}
\def\tableau#1{\vcenter{\offinterlineskip
		\tableaustep=\tableauside\advance\tableaustep by-\tableaurule
		\kern\normallineskip\hbox
		{\kern\normallineskip\vbox
			{\gettableau#1 0 }%
			\kern\normallineskip\kern\tableaurule}%
		\kern\normallineskip\kern\tableaurule}}
\def\gettableau#1 {\ifnum#1=0\let\next=\null\else
	\squares{#1}\let\next=\gettableau\fi\next}
\newcommand{\Yfund}{\tableau{1}}
\newcommand{\Ysymm}{\tableau{2}}
\newcommand{\Yasymm}{\tableau{1 1}}

\allowdisplaybreaks

\begin{document}
\begin{titlepage}
\vspace*{10mm}
\begin{center}
{\LARGE \bf 
$\cN=2$ Conformal SYM theories
at large $N$
}

\vspace*{15mm}

{\Large M. Beccaria${}^{\,a}$, M. Bill\`o${}^{\,b}$, F. Galvagno${}^{\,b}$, 
	A. Hasan${}^{\,a}$, A. Lerda${}^{\,c}$}

\vspace*{8mm}
	
${}^a$ Università del Salento, Dipartimento di Matematica e Fisica ``Ennio De Giorgi'',\\ 
		and I.N.F.N. - sezione di Lecce, \\Via Arnesano, I-73100 Lecce, Italy
			\vskip 0.3cm
			
${}^b$ Universit\`a di Torino, Dipartimento di Fisica,\\
           and I.N.F.N. - sezione di Torino,\\
			Via P. Giuria 1, I-10125 Torino, Italy
			\vskip 0.3cm
			
${}^c$  Universit\`a del Piemonte Orientale,\\
			Dipartimento di Scienze e Innovazione Tecnologica\\
			Viale T. Michel 11, I-15121 Alessandria, Italy\\[1mm]
			and I.N.F.N. - sezione di Torino,\\
			Via P. Giuria 1, I-10125 Torino, Italy 

\vskip 0.8cm
	{\small
		E-mail:
		\texttt{matteo.beccaria@le.infn.it, billo,galvagno,lerda@to.infn.it, ahasan@gradcenter.cuny.edu}
	}
\vspace*{0.8cm}
\end{center}

\begin{abstract}
We consider a class of $\cN=2$ conformal SU$(N)$ SYM theories in four dimensions with 
matter in the fundamental, two-index symmetric and anti-symmetric representations,
and study the corresponding matrix model provided by localization on a sphere $S^4$, which also encodes information on flat-space observables involving chiral operators and circular BPS Wilson loops. We review and improve known techniques for studying the matrix model in the large-$N$ limit, deriving explicit expressions in perturbation theory for these observables. We exploit both recursive methods in the so-called full Lie algebra approach and the more standard Cartan sub-algebra approach based on the eigenvalue distribution. The sub-class of conformal theories for which the number of fundamental hypermultiplets does not scale with $N$ differs in the planar limit from the $\cN=4$ SYM theory only in observables involving chiral operators of odd dimension. In this case we are able to derive compact expressions which allow to push the small 't Hooft coupling expansion to very high orders. We argue that the perturbative series have a finite radius of convergence and extrapolate them numerically to intermediate couplings. This is preliminary to an analytic investigation of the strong coupling behavior, which would be very interesting given that for such theories holographic duals have been proposed.  
\end{abstract}
\end{titlepage}
\setcounter{tocdepth}{2}
\tableofcontents
\vspace{1cm}

\section{Introduction}
\label{sec:intro}
One ambitious but necessary goal in theoretical physics is to understand the dynamics of interacting Quantum Field Theories (QFTs) at strong coupling. Many ideas have been proposed and investigated, often involving the use of duality relations. 
Among them, a prominent role is played by the Anti-de Sitter/Conformal Field Theory (AdS/CFT)
correspondence \cite{Maldacena:1997re,Gubser:1998bc,Witten:1998qj}.

$\cN=4$ Super Yang-Mills (SYM) theory represents a benchmark for exact computations in QFTs 
and for an explicit realization of the AdS/CFT correspondence. In a way, this theory is the simplest interacting four-dimensional QFT, since it enjoys the highest possible amount of superconformal symmetry \cite{Sohnius:1981sn} for a theory with at most spin-one fields, and $S$-duality. This large symmetry constrains so much of its dynamics that many sectors can be described in an exact way by resorting to powerful techniques, among which we can mention localization and the relation to integrable models. 
Furthermore, the $\cN=4$ SU$(N)$ SYM theory admits an holographic description as type II B strings 
on $\mathrm{AdS}_5\times S^5$ which is the prototype of all AdS/CFT relations. In this context, 
the 't Hooft large-$N$ limit singles out the planar diagrams on the field theory side and suppresses string loop effects.

Important achievements have been obtained in this highly symmetric context also in presence of extended objects, like the BPS Wilson loops, that preserve part of the $\cN=4$ superconformal symmetry \cite{Maldacena:1998im,Erickson:2000af,Drukker:2000rr,Semenoff:2001xp} and are examples of conformal defects \cite{Billo:2016cpy,Beccaria:2017rbe,Giombi:2017cqn,Giombi:2018qox,Giombi:2018hsx,Beccaria:2019dws,Komatsu:2020sup,Giombi:2020amn}. Many of these results can be efficiently derived using supersymmetric localization \cite{Pestun:2007rz,Pestun:2016zxk}, which allows to reduce the calculation of the partition function on a sphere $S^4$ and of other observables to a computation in a Gaussian matrix model.

Many efforts have been devoted over the years to extend as much as possible these results to less symmetric theories. In this perspective, the $\cN=2$ SYM theories play an outstanding role. For such theories the localization procedure is available \cite{Pestun:2007rz}. It expresses a class of observables on $S^4$ - including chiral operators and BPS Wilson loops -  in terms of a matrix model. This matrix model is no longer Gaussian as in the $\cN=4$ case, and contains both perturbative and non-perturbative contributions. When the $\cN=2$ theory is conformal\,%
\footnote{Large classes of $\cN=2$ superconformal theories were early found in \cite{Howe:1983wj}.}, from these localization
results it is possible to obtain information also about the analogous observables in flat space\,%
\footnote{In the case of chiral/anti-chiral two-point functions, it has been argued in \cite{Billo:2019job} that the matrix model reproduces to a large extent the flat-space result also in theories
with a non-vanishing $\beta$-function. On the contrary, when conformal invariance is explicitly broken by mass terms as in the $\cN=2^*$ theories, the observables computed from the matrix model differ from those computed in flat space \cite{Belitsky:2020hzs}.}
\cite{Baggio:2014sna,Fiol:2015mrp,Mitev:2015oty,Gerchkovitz:2016gxx}. 

It is obviously of great importance to study $\cN=2$ SYM theories in the large $N$ limit and at strong coupling and to understand if and how some analogue of the AdS/CFT duality applies \cite{Ennes:2000fu,Gadde:2009dj,Rey:2010ry,Russo:2013sba}. In this paper we provide some 
contributions to this long term goal by exploiting the localization matrix model to extract in a rather efficient way the expression of protected observables in the large-$N$ limit. We do this for a certain class of superconformal theories with matter in the fundamental, symmetric and anti-symmetric representations. 

For a sub-class of these theories a dual holographic description, built out as an appropriate orientifold projection of the AdS$_5 \times S^5$ geometry, has been proposed in \cite{Ennes:2000fu}. These theories are extremely close to the $\cN=4$ SYM theory: many observables coincide at large $N$ with the $\cN=4$ results. However, observables involving chiral operators built with traces of odd powers of the scalar fields do not; in the holographic correspondence of \cite{Ennes:2000fu} these odd-dimensional observables are related to twisted sectors of the orientifold. If we regard the $\cN=4$ SYM theory as the simplest non-trivial QFT, these theories represent the next-to-simplest ones. One could hope to be able to explicitly study them beyond the perturbative regime, at least in the 't Hooft large-$N$ limit, and to match the field-theoretic description with its holographic dual. In this paper, from the matrix model we obtain, through an effective description in terms of free variables, closed forms of the perturbative series for the odd observables
which appear to have a finite radius of convergence. 
Although we are not yet able at this stage to infer analytically the strong coupling behavior, these expressions allow for reliable numerical extrapolation to the intermediate coupling regime and indicate that the strong coupling regime might not be out of reach in a nearby future.  

Let us now be more specific about the content of this work which is divided in two parts.
In the first part we review the matrix model methods for $\cN=2$ conformal models at large values of the rank $N$ of the gauge group SU($N$). We distinguish two matrix model approaches.
In the original Pestun derivation \cite{Pestun:2007rz} the matrix model was written as an integral over the Cartan sub-algebra variables, {\it{i.e.}} over the matrix eigenvalues. 
In this framework the large-$N$ limit is described by the asymptotic eigenvalue distribution which satisfies an integral equation obtained with a saddle-point approximation \cite{Rey:2010ry,Passerini:2011fe,Bourgine:2011ie,Russo:2012ay,Mitev:2014yba,Baggio:2014sna,Fiol:2015mrp,Mitev:2015oty,Gerchkovitz:2016gxx,Baggio:2016skg,Rodriguez-Gomez:2016cem,Rodriguez-Gomez:2016ijh,Zarembo:2016bbk,Zarembo:2020tpf}. For the
$\cN=2$ conformal theories we are considering, this integral equation depends on the matter 
content only through a single parameter $\n$, which counts the fraction of hypermultiplets transforming in the fundamental representation:
\begin{equation}
	\label{defnu1}
		\nu = \lim_{N\to\infty} \frac{N_F}{2N}~.
\end{equation}
The second approach has been developed more recently, and it has been named the ``full Lie algebra approach'' in \cite{Fiol:2020bhf}. It consists of keeping the matrix integral over the full Lie algebra and developing a series of recursive rules \cite{Billo:2017glv,Billo:2018oog,Billo:2019fbi} to evaluate correlation functions. These techniques are very efficient in the perturbative regime at finite $N$. 
They allow to explore different sectors of the gauge theory \cite{Bourget:2018obm,Bourget:2018fhe,Beccaria:2018xxl,Beccaria:2018owt,Grassi:2019txd,Beccaria:2020azj} and can be used also in
non-conformal cases \cite{Billo:2019job}.
Similar methods have been used also in \cite{Fiol:2018yuc,Fiol:2020bhf}.
We shall see that in fact also the large-$N$ regime is easily accessible within the 
full Lie algebra approach by exploiting the recursion relations in a suitable way. 

We present the main technical points of the two approaches and present a thorough evaluation of the
following observables:
\begin{itemize}
\item
the vacuum expectation value of the 1/2 BPS circular Wilson loop;
\item
the two-point functions of chiral/anti-chiral operators;
\item
the one-point function of chiral operators in presence of a Wilson loop.
\end{itemize} 
In doing so, we address also some specific issues related to the computation of correlators with operators of odd dimensions, which play a crucial role throughout the paper.

In the second part of the paper we concentrate on a set of $\cN=2$ theories whose fundamental matter content does not scale with $N$, so that they have $\n=0$. As noted above, these models have a holographic dual \cite{Ennes:2000fu} and are very close to the $\cN=4$ SYM theory, as  confirmed by the fact that some observables, such as the vacuum expectation value of the
Wilson loop \cite{Billo:2019fbi} and the Bremsstrahlung function \cite{Fiol:2015mrp, Bianchi:2019dlw}, do not deviate from the $\cN=4$ result in the large-$N$ limit.
In the present paper we compute the set of observables listed above for the $\nu=0$ theories, 
using matrix model techniques, and clarify which observables are different with respect to $\cN=4$ 
in the planar limit. It turns out that the correlation functions involving only chiral operators made of traces of even order have the same behavior as in the $\cN=4$ SYM theory; this applies to both chiral/anti-chiral correlators and one-point functions in presence of the Wilson loop.  
Correlation functions with traces of odd order, instead, do deviate from the $\cN=4$ results through an infinite perturbative series. This analysis allows to identify a sort of ``twisted'' sector of operators that, in the holographic correspondence, feel the presence of the orientifold, consistently with the analysis of \cite{Ennes:2000fu}.

We also investigate the difference between even and odd correlators with a perturbative expansion of the $\cN=2$ field theory directly in flat space. Using the $\cN=1$ super-space formalism and the diagrammatic difference between $\cN=2$ and $\cN=4$ \cite{Andree:2010na,Pomoni:2011jj,Pomoni:2013poa,Fiol:2015spa,Mitev:2015oty,Billo:2017glv,Billo:2018oog,Billo:2019fbi} we are able to perform an explicit large-$N$ analysis of the two-point functions of operators with 
low dimensions up to three-loops in the $\nu=0$ models. This allows us to understand at the diagrammatic level the origin of the different behavior of the even and odd correlators in the planar limit, and to infer the general structure of the leading term in the two-point correlator of operators with arbitrary odd dimension.

Finally, we develop some new matrix model techniques which are particularly efficient when $\n=0$ and make it very easy to obtain for any odd correlator expansions at any desired order in perturbation theory. By applying some numerical resummation technique to these long expansions, we produce a first attempt in going beyond perturbation theory. Even if this still falls short of an analytic treatment of the strong coupling regime, we think that our results for this special class of $\cN=2$ theories
may represent a first step towards this important goal.

Several technical details, which are useful to reproduce our results, and some explicit high-order expansions are collected in the appendices.

\part*{Part I}
\label{part:1}
In the first part of this paper we study $\cN=2$ conformal SYM 
theories and several of their physical observables in the planar limit using matrix model
techniques.

\section{$\cN=2$ CFT theories}
\label{sec:N2scft}
We consider $\cN=2$ SYM theories in 4$d$ with gauge group SU$(N)$\,%
\footnote{We mainly concentrate on the large-$N$ limit, where SU$(N)$ yields the same results as
U$(N)$; we will comment about the relation between the two cases in Appendix \ref{app:UN}.}. The field content of these theories consists of 
one $\mathcal{N}=2$ vector multiplet, which contains the gauge vector 
$A_\mu(x)$ and a complex scalar field $\varphi(x)$ plus their fermionic partners, 
all transforming in the adjoint representation, and several
$\mathcal{N}=2$ matter hypermultiplets, each containing two complex scalars plus 
their fermionic partners
transforming in a representation $\cR$. Given this field content, 
the gauge coupling constant $g$ receives contributions only at one
loop, and the coefficient $\beta_0$ of the $\beta$-function is 
\begin{align}
	\label{beta0is}
		\beta_0 = 2N-2 i_\cR ~,
\end{align}
where $i_\cR$ is the index of $\cR$. 
In the following we will focus on conformal theories, for which $\beta_0$ vanishes. This condition is clearly satisfied if $\cR$ is the adjoint representation, in which case we have a SYM theory
with $\cN=4$ supersymmetry.

An important set of local operators in these theories is provided by the multitraces  
\begin{align}
	\label{defOn}
		O_{\mathbf{n}}(x) \equiv \tr \varphi^{n_1}(x)\, \tr \varphi^{n_2}(x) \ldots
\end{align}
where $\mathbf{n}=\{n_i\}$.
These operators are chiral, {\it{i.e.}} they are annihilated by half of the supercharges.
Their $R$-charge is $n = \sum_i n_i$ and they are automatically normal-ordered because of 
$R$-charge conservation. The analogous anti-chiral operators, constructed with the conjugate field 
$\overbar \varphi(x)$, are denoted by $\overbar O_{\mathbf{n}}(x)$. We will study 
the two-point functions between chiral and anti-chiral operators, which in conformal theories take 
the general form
\begin{equation}
	\label{twopointdef}
		\big\langle O_{\mathbf{n}}(x) \,\overbar O_{\mathbf{m}}(0)\big\rangle 
		= \frac{G_{\mathbf{n},\mathbf{m}}(g,N)}{(4\pi^2 x^2)^{m+n}\phantom{\big|}}\,\delta_{m,n}~.
\end{equation}
In the following we will often consider diagonal cases, for which we will employ the streamlined notation $G_{\mathbf{n}}\equiv G_{\mathbf{n},\mathbf{n}}$.

Another operator that we will consider is the half-BPS Wilson loop 
in the fundamental representation on a circle $C$ of radius $R$: 
\begin{equation}
	\label{WLdef}
		W_C=\frac{1}{N}\tr \mathcal{P}
		\exp \bigg\{g \oint_C d\tau \Big[\ii \,A_{\mu}(x)\,\dot{x}^{\mu}(\tau)
		+\frac{R}{\sqrt{2}}\big(\varphi(x) + \varphi^\dagger(x)\big)\Big]\bigg\}
\end{equation}
where $\mathcal{P}$ denotes the path-ordering. In particular, we will study its vacuum expectation
value (v.e.v.)
\begin{equation}
	\label{Wvev}
	\big\langle W_C\big\rangle=
		w(g,N) ~,
\end{equation}
and its one-point functions with the chiral operators, whose form is fixed by
conformal invariance to be \cite{Billo:2016cpy,Billo:2018oog}
\begin{equation}
	\label{key}
		\big\langle O_{\mathbf{n}}(0)\, W_C\big\rangle = \frac{w_{\mathbf{n}}(g,N)}{(2\pi R)^n}~.
\end{equation}

The sets of functions $G_{\mathbf{n},\mathbf{m}}(g,N)$, $w(g,N)$ and $w_{\mathbf{n}}(g,N)$ are the main subject of our analysis. In particular we will study these functions in the large-$N$ 't Hooft limit in which $N\to\infty$ with $\lambda=g^2N$ held fixed. Notice that all these observables 
involve operators constructed entirely with fields of the gauge multiplet. Therefore, in 
their perturbative evaluation, the matter hypermultiplets run only inside the loops.  

\subsection{The ABCDE theories}
\label{subsec:ABCDE}
To be specific, we will consider theories whose matter fields transform in the following representation of SU($N$):
\begin{equation}
	\label{RNFNASNA}
		\cR = N_F\, \Yfund \oplus N_S\, \Ysymm \oplus N_A\, \Yasymm~,
\end{equation}
corresponding to $N_F$ fundamental, $N_S$ symmetric and $N_A$ anti-symmetric hypermultiplets.
For these theories the $\beta$-function coefficient (\ref{beta0is}) reads
\begin{align}
 	\label{beta0}
 		\beta_0 = 2N-N_F-N_S(N+2)-N_A(N-2)~. 
\end{align}
The condition $\beta_0=0$ leads to five families of $\mathcal{N}=2$ 
superconformal theories, whose field content is displayed in Table~\ref{tab:scft}.
\begin{table}[ht]
	\begin{center}
		{\small
			\begin{tabular}{cccc|c}
				\toprule
				\,\,theory \phantom{\Big|}& $N_F$ & $N_S$ & $N_A$ & ~$\nu$~ \\
				\midrule
				$\mathbf{A}\phantom{\Big|}$& $~~2N~~$ & $~~0~~$ & $~~0~~$ & $1$ \\
				$\mathbf{B}\phantom{\Big|}$& $~~N-2~~$ & $~~1~~$ & $~~0~~$ & $\frac 12 $ \\
				$\mathbf{C}\phantom{\Big|}$& $~~N+2~~$ & $~~0~~$ & $~~1~~$ & $\frac 12 $ \\
				$\mathbf{D}\phantom{\Big|}$& $~~4~~$ & $~~0~~$ & $~~2~~$ & $ 0 $ \\
				$\mathbf{E}\phantom{\Big|}$& $~~0~~$ & $~~1~~$ & $~~1~~$ & $ 0 $ \\
				\bottomrule
			\end{tabular}
		}
	\end{center}
	\caption{The five families of $\cN=2$ superconformal theories with SU($N$) gauge 
		group and matter in fundamental, symmetric and anti-symmetric representations.
		These theories have been identified long ago in \cite{Koh:1983ir}, and more recently they have been considered in \cite{Fiol:2015mrp,Bourget:2018fhe}.}
	\label{tab:scft}
\end{table}

Theory $\mathbf{A}$ is the $\mathcal{N}=2$ conformal QCD. 
Theories $\mathbf{D}$ and $\mathbf{E}$ are superconformal models for which a holographic dual 
of the form $\text{AdS}_5 \times S^5/\Gamma$ has been identified \cite{Ennes:2000fu}. In the last column of the table we have written the values of the variable $\nu$ defined in (\ref{defnu1}), which in the specific case becomes
\begin{equation}
	\label{defnu}
		\nu 
		= 1 - \frac{N_S + N_A}{2}~.
\end{equation}
As mentioned in the Introduction, 
this quantity determines the large-$N$ behavior of the theory, so that
theories $\mathbf{B}$ and $\mathbf{C}$ become equivalent in the large-$N$ limit, and
the same is true for theories $\mathbf{D}$ and $\mathbf{E}$.

\subsection{Matrix model from localization}
Exploiting localization, it is possible to prove that 
certain protected observables of the $\cN=2$ SYM theories
can be exactly reduced to a matrix model computation \cite{Pestun:2007rz,Pestun:2016zxk}. 
Among these observables there are the partition function on a four-sphere $S^4$, the 
correlators between chiral and anti-chiral operators, as well as the v.e.v. of a circular BPS Wilson loop and the one-point functions of chiral operators in presence of the loop. 

\paragraph{The sphere partition function:}
The partition function of a $\cN=2$ SYM theory with gauge group SU$(N)$ on a sphere $S^4$ 
of unit radius can be expressed as follows \cite{Pestun:2007rz}:
\begin{align}
	\label{partf1}
		\cZ_{S^4} = \int \!\prod_{u=1}^N dm_u~ \Delta(m) \,\big| Z(\ii m,g)\big|^2\,
		\delta\Big(\sum_u m_u\Big)~.
\end{align} 
Here $m_u$ are the eigenvalues of a Hermitean traceless $(N\times N)$ matrix $M$
and $\Delta(m)$ is their Vandermonde determinant
\begin{align}
	\label{Vandermonde}
		\Delta(m) = \prod_{u<v=1}^N (m_u - m_v)^2~. 
\end{align} 
Moreover $Z(\ii m,g)$ is the partition function for the theory on $\mathbb{R}^4$ with gauge coupling $g$ evaluated at a point in the Coulomb moduli-space parametrized by the eigenvalues $m_u$. 
It consists of a classical, a one-loop and an instanton factor:
\begin{align}
	\label{Zthreefactors}
		Z(\ii m,g) = Z_{\mathrm{tree}}\, Z_{\mathrm{1-loop}}\, Z_{\mathrm{inst}}~.
\end{align}
The classical part is simply
\begin{align}
	\label{zcl2}
		\big| Z_{\mathrm{tree}} \big|^2 
		= \rme^{- \frac{8\pi^2}{g^2} \sum_u m_u^2} = \rme^{-\frac{8\pi^2}{g^2} \tr M^2}~.
\end{align}
The instanton part can be neglected when working in perturbation theory; moreover it does not contribute in the large-$N$ 't Hooft limit, and thus in the following we set 
$Z_{\mathrm{inst}}= 1$.

The one-loop part depends on the matter representation $\cR$. Denoting by 
$\mathbf{m}$ the $N$-dimensional vector of 
components $m_u$, by $W(\cR)$ the set of the weights $\mathbf{w}$ of the 
representation $\cR$ and by
$W(\mathrm{adj})$ that of the adjoint representation, we have 
\begin{equation}
	\label{Z1loop}
		\big|Z_{\mathrm{1-loop}}\big|^2 =
		\frac{\prod_{\mathbf{w}\in W(\mathrm{adj}) } H(\ii \mathbf{w}\cdot\mathbf{m})}
		{\prod_{\mathbf{w}\in W(\mathcal{R})} H(\ii \mathbf{w}\cdot\mathbf{m})}~,
\end{equation} 
where
\begin{equation}
H(x)=G(1+x)\,G(1-x)~,
\label{His}
\end{equation}
with $G$ being the Barnes $G$-function. Writing 
\begin{align}
	\label{intac1}
		\big|Z_{\mathrm{1-loop}}\big|^2 = \rme^{-S_{\mathrm{int}}}		
\end{align}
we deduce from (\ref{Z1loop}) that the interacting action is \cite{Billo:2019fbi}
\begin{align}
	\label{StologH}
		S_{\mathrm{int}}
		&= \Tr_{\cR} \log H(\ii M) - \Tr_{\mathrm{adj}} \log H(\ii M)  \,=\, \trp \log H(\ii M)
\end{align}
where we have introduced the combination of traces
\begin{align}
	\label{deftrp}
		\Tr_{\cR}^\prime \bullet \,= \,\Tr_{\cR} \bullet - \Tr_{\mathrm{adj}} \bullet~.
\end{align}
In the $\cN=4$ theory, where $\cR$ is the adjoint representation, this combination clearly vanishes
while in the $\cN=2$ theories it accounts for the matter content of the so-called ``difference theory'' which is often used in field theory computations \cite{Andree:2010na}, 
where one removes from the $\cN=4$ result the contributions of the adjoint hypermultiplets 
and replaces them with the contributions of hypermultiplets in the representation $\cR$.

For the class of theories listed in Table~\ref{tab:scft}, by 
combining the tree-level and one-loop factors, we obtain
\begin{align}
	\label{Zexpl}
		\big| Z (\ii m,g)\big|^2 = \rme^{-S} 
		= \rme^{-\frac{8\pi^2}{g^2} \tr M^2- S_{\mathrm{int}}} ~.
\end{align}
In terms of the eigenvalues $m_u$, the matrix model action $S$ is explicitly given by
\begin{align}
	\label{Sexpl}
		S &=\sum_u \Big[
		\frac{8\pi^2}{g^2} m_u^2 + N_F \log H(\ii m_u) + N_S \log H(2\ii m_u)\Big]
		+ (N_S + N_A) \sum_{u<v} H(\ii m_u + \ii m_v)\nonumber\\[1mm]
		&\quad - \sum_{u<v} \big[ \log (\ii m_u-\ii m_v)^2 + 2 \log H(\ii m_u - \ii m_v)\Big]~.
\end{align}

\paragraph{The BPS Wilson loop:}
In \cite{Pestun:2007rz} it was shown that also the v.e.v. of the supersymmetric circular Wilson loop (\ref{WLdef})
can evaluated exactly using localization. The result, which was anticipated by direct diagrammatic computations in \cite{Erickson:2000af,Drukker:2000rr}, is  
\begin{align}
	\label{vevWpestun}
		\big\langle W_C\big\rangle_{S^4} 
		= \frac{1}{N} \int \!\prod_{u=1}^N dm_u~ \Delta(m) 
		\,\Big(\sum_u \,\rme^{2\pi m_u}\Big)\,
		 \big| Z(\ii m,g)\big|^2\,
		 \delta\Big(\sum_u m_u\Big)
\end{align} 
when the Wilson loop is in the fundamental representation. This means that in the matrix model
$W_C$ is represented by the following operator
\begin{equation}
\label{WCM}
W_C=\frac{1}{N}\,\tr\exp\!\big(2\pi M\big)~.
\end{equation}
For conformal theories the v.e.v. of $W_C$ on $S^4$ coincides with the flat-space observable $w(g,N)$ defined in (\ref{Wvev}).

\subsection{The full Lie algebra approach}
\label{subsec:fla}
The sphere partition function (\ref{partf1}) is expressed in terms of the eigenvalues $m_u$ of $M$. The integration measure over these eigenvalues with the Vandermonde determinant 
is precisely the one that arises from diagonalizing the flat measure 
over the entire matrix $M$; moreover, the exponential weight, $\rme^{-S}$,
can be written in terms of the matrix $M$ using (\ref{zcl2}) and (\ref{StologH}). Thus, 
the partition function can be recast in the form 
\begin{equation}
	\label{intdM}
		\cZ_{S^4} = \int dM\, \rme^{-S(M)}\,\delta\big(\tr M\big)
\end{equation}
namely as an integral over all elements of $M$.

In the following we will use the conventions of \cite{Billo:2017glv,Billo:2018oog} and rescale the matrix $M$ so as to get a tree-level term in the matrix model action 
with unit weight. We then introduce the matrix
\begin{align}
	\label{atoM}
		a = \sqrt{\frac{8\pi^2}{g^2}}\, M
\end{align} 
understanding, from now on, that it is traceless. The partition function (\ref{intdM}) reads
\begin{align}
	\label{defZa}	
		\cZ_{S^4} & = \Big(\frac{g^2}{8\pi^2}\Big)^{\frac{N^2-1}{2}} 
		\int da\, \rme^{- \tr a^2 - S_{\mathrm{int}}(a)}
		~	= ~
		\Big(\frac{g^2}{8\pi^2}\Big)^{\frac{N^2-1}{2}}\, \big\langle \rme^{- S_\mathrm{int}(a)}
		\big\rangle_{(0)}~,
\end{align} 
where in the second step we used the notation
\begin{align}
	\label{defvev0}
		\big\langle f(a)\big\rangle_{(0)} \,\equiv\, \int da\, \rme^{-\tr a^2}\, f(a)
\end{align}
for any function of $a$. This shows that $\cZ_{S^4}$ can be regarded as the 
expectation value of $\rme^{-S_{\mathrm{int}}(a)}$ in the free Gaussian model.

We now consider a basis of $\mathfrak{su}(N)$ generators $T_b$, with
$b = 1,\ldots, N^2-1$, normalized as
\begin{equation}
\tr \,T_b \,T_c=\frac{1}{2}\,\delta_{bc}~,
\label{normtrace}
\end{equation}
and write $a=a^b\,T_b$. Then, the flat integration measure appearing above becomes
\begin{align}
	\label{defda}
		da = \prod_b \frac{da^b}{\sqrt{2\pi}}
\end{align}
where the normalization has been chosen in such a way that 
$\big\langle\mathbb{1}\big\rangle_{(0)}=1$ 
and the ``propagator'' for the components of $a$ is simply
\begin{align}
\label{propa}
\big\langle a^b\, a^c\big\rangle_{(0)} = \delta^{bc}~.
\end{align} 

Let us now discuss the interacting part of the matrix model in this full Lie algebra approach.
{From} (\ref{StologH}) and (\ref{atoM}) we see that
\begin{equation}
S_{\mathrm{int}}(a)=\trp \log H\Big(\ii\sqrt{\frac{g^2}{8\pi^2}}\,a\Big)~.
\end{equation}
If $g$ is small, we can use the expansion
\begin{equation}
	\label{logHexp}
		\log H(x)=-(1+\gamma_{\mathrm{E}})\,x^2-
		\sum_{p=1}^\infty \frac{\zeta(2p+1)}{p+1}\, x^{2p+2}~,
\end{equation}
where $\gamma_{\mathrm{E}}$ is the Euler-Mascheroni constant, and obtain
\begin{align}
	\label{Saexp}
		S_\tmb{int}(a) = 
		\sum_{p=1}^\infty (-1)^p	 \Big(\frac{g^2}{8\pi^2}\Big)^{\!p+1}\,
		\frac{\zeta(2p+1)}{p+1} \trp a^{2p+2}~.
\end{align}
Notice that the quadratic term, proportional to $(1+\gamma_{\mathrm{E}})$, drops out since
it is proportional to the coefficient $\beta_0$ which vanishes for the theories we are considering; indeed
\begin{align}
	\label{trpa2}
	 \trp a^2 = \Tr_{\cR} a^2 - \Tr_{\mathrm{adj}} a^2 = -\beta_0\,\tr a^2= 0~.
\end{align}
The higher traces $\trp a^{2k}$ with $k>1$ appearing in the interacting action 
can be re-expressed in terms of traces in the fundamental representation as follows
\cite{Billo:2019fbi}:
\begin{equation}
\begin{aligned}
	\label{trpa2k}
		\trp a^{2k} & = \frac{1}{2} \sum_{\ell=2}^{2k-2} \binom{2k}{\ell} 
		\big(N_S + N_A - 2 (-1)^\ell\big) \tr a^\ell\, \tr a^{2k-\ell}\\[1mm]
		& 	\quad + \big(2^{2k-1}-2\big)\big(N _S-N_A\big) \tr a^{2k}~.
\end{aligned}
\end{equation}
Therefore, in the full Lie algebra approach 
the action $S_{\mathrm{int}}$ of the interacting matrix model is a linear combination of single traces and double traces of powers of $a$ in the fundamental representation
with coefficients depending on the gauge coupling $g$ and on $\zeta$-values. 

Given this structure, the expectation value of a generic function $f(a)$ in the $\cN=2$ matrix model is defined by
\begin{align}
	\label{vevmat}
		\big\langle f(a) \big\rangle 
		= \frac{\displaystyle{ \int \!da ~f(a)\,\rme^{-\tr a^2-S_\mathrm{int}(a)}}}
		{ \displaystyle{\int \!da~\rme^{-\tr a^2-S_\mathrm{int}(a)}}}	
		= \frac{\displaystyle{\big\langle\,
		f(a)\,\rme^{- S_\mathrm{int}(a)}\,\big\rangle_{(0)}\phantom{\Big|}}}{\displaystyle{\big\langle\,
		\rme^{- S_\mathrm{int}(a)}\,\big\rangle_{(0)}\phantom{\Big|}}}~.
\end{align} 
Evaluating this v.e.v. in perturbation theory is therefore just a matter of computing
expectation values in the free Gaussian model using the propagator (\ref{propa}). 

\paragraph{The BPS Wilson loop:}
In the full Lie algebra approach, the field theoretic BPS Wilson loop expectation value is exactly captured by the v.e.v. of the following operator in the matrix model
\begin{align}
	\label{defWa}
		W_C = \frac{1}{N} \tr \exp\Big(\frac{g}{\sqrt{2}} \,a\Big)
\end{align} 
which simply follows from (\ref{WCM}) upon using the redefinition (\ref{atoM}).

\paragraph{Chiral operators and normal ordering:} 
As shown in \cite{Baggio:2014ioa,Baggio:2014sna,Gerchkovitz:2014gta,Baggio:2015vxa,Baggio:2016skg,Gerchkovitz:2016gxx,Rodriguez-Gomez:2016ijh,Rodriguez-Gomez:2016cem,Billo:2017glv}
the matrix model encodes information also about the flat space correlators of chiral/anti-chiral operators in conformal theories\,%
\footnote{Despite the presence of a conformal anomaly, even in non-conformal cases the interacting matrix model contains a lot of information about the perturbative expansion of such correlators in flat space \cite{Billo:2017glv,Billo:2019job}.}. To obtain this information one maps the multitrace 
operators $O_\mathbf{n}(x)$ introduced in (\ref{defOn}) to suitable matrix 
operators $O_{\mathbf{n}}(a)$ such that their two-point functions correspond to the
quantities $G_{\mathbf{n},\mathbf{m}}$ defined in (\ref{twopointdef}), namely
\begin{align}
	\label{Gtooa}
		G_{\mathbf{n},\mathbf{m}} = \big\langle O_{\mathbf{n}}(a) \,
		O_{\mathbf{m}}(a)\big\rangle~,
\end{align}
where the v.e.v. in the right hand side is computed according to (\ref{vevmat}). 
Na\"ively one would associate $O_\mathbf{n} (x)$ to
matrix operators with the same trace structure, that is
\begin{align}
	\label{defOno} 
	\Omega_\mathbf{n}(a) = \tr a^{n_1}\,\tr a^{n_2}\,\ldots~.
\end{align}
These matrix operators, however, contrarily to their field theoretic counterparts, 
are not normal ordered and their two-point functions 
\begin{align}
	\label{defC2p}
		C_{\mathbf{n},\mathbf{m}} =
		\big\langle \Omega_\mathbf{n}(a)\, \Omega_\mathbf{m}(a)\big\rangle
\end{align}
are not diagonal.
Therefore, as argued in \cite{Gerchkovitz:2016gxx,Billo:2017glv}, one has apply the
Gram-Schmidt orthogonalization procedure with respect to $C_{\mathbf{n},\mathbf{m}}$, and
construct the normal-ordered version of $\Omega_{\mathbf{n}}(a)$, which we denote by
$O_{\mathbf{n}}(a)$.
If $C_{(n)}$ is the matrix of two-point functions among the operators of dimension lower than $n$, then one finds
\begin{align}
	\label{Omegs}
		O_\mathbf{n}(a) = \Omega_\mathbf{n}(a) - \sum_{\substack{\mathbf{p},\mathbf{q}\\p,q < n}}
		C_{\mathbf{n},\mathbf{p}}\, \big(C^{-1}_{(n)}\big)^{\mathbf{p},\mathbf{q}} \,\Omega_\mathbf{q}(a)~.  
\end{align}
With this definition, the correlator of $O_{\mathbf{n}}(a)$ with any operator of lower dimensions is zero and the two-point functions $G_{\mathbf{n},\mathbf{m}}$ vanish 
for $n\not= m$, as required. Using  (\ref{Omegs}), these two-point functions
can be expressed in terms of the correlators $C_{\mathbf{n},\mathbf{m}}$. It turns out that
they have a particularly simple expression when there is a single independent operator for each dimension $n$, as in the SU$(2)$ case \cite{Gerchkovitz:2016gxx}. 
We will see that in the large-$N$ limit also the set of single-trace operators is closed under normal ordering and we will restrict our attention to this set. Since there is one single-trace operator for each dimension $n$, then the matrix $G$ has a simple expression in terms of the matrix $C$.

\paragraph{Recursion relations:}
From what we have reviewed above, it is clear that the basic ingredients for the
calculation of the various observables in the $\cN=2$ matrix model
are the expectation values of the multitrace operators in the Gaussian theory, 
which we denote as
\begin{equation}
	\label{tn}
		t_\mathbf{n} = \big\langle \Omega_\mathbf{n}(a)\big\rangle_{(0)}~.
\end{equation}
One obvious fact is that
\begin{align}
	\label{tnodd}
		t_{\mathbf{n}} = 0~~~\text{for $n$ odd}~.
\end{align}
For $n$ even, instead, they are non-vanishing and one can explicitly evaluate their expressions starting from the initial condition $t_0 = N$ and a set of recursion relations of the form
\begin{subequations}
\label{rrecursion}
\begin{align}	
		t_{n}    &=   \frac12 \sum_{m=0}^{n-2}  \Big( t_{m,n-m-2}
		-\frac{1}{N}\, t_{n-2}  \Big)  ~,\label{rec1}\\
		t_{n,n_1} &=  \frac12 \sum_{m=0}^{n-2}  \Big( t_{m,n-m-2,n_1}
		-\frac{1}{N}\,   t_{n-2,n_1}  \Big)  
		+ \frac{n_1}{2} \,\Big(  t_{n+n_1-2 } -\frac{1}{N} \,t_{n-1,n_1-1} \Big)~,
		\label{rec2}
\end{align}%
\end{subequations}
and so on. These relations follow \cite{Billo:2017glv} from the fusion/fission identities satisfied by the $\mathfrak{su}(N)$ generators $T_b$, namely
\begin{equation}
\label{fussion}
\begin{aligned}	
		\tr\big(T_b\, A\, T_b\, B\big) & = \frac{1}{2}\,\tr A\, \tr B
    	-\frac{1}{2N}\,\tr \big(A\, B\big)~,\\
		\tr \big(T_b\, A \big) \,\tr\big(T_b \,B\big) & 
		= \frac{1}{2}\,\tr\big(A\,B\big) -\frac{1}{2N}\,\tr A \,\tr B~,
\end{aligned}
\end{equation}
for two arbitrary $(N\times N)$ matrices $A$ and $B$. 
In fact, using these identities one can recursively relate any correlator $t_\mathbf{n}$
to the combination of correlators obtained after a single Wick contraction with the propagator
(\ref{propa}).

\section{Large-$N$ limit from the recursion relations}
\label{sec:3}
We now consider the 't Hooft limit in which $N\to\infty$ with
\begin{align}
	\label{deflambda}
		\lambda = g^2 N
\end{align}
kept fixed. As argued in the previous section, all relevant observables can be expressed in terms of the quantities $t_\mathbf{n}$ defined in (\ref{tn}). 
Therefore, as a preliminary step, we study the large-$N$ limit of the latter. 

\subsection{Basic ingredients}
\label{sunsec:largeNgaussian}

\paragraph{Single traces:}
Eq. (\ref{tnodd}) implies that the odd single traces have an identically vanishing v.e.v.:
\begin{equation}
t_{2k+1}=0~.
\label{todd}
\end{equation}
The v.e.v. of the even single traces, $t_{2k}$, can be computed by applying Wick's theorem and 
taking into account that all contractions in which some propagators cross,
are suppressed in the large-$N$ limit. In other words, only rainbow diagrams count \cite{Erickson:2000af} and, up to subleading terms in the $1/N$ expansion, one finds
\begin{equation}
	\label{t2kres1}
		t_{2k} = N^{k+1}\, \frac{C_k}{2^k}
\end{equation}
where $C_k$ are the Catalan numbers
\begin{equation}
	\label{Catexpl}
		C_k 	= \frac{1}{k+1} \binom{2k}{k}
\end{equation}
which enumerate the distinct rainbow diagrams\,%
\footnote{The generating function of the Catalan numbers is 
	\begin{align*}
	\label{gencat}
	f(x) = \sum_{k=0}^\infty C_k\, x^k = \frac{1 - \sqrt{1 - 4 x}}{2x}~.
	\end{align*}}.

\paragraph{Double traces:}
The mixed even/odd double traces are clearly vanishing due to (\ref{tnodd}):
\begin{equation}
t_{2k_1,2k_2+1}=0~.
\label{mixed}
\end{equation}
Thus we have to consider only the even and the odd double traces.
At the leading order in the large-$N$ limit, the even double traces factorize: 
\begin{equation}
	\label{t2k2k}
		t_{2k_1,2k_2} = t_{2k_1}\, t_{2k_2}~.
\end{equation}
To show this, one can consider the relation (\ref{rec2}) and observe that the term in the right-hand
side proportional to $n_1$, obtained when a propagator connects the first component with the second component, is subleading with respect to the first term in which the propagator remains within the first component.  This fact leads to the factorized result (\ref{t2k2k}).
 
The next-to-leading terms determine the connected part of the double trace v.e.v.'s. They are
defined as
\begin{align}
	\label{defct}
		t_{2k_1,2k_2}^{\mathrm{c}}
		& \equiv t_{2k_1,2k_2}-t_{2k_1}\,t_{2k_2}~,
\end{align}
and at large $N$ they behave as
\begin{align}
	\label{t2evenc}
		t_{2k_1,2k_2}^{\mathrm{c}} =\frac{\alpha_{k_1} \alpha_{k_2}}{k_1 + k_2} 
		~~~~ \text{with}~~
		\alpha_k = N^k \frac{(2k-1)!!}{(k-1)!}~.
\end{align}
Comparing (\ref{t2k2k}) and (\ref{t2evenc}), we see that the connected v.e.v. 
$t_{2k_1,2k_2}^{\mathrm{c}}$ is suppressed by a factor of $1/N^2$ 
with respect to $t_{2k_1,2k_2}$.

Also in the odd case we can use the recursion relations (\ref{rrecursion}), but we cannot discard 
the terms that superficially look subleading. Clearly, the odd double traces
coincide with their connected part because of (\ref{tnodd}), and at large $N$ they are given by\,%
\footnote{The U($N$) analogue of this result was given in 
\cite{Rodriguez-Gomez:2016ijh,Rodriguez-Gomez:2016cem}
where it was obtained from the eigenvalue distribution approach.}
\begin{align}
	\label{t2oddc}
		t_{2k_1+1,2k_2+1}
		& = \frac{\beta_{k_1} \beta_{k_2}}{k_1+k_2+1} 
		~~~~ \text{with}~~
		\beta_k = \frac{N^{k+1/2}}{\sqrt{2}} \frac{k(2k+1)!!}{(k+1)!}~.
\end{align}
In what follows it will turn out to be convenient to write the odd double trace v.e.v. as
\begin{equation}
t_{2k_1+1,2k_2+1} = H_{k_1,k_2}\,\beta_{k_1} \beta_{k_2}
\end{equation}
where
\begin{align}
	\label{defHilb}
		H_{k_1,k_2} = \frac{1}{k_1 + k_2 + 1}
\end{align}
is closely related to the so-called Hilbert matrix.

\paragraph{Multitraces:} Let us now consider the v.e.v. of the multitraces. When there is an even trace
component, at large $N$ this factorizes as follows
\begin{equation}
	\label{factn1even}
		t_{2k_1,n_2,n_3,\ldots} = t_{2k_1}\, t_{n_2,n_3,\dots} ~,
\end{equation}
and the subleading contributions are suppressed by a factor of $1/N^2$.

When all components are odd (with a total even number of factors), the
recursion relations at large $N$ imply that the full result is obtained by pairing all
components in all possible ways, and replacing each pair by its expectation value (\ref{t2oddc}).
In other words, we have
\begin{align}
	\label{tracewick}
		t_{2k_1+1,2k_2+1,2k_3+1,2k_4+1,\ldots} 
		& = \cH(k_1,k_2,k_3,k_4,\ldots)\times \prod_i \beta_{k_i}~, 
\end{align}
where $\cH(k_1,k_2,k_3,k_4,\ldots)$ represents the total Wick contraction computed 
with the ``propagator'' $H_{k_i,k_j}$.  For instance, with 4 components we have
\begin{align}
	\label{wick4}
		\cH(k_1,k_2,k_3,k_4) = 
		H_{k_1,k_2} H_{k_3,k_4} + H_{k_1,k_3} H_{k_2,k_4} + H_{k_1,k_4} H_{k_2,k_3}~.
\end{align}
Similarly, with 6 components we have the sum of the 15 possible ways 
to make a complete contraction, and so on.

Repeatedly using the above results, one can obtain the large-$N$ expansion of all observables
in a quite explicit and detailed form.

\subsection{The $\cN=4$ SYM theory}
\label{subsec:resN4}
Let us begin by illustrating the procedure for the $\cN=4$ SYM theory. 
In this case the matrix model is purely Gaussian and thus, following the convention introduced above, we use a label $(0)$ to distinguish its observables. The results we present in this subsection
are well-known, but it is convenient to briefly review them before moving to the $\cN=2$ cases of our interest.
  
\paragraph{The BPS Wilson loop:} The v.e.v. of the Wilson loop (\ref{defWa}) in the $\cN=4$ theory
is
\begin{align}
	\label{resw0}
		w^{(0)}\,\equiv\, \frac{1}{N}\, \Big\langle \tr \exp\Big(\frac{g}{\sqrt{2}} \,a\Big)\Big\rangle_{(0)} 
		= \frac{1}{N} \sum_{k=0}^\infty \frac{1}{(2k)!}\,
		\frac{g^{2k}}{2^k}\, t_{2k}~.
\end{align} 
Using (\ref{t2kres1}), in the large-$N$ limit one gets:
\begin{align}
	\label{resw02}
		w^{(0)} = \sum_{k=0}^\infty  \frac{C_k}{(2k)!} \Big(\frac{\lambda}{4}\Big)^k
		= \frac{2}{\sqrt{\lambda}} I_1(\sqrt{\lambda})~,
\end{align}	
where $I_\ell$ is the modified Bessel function of the first kind. This well-known result was first obtained in \cite{Erickson:2000af} by resumming ladder diagrams in perturbation theory.
	
\paragraph{Single-trace operators and their mixing:}
Let us consider the single-trace operators 
\begin{equation}
\Omega_n(a) = \tr a^n
\label{Ona}
\end{equation}
with $n\geq 2$, and subtract their v.e.v., so as to work with a basis formed by the vevless operators
\begin{align}
	\label{defohat}
		\widehat{\Omega}_n(a) =
		\Omega_n(a) - t_n \,\mathbb{1}~.
\end{align} 
This is convenient because, after this redefinition, which is trivial when $n$ is odd, 
the even and odd cases are on the same footing. Indeed, in both cases their two-point functions coincide with the connected correlators
we introduced above:
\begin{align}
	\label{defChat}
		\widehat{C}^{\,{(0)}}_{nm} 
		\equiv \big\langle \widehat{\Omega}_n(a)\, \widehat{\Omega}_m(a)\big\rangle_{(0)}
		= t_{n,m} - t_n\,t_m  = t^{\mathrm{c}}_{n,m}
\end{align} 
whose expression in the large-$N$ limit is given in (\ref{t2evenc}) and (\ref{t2oddc}).
Our goal is to apply the Gram-Schmidt diagonalization procedure with respect to the pairing
(\ref{defChat}) and
find the normal-ordered version of the single-trace operators $\widehat{\Omega}_n(a)$,
which we denote by $O_n^{(0)}(a)$. 

The set of operators of dimension lower than $n$ comprises 
both $\widehat \Omega_p(a)$ with $p<n$, and multitrace operators 
with a total dimension less than $n$. However, in the large-$N$ limit these multitrace
operators do not play any role, because the factorization and the Wick-like expansion properties 
of the expectation values (\ref{factn1even}) and (\ref{tracewick}) imply that 
if one imposes the vanishing of the correlators between
$O_n^{(0)}(a)$ and all single-trace operators $\widehat \Omega_p(a)$ with $p<n$, then one automatically imposes also the vanishing of the correlators between $O_n^{(0)}(a)$
and all lower multitrace operators. Therefore, to obtain the explicit expression of 
$O_n^{(0)}(a)$ it is enough to run the normal-ordering procedure considering only the 
single-trace operators. In this way, (\ref{Omegs}) reduces to
\begin{align}
	\label{Omegs0}
		O^{(0)}_n(a) = \widehat{\Omega}_n(a) - \sum_{p,q< n}
		\widehat{C}_{np}^{\,{(0)}} \,\Big(\widehat{C}_{(n)}^{\,{(0)}~-1}\Big)^{pq} \,\widehat\Omega_q(a)~.  
\end{align}
Given the structure of the two-point functions (\ref{defChat}), the even
operators $O_{2k}^{(0)}(a)$ are expressed entirely in terms of even trace 
operators, while the expansion of the odd operators 
$O_{2k+1}^{(0)}(a)$ only contains odd traces. 
The first few cases are:
\begin{equation}
\label{firstome0}
\begin{aligned}	
		O^{(0)}_2(a)  & = \widehat{\Omega}_2(a)~, & ~~
		O^{(0)}_3(a)  & = \widehat{\Omega}_3(a)~,\\[1mm]
		O^{(0)}_4(a)  & = \widehat{\Omega}_4(a) - 2 N \widehat{\Omega}_2(a)~, & ~~
		O^{(0)}_5(a)  & = \widehat{\Omega}_5(a) - \frac 52 N \widehat{\Omega}_3(a)~,\\[1mm]
		O^{(0)}_6(a)  & =\widehat{\Omega}_6(a) - 3 N \widehat{\Omega}_4(a) + \frac 94 N^2 
		\widehat{\Omega}_2(a)~, &~~
		O^{(0)}_7(a)  & = \widehat{\Omega}_7(a) - \frac{7}{2} N \widehat{\Omega}_5(a) + 
		\frac{7}{2} N^2 \widehat{\Omega}_3(a)~.
\end{aligned}
\end{equation}
Actually, the result can be given in closed form for any $n$ as follows:
\begin{align}
	\label{omega0isC}
		O_n^{(0)}(a) = n \sum_{k=0}^{\lfloor \frac{n-1}{2}\rfloor}
		(-1)^k \frac{N^k\,(n-k-1)!}{2^k\,k!(n-2k)!} \,\widehat{\Omega}_{n-2k}(a)
\end{align}
with the understanding that $\widehat{\Omega}_1(a)=0$, which is true in the SU($N$) theory.

Using (\ref{Ona}), (\ref{defohat}) and (\ref{Omegs0}), we easily deduce that $O_n^{(0)}(a)$ 
is the trace of a degree $n$ monic polynomial $p_n(a)$, namely
\begin{align}
	\label{Oisp}
		O_n^{(0)}(a) = \tr p_n(a)~~~\mbox{with}~~~ p_n(a) = a^n + \ldots~.
\end{align} 
The coefficients of the expansion of $O_n^{(0)}(a)$, or equivalently of $p_n(a)$, are related to the ones appearing in the expansion of the Chebyshev polynomials of the first kind $T_n(x)$. Indeed, one has
\begin{align}
	\label{omega0isC2}
		p_n(a) = 2 \Big(\frac{N}{2}\Big)^{\frac{n}{2}}\, 
		T_n\Big(\frac{a}{\sqrt{2N}}\Big) + \delta_{n,2}\,\frac{N}{2}\,\mathbb{1} 
\end{align}
in agreement with the results of \cite{Rodriguez-Gomez:2016cem}. 

\paragraph{Two-point functions:}
The two-point functions of the normal-ordered operators are indeed diagonal:
\begin{align}
	\label{G0}
	G_{n,m}^{(0)} \equiv \big\langle O_n^{(0)}(a)\,O_m^{(0)}(a)\big\rangle_{(0)} =
		G_n^{(0)}\,\delta_{n,m}~
		\quad\text{with}~~~~
		G_n^{(0)}	= n\Big(\frac{N}{2}\Big)^n~,
\end{align}
as it can be proven by using (\ref{omega0isC}) and (\ref{defChat}). 
We note that the Gram-Schmidt procedure yields an expression of 
$G_{n}^{(0)}$ directly in terms of the elements of $\widehat{C}^{\,{(0)}}$:
\begin{align}
 	\label{ratiodetG0}
		G_{n}^{(0)} = \frac{\det \widehat C^{\,{(0)}}_{(n+1)}}{\det \widehat C^{\,{(0)}}_{(n)}}~.
\end{align}

\paragraph{One-point functions in presence of the Wilson loop:} Also the one-point functions
of the operators $O_n^{(0)}(a)$ with the Wilson loop $W_C$
can be easily computed in the large-$N$ limit. 
Indeed, using (\ref{omega0isC}) and expanding the Wilson loop operator (\ref{defWa}),
we find
\begin{equation}
w_n^{(0)} \equiv \big\langle O_n^{(0)}(a)\,
W_C\big\rangle_{(0)} \simeq \frac{n}{N}\sum_{\ell=0}^\infty\sum_{k=0}^{\lfloor \frac{n-1}{2}\rfloor}
(-1)^k \frac{N^k\,(n-k-1)!}{2^k\,k!(n-2k)!} \,\frac{1}{\ell!}\Big(\frac{\lambda}{2N}\Big)^{\!\frac{\ell}{2}}\,t^c_{n-2k,\ell}~.
\end{equation}
Inserting the large-$N$ behavior of the connected correlators given in (\ref{t2evenc}) 
and (\ref{t2oddc}) 
and retaining the leading contributions for $N\to\infty$, after simple algebra we can recast 
the above sum as an expansion in powers of $\lambda$ which can be resummed into a modified
Bessel function $I_n$. More precisely, we have
\begin{equation}
w_n^{(0)} = \frac{n}{N}\,\Big(\frac{N}{2}\Big)^{\frac{n}{2}} \,I_n(\sqrt{\lambda})~.
\label{wnN}
\end{equation}
This result was originally obtained in \cite{Semenoff:2001xp} by resumming rainbow diagrams
in the planar limit.

\subsection{The ABCDE theories}
\label{subsec:resABCDE}
The matrix model for the $\cN=2$ conformal theories of Table~\ref{tab:scft} contains
an interacting action $S_{\mathrm{int}}(a)$ given in (\ref{Saexp}), and the v.e.v. of the various observables are computed according to (\ref{vevmat}).

Using (\ref{trpa2k}), $S_{\mathrm{int}}(a)$ can be written as a sum of terms that are either quadratic or linear in the single-trace operators (\ref{Ona}). We find convenient to split this sum
into three parts as follows:
\begin{align}
	\label{Seos}
		S_{\mathrm{int}}(a) = S_{\mathrm{odd}}(a) + S_{\mathrm{even}}(a) + S_{\mathrm{s.t.}}(a)~. 		
\end{align}	
Here $S_{\mathrm{odd}}(a)$ contains odd double traces:
\begin{align}
	\label{Soddis}
		S_{\mathrm{odd}}(a) = (2-\nu)
		\sum_{p=1}^\infty (-1)^p\Big(\frac{\widehat\lambda}{N}\Big)^{\!p+1}
		\,\frac{\zeta(2p+1)}{p+1}~ \sum_{k=1}^{p-1}\binom{2p+2}{2k+1}\, 
		\Omega_{2k+1}(a)\, \Omega_{2p-2k+1}(a)~,
\end{align} 
$S_{\mathrm{even}}(a)$ contains even double traces:
\begin{align}
	\label{Sevenis}
		S_{\mathrm{even}}(a) = -\nu
		\sum_{p=1}^\infty (-1)^p\Big(\frac{\widehat\lambda}{N}\Big)^{\!p+1}
		\,\frac{\zeta(2p+1)}{p+1}~
		\sum_{k=1}^{p}\binom{2p+2}{2k}\, \Omega_{2k}(a) \,\Omega_{2p-2k+2}(a)~,
\end{align} 
while $S_{\mathrm{s.t.}}(a)$ contains single-trace operators, all even:
\begin{align}
	\label{Sstis}
		S_{\mathrm{s.t.}}(a) = 2(N_S - N_A) 
		\sum_{p=1}^\infty (-1)^p\Big(\frac{\widehat\lambda}{N}\Big)^{p+1}
		\,\frac{\zeta(2p+1)}{p+1}\,\big(2^{2p}-1\big) \Omega_{2p+2}(a)~.
\end{align}	
Finally, we have introduced the rescaled 't Hooft coupling
\begin{align}
	\label{lambdahatis}
		\widehat{\lambda} = \frac{\lambda}{8\pi^2}
\end{align}
to make the end results more compact. 

\paragraph{Single-trace operators and their mixing:} To obtain the normal-ordered operators
$O_n(a)$ we have to repeat the Gram-Schmidt procedure and diagonalize the matrix of the two-point functions of the single-trace operators $\Omega_n(a)$ computed
in the interacting matrix model. To take advantage of the calculations already performed, we 
proceed in two steps and start from the operators $O_n^{(0)}(a)$ introduced in (\ref{Omegs0}) that realize the normal ordering in the $\cN=4$ theory. They can be considered the tree-level approximation of those in the $\cN=2$ theories. The two-point functions of these operators are
\begin{align}
	\label{2pf0n0}
		\big\langle O_n^{(0)}(a) \, O_m^{(0)}(a)\big\rangle & = 
		\frac{\big\langle O_n^{(0)}(a)\, O_m^{(0)}(a) \,
		\rme^{-S_{\mathrm{int}}(a)}\big\rangle_{(0)}\phantom{\Big|}}{\big\langle
		\rme^{-S_{\mathrm{int}}(a)}\big\rangle_{(0)}\phantom{\Big|}}
		= G_{n,m}^{(0)} + G^{(1)}_{n,m} + G^{(2)}_{n,m} + \ldots~,
\end{align}
where $G^{(k)}_{n,m}$ is the term of order $k$ in $S_{\mathrm{int}}$.
In particular, at the first order we have
\begin{align}
	\label{cG1is}		
		G^{(1)}_{n,m} 
		= - \big\langle O_n^{(0)}(a)\, O_m^{(0)}(a)\, S_{\mathrm{int}}(a)\big\rangle_{(0)}
		+\big\langle O_n^{(0)}(a)\, O_m^{(0)}(a)\big\rangle_{(0)} \,\big\langle
		S_{\mathrm{int}}(a)\big\rangle_{(0)}~.
\end{align}    
The corrections terms $G^{(k)}_{n,m}$ make the two-point functions (\ref{2pf0n0}) 
non-diagonal.  We have therefore to rerun the Gram-Schmidt procedure and redefine the
operators. At large $N$ several important simplifications occur. We illustrate them
by considering the $N$ dependence of the various terms contributing to $G_{n,m}^{(1)}$, but
these arguments can be readily extended also to the higher correction terms $G_{n,m}^{(k)}$ with $k>1$.

Using the general formula (\ref{omega0isC2}), the product $O_n^{(0)}(a)\, O_m^{(0)}(a)$
decomposes into a sum of terms which contain a couple of single-trace operators 
$\Omega_{r}(a)\,\Omega_{s}(a)$. The tree-level contribution of any such term to the two-point function
$G_{n,m}$ is proportional to $t_{r,s}$. Its contribution to the first correction 
$G_{n,m}^{(1)}$ depends on which part of the interacting action one considers.
The single-trace part $S_{\mathrm{s.t.}}$ given in (\ref{Sstis}) corresponds to an insertion of
$\Omega_{2p+2}(a)$ accompanied by a factor of $1/N^{p+1}$, so that 
the contribution of $\Omega_{r}(a)\,\Omega_{s}(a)$ is proportional to
\begin{align}
	\label{stcont}
		\frac{1}{N^{p+1}}\,\big(t_{r,s,2p+2} - t_{r,s} \,t_{2p+2}\big)~
		\propto~ \frac{1}{N}\, t_{r,s}
\end{align}
where the second step follows from (\ref{factn1even}) and (\ref{t2kres1}). This correction
is therefore subleading in $N$ with respect to the tree-level result $t_{r,s}$. This fact means that
all contributions arising from the single-trace part of the interacting action can be ignored in the
large-$N$ limit. Since $S_{\mathrm{s.t.}}$ is the only part of $S_{\mathrm{int}}$ that does 
not depend on the parameter $\nu$, it follows that in the large-$N$ limit the two-point 
correlators will only depend on $\nu$ and not on the more specific matter content of the $\cN=2$ theory. Thus, $G_{n,m}$ will be equal for the $\mathbf{B}$ and $\mathbf{C}$ models, 
and for the $\mathbf{D}$ and $\mathbf{E}$ models.

Let us now consider the corrections coming from the even double-trace part $S_{\mathrm{even}}$
of the interaction action given in (\ref{Sevenis}). In this case, the interaction produces an insertion of $\Omega_{2k}(a)\,\Omega_{2p-2k+2}(a)$ accompanied by a factor of $1/N^{p+1}$, so that the typical term goes like 
\begin{align}
	\label{sevencont}
		\frac{1}{N^{p+1}}\,\big( t_{r,s,2k,2p-2k+2} - t_{r,s} \,t_{2k,2p-2k+2}\big)~
		\propto~ t_{r,s}~.
\end{align}
Again we have used again the factorization property (\ref{factn1even}) and the expression of the one-point function (\ref{t2kres1}). We see that now the correction scales in the large-$N$ limit just like the tree level result. So the even part $S_{\mathrm{even}}$, which is proportional to $\nu$, will always contribute to the corrections of the two-point functions.

Finally, we consider the odd double-trace part $S_{\mathrm{odd}}$ defined in (\ref{Soddis}). Here
we have to distinguish two cases: when the operators $\Omega_r(a)\,\Omega_s(a)$ are both even and when they are both odd. In the first case, which occurs in the two-point functions $G_{n,m}$
of two even operators, the typical contribution due to $S_{\mathrm{odd}}$ has the following 
behavior at large $N$
\begin{align}
	\label{sevenodd}
		\frac{1}{N^{p+1}}\,\big(t_{r,s,2k+1,2p-2k+1} - t_{r,s} t_{2k+1,2p-2k+1}\big)~
		\propto ~\frac{1}{N^2}\, t_{r,s}
\end{align}
for $r$ and $s$ even. This result follows again upon using (\ref{factn1even}) and  (\ref{t2oddc}).
Thus, being subleading with respect to the tree-level term, the odd part $S_{\mathrm{odd}}$ 
has no effect on the correlators of two even operators in the large-$N$ limit. Instead, it corrects 
the correlators of two odd operators. Indeed, in
this case the typical contribution due to $S_{\mathrm{odd}}$ is of the type
\begin{align}
	\label{soddodd}
		\frac{1}{N^{p+1}}\,\big(t_{r,s,2k+1,2p-2k+1} - t_{r,s} t_{2k+1,2p-2k+1}\big)~
		\propto ~ t_{r,s}
\end{align}
for $r$ and $s$ odd. Here we have taken into account the structure of odd multitraces 
described in (\ref{tracewick}) and (\ref{wick4}). This correction has the same $N$-dependence of the
tree-level term and survives in the large-$N$ limit.

To summarize, we have shown that the correlators of two even operators in the large-$N$ limit
only receive corrections from $S_{\mathrm{even}}$, while those of two odd operators are corrected
both by $S_{\mathrm{even}}$ and by $S_{\mathrm{odd}}$. This means that the $\cN=2$ correlators 
of two even operators differ from the $\cN=4$ ones by terms which are proportional to $\nu$.
In particular, for the $\mathbf{D}$ and $\mathbf{E}$ theories, for which $\nu=0$, this difference
vanishes. Thus, in the sector of the even operators, these two $\cN=2$ models are indistinguishable
from the $\cN=4$ SYM theory in the planar limit. On the contrary, the correlators of odd operators
have corrections proportional to $(2-\nu)$ arising from $S_{\mathrm{odd}}$ and corrections
proportional to $\nu$ arising from $S_{\mathrm{even}}$, and these are non-trivial even for
the $\mathbf{D}$ and $\mathbf{E}$ theories.

\paragraph{Some explicit examples:}
We have at our disposal all elements to carry out explicitly the Gram-Schmidt procedure starting from the operators $O^{(0)}_n(a)$ and their correlation matrix (\ref{2pf0n0}), and find the 
large-$N$ expression of the normal-ordered operators $O_n(a)$ and of their two-point functions $G_{n,m}$. 
On very general grounds, applying the analogue of (\ref{Omegs0}), we find
\begin{align}
	\label{defDeltaOme}
	  	O_n(a) = O^{(0)}_n(a) + \Delta O_n(a)
\end{align}
while, applying the analogue of (\ref{ratiodetG0}), we obtain the diagonal two-point functions
\begin{align}
	\label{defGN2}
		G_{n,m}\equiv \big\langle O_n(a)\, O_m(a)\big\rangle 
		= G_n \, \delta_{n,m}~,
\end{align}
which we parametrize as follows:
\begin{align}
	\label{Gnis}  
		G_n = n\Big(\frac{N}{2}\Big)^n\, \gamma_n~.
\end{align}
Here we have factorized the $\cN=4$ result (\ref{G0}), so that $\gamma_n$, which is a function
of the 't Hooft coupling, reduces to 1 in the limit $\widehat{\lambda}\to 0$.

This procedure is entirely algorithmic and it is easy to reach quite high orders in the coupling and in transcendentality. Here we report the explicit results for the lowest dimensional operators, showing only the the lowest terms in their expansion to avoid excessive clutter. 
\begin{itemize}
\item[$\bullet$]
At dimension 2 the operator $O_2^{(0)}(a)$ can only mix with the identity and we find that the
corresponding correction is
		\begin{align}
			\label{Ome2is}
				\Delta O_2(a) = \frac{N^2\,\nu}{2}\,
				\Big[3\,\zeta(3)\,\widehat{\lambda}^{\,2}
				- 15\,\zeta(5)\,\widehat{\lambda}^{\,3} + \big(70 \,\zeta(7)-18\,\zeta(3)^2\,\nu
				\big)\widehat{\lambda}^{\,4}
				+ \ldots \Big]
				\, \mathbb{1}~,
		\end{align} 
where the ellipses stand for terms of higher order in $\widehat{\lambda}$.
Computing the two-point function of $O_2(a)$, we find that the correction factor 
$\gamma_2$ is 	
\begin{align}
			\label{gamma2is}
				\gamma_2 = 1 - \nu\,
				\Big[9\,\zeta(3)\,\widehat{\lambda}^{\,2}  - 60\,\zeta(5)\,\widehat{\lambda}^{\,3}
				+ \big(350 \,\zeta(7) -90\,\zeta(3)^2\,\nu
				\big)\widehat{\lambda}^{\,4} 
				+ \ldots \Big]~.
		\end{align}
		
\item[$\bullet$]
The operator $O_3^{(0)}(a)$ of dimension 3 receives no corrections 
because for SU$(N)$ there is no single-trace operator of lower dimension with which it can mix. 
Thus
\begin{equation}
\Delta O_3(a)=0~.
\label{Ome3is}
\end{equation}
The corresponding two-point coefficient $\gamma_3$ is
\begin{equation}
\label{gamma3is}
\gamma_3= 1 - \bigg[9\,\zeta(3)\,\nu\widehat{\lambda}^{\,2} -
				5\,\zeta(5) (13\nu -2)\widehat{\lambda}^{\,3} 
				+\Big( \frac{105}{4}\,\zeta(7) (15\nu-4)-81\,\zeta(3)^2\,\nu^2\Big)
				\widehat{\lambda}^{\,4}
				+ \ldots \bigg]~.
\end{equation}

\item[$\bullet$]
At dimension 4 the operator $O_4^{(0)}(a)$ can mix with $O_2^{(0)}(a)$
and with the identity. Computing the corresponding mixing, we find
\begin{equation}
\begin{aligned}
\Delta O_4(a)&=2N \nu
\bigg[3\,\zeta(3)\,\widehat{\lambda}^{\,2}
				- \frac{85}{4}\,\zeta(5)\,\widehat{\lambda}^{\,3} +
				\Big( \frac{511}{4} \,\zeta(7)-18\,\zeta(3)^2\,\nu
				\Big)\widehat{\lambda}^{\,4}
				+ \ldots \bigg]\,O_2^{(0)}(a)\\
				&\quad~-\frac{N^3\,\nu}{2}\bigg[10\,\zeta(5)\,\widehat{\lambda}^{\,3} 
				-\Big(\frac{77}{4} \,\zeta(7)+9\zeta(3)^2\,\nu\Big)\widehat{\lambda}^{\,4}
				+ \ldots \bigg]\,\mathbb{1}~.
\end{aligned}
\label{Ome4is}
\end{equation}
The two-point function is captured by
\begin{align}
	\label{gamma4is}
		\gamma_4= 1 - \nu
		\,\bigg[12\,\zeta(3)\,\widehat{\lambda}^{\,2}  - 80\,\zeta(5)\,\widehat{\lambda}^{\,3}
		+ \Big(\frac{1855}{4} \,\zeta(7)-126\,\zeta(3)^2\,\nu\Big)\widehat{\lambda}^{\,4} 
		+ \ldots \bigg]~.
\end{align}

\item[$\bullet$]
The operator $O_5(a)$ of dimension 5 can only mix with 
$O_3^{(0)}(a)$ and the mixing term is given by
\begin{equation}
\begin{aligned}
\Delta O_5(a)&=\frac{5N}{2}\bigg[
                3\,\zeta(3)\,\nu\widehat{\lambda}^{\,2} -
				20\,\zeta(5) \,\nu \widehat{\lambda}^{\,3} 
				+\Big( \frac{7}{2}\,\zeta(7) (33\nu-1)-18\,\zeta(3)^2\,\nu\Big)\widehat{\lambda}^{\,4}
				\!+\! \ldots \bigg] O_3^{(0)}(a)
\end{aligned}
				\label{Ome5is}
\end{equation}
while the two-point function coefficient is\,%
		\footnote{We report this result to a higher order in $\widehat{\lambda}$ with respect to the previous ones to exhibit the fact that it does not vanish at $\nu=0$.} 
\begin{equation}
\begin{aligned}
\gamma_5&=1 - \bigg[15\,\zeta(3)\,\nu \widehat{\lambda}^{\,2} -
				100\,\zeta(5) \,\nu \widehat{\lambda}^{\,3} 
				+ \Big(\frac{2275}{4}\,\zeta(7)\, \nu -180\,\zeta(3)^2\,\nu^2\Big)
				\widehat{\lambda}^{\,4}\\
&\qquad\quad	-  \Big(\frac{1155}{8}\,\zeta(9) (133\nu-4)-2625\,\zeta(3)\zeta(5)\,\nu^2\Big)\widehat{\lambda}^{\,5} 
				+ \ldots \bigg]~.
\end{aligned}
\label{gamma5is}
\end{equation}
\end{itemize}
All these results explicitly show the pattern discussed above, namely that the even operators 
and their two-point functions are not corrected with respect to the $\cN=4$ theory when 
$\nu=0$. On the contrary the odd operators and their two-point functions are different from the
corresponding ones in the $\cN=4$ theory even when $\nu=0$. In particular we point out that
in the $\mathbf{D}$ and $\mathbf{E}$ theories, the first correction to $\gamma_3$ is
proportional to $\zeta(5)\,\widehat{\lambda}^{\,3}$, while the first correction to 
$\gamma_5$ is proportional to $\zeta(9)\,\widehat{\lambda}^{\,5}$. 
In Section~\ref{sec:diagrams} we will give a diagrammatic explanation of this fact and of 
its generalization to the two-point functions $\gamma_{2k+1}$ with $k>2$.

\paragraph{The BPS Wilson loop:} 
We now consider the Wilson loop operator (\ref{defWa}). Its v.e.v. 
in the $\cN=2$ theories is given by
\begin{align}
w&\equiv \big\langle W_C\big\rangle=\frac{1}{N}\,\sum_{k=0}^\infty
\frac{1}{k!}\Big(\frac{\lambda}{2N}\Big)^{\!\frac{k}{2}}~
\frac{\displaystyle{\big\langle \tr a^k\,\rme^{- S_\mathrm{int}(a)}\big\rangle_{(0)}}}{\displaystyle{\big\langle \rme^{- S_\mathrm{int}(a)}\big\rangle_{(0)}}}~.
\label{WCN2}
\end{align}
Expanding the exponentials and using the explicit form of the interacting action
(\ref{Seos}), one gets a non-vanishing contribution only when $k$ is even. 
Thus, the difference with respect to the $\cN=4$ result (\ref{resw0}) can be written as
\begin{equation}
\begin{aligned}
\Delta w &\equiv
w -w^{(0)}
=-\frac{1}{N}\,\sum_{k=0}^\infty
\frac{1}{(2k)!}\Big(\frac{\lambda}{2N}\Big)^{\!k}\,\Big[\big\langle \tr a^{2k}\,S_\mathrm{int}(a)\big\rangle_{(0)}
- t_{2k}\,\big\langle S_\mathrm{int}(a) \big\rangle_{(0)}\Big]+\cO\big(S^2_{\mathrm{int}}\big)~.
\end{aligned}
\label{DeltaWC}
\end{equation}
This quantity greatly simplifies in the large-$N$ limit. In fact, using the same arguments explained above, one can show that the single-trace part of the interacting action (\ref{Sstis}) does not contribute at leading order and that the same thing happens for $S_{\mathrm{odd}}$.
Thus, one is left only with the contributions arising from the even piece of the action, $S_{\mathrm{even}}$, which can be evaluated using the large-$N$
factorization property (\ref{factn1even}) together with (\ref{t2evenc}). Collecting all terms contributing to a given $\zeta$-value, one finds a series in $\lambda$ which can be resummed into
a combination of Bessel functions $I_n$. Explicitly, the very first few terms are\,%
\footnote{The term proportional to $\zeta(3)$ was already obtained in \cite{Sysoeva:2017fhr,Billo:2018oog} for the $\mathbf{A}$ theory, corresponding to $\nu=1$.}
\begin{equation}
\begin{aligned}
\Delta w&= -\nu\,\bigg[3\,\zeta(3) \bI_2\,\widehat{\lambda}^{\,2}-
10\,\zeta(5)\big(2\bI_2-3\bI_3\big)\,\widehat{\lambda}^{\,3} 
\\&\qquad\qquad
+\Big(\frac{35}{4}\,\zeta(7)\big(13\bI_2-36\bI_3+48\bI_4\big)
-\frac{9}{4}\,\zeta(3)^2\big(\bI_1+4\bI_2\big)\Big)
\widehat{\lambda}^{\,4}+\ldots\bigg]
\end{aligned}
\label{DeltaWCexpl}
\end{equation}
where for brevity we have defined
\begin{equation}
\bI_{n}\equiv \big(\sqrt{\lambda}\big)^{2-n}\,I_{n}(\sqrt\lambda)~.
\label{bI}
\end{equation}
Terms with higher powers of $\widehat{\lambda}$ for which the ellipses in (\ref{DeltaWCexpl}) stand, 
can be systematically computed without any difficulty.

\paragraph{One-point functions in presence of the Wilson loop:}
In presence of the Wilson loop $W_C$, 
the normal-ordered operators $O_n(a)$ have a non-trivial one-point function 
which is given by
\begin{align}
w_n&\equiv \big\langle O_n(a)\,W_C\big\rangle=\frac{1}{N}\,\sum_{k=0}^\infty
\frac{1}{k!}\Big(\frac{\lambda}{2N}\Big)^{\!\frac{k}{2}}~
\frac{\displaystyle{\big\langle O_n(a)\,\tr a^k\,\rme^{- S_\mathrm{int}(a)}\big\rangle_{(0)}}}{\displaystyle{\big\langle \rme^{- S_\mathrm{int}(a)}\big\rangle_{(0)}}}~.
\label{1pointWL}
\end{align}
To compute the difference of these quantities with respect to the $\cN=4$ theory, we have to expand the exponentials and also to take into account that the operators bear a dependence on the interacting action because of the normal ordering that we discussed in the previous part of this
section. We then find
\begin{align}
\Delta w_n&\equiv
w_n-w_n^{(0)}
=-\frac{1}{N}\,\sum_{k=0}^\infty
\frac{1}{k!}\Big(\frac{\lambda}{2N}\Big)^{\!\frac{k}{2}}\,\Big[\big\langle 
O_n^{(0)}(a)\,\tr a^{k}\,S_\mathrm{int}(a)\big\rangle_{(0)}
-\big\langle O_n^{(0)}(a)\,\tr a^{k} \big\rangle_{(0)}\,\big\langle S_\mathrm{int}(a) \big\rangle_{(0)}
\notag\\[1mm]
&\hspace{6cm}-\big\langle \Delta O_n(a) \,\tr a^k\big\rangle_{(0)}\Big]
+\cO\big(S_{\mathrm{int}}^2\big)~.
\label{DeltanWC}
\end{align}
When we take the large-$N$ limit, drastic simplification occur in these quantities and the results
can be compactly written in terms of the rescaled Bessel functions (\ref{bI}). For example, for
$n=2,3$ we find 
\begin{align}
\Delta w_2&= -\frac{\nu}{2}\,\bigg[
3\,\zeta(3)\,(\bI_1+2\bI_2)\widehat{\lambda}^{\,2}
-10\,\zeta(5)\,(2\bI_1+5\bI_2-6\bI_3)\,\widehat{\lambda}^{\,3}  \label{DW2is}\\[1mm]
&\qquad~~+\Big(\frac{35}{4}\,\zeta(7)\,(13\bI_1+42\bI_2-96\bI_3+96\bI_4)
-\frac{9\,\nu}{4}\,\zeta(3)^2\,(\bI_0+12\bI_1+24\bI_2)\Big)
\widehat{\lambda}^{\,4}+\ldots\bigg]~,\notag \\[2mm] 
\Delta w_3&= -\frac{3\pi\sqrt{N}}{2} \,\bigg[
3\,\zeta(3)\,\nu\bI_2\,\widehat{\lambda}^{\,\frac{5}{2}}
-10\,\zeta(5)\big(2\,\nu\bI_2-(2-\nu)\bI_3\big)\,\widehat{\lambda}^{\,\frac{7}{2}} \label{DW3is} \\[1mm]
&\qquad~~+\Big(\frac{35}{4}\,\zeta(7)\big(13\nu\bI_2-(2-\nu)(14\bI_3-16\bI_4)\big)
+\frac{9\,\nu^2}{4}\,\zeta(3)^2\,(\bI_1+2\bI_2)
\Big)\widehat{\lambda}^{\,\frac{9}{2}}+\ldots\bigg]~.\notag 
\end{align}
Similar expressions for higher values of $n$, as well as the contributions at
higher orders in $\widehat{\lambda}$, can be obtained without any problem since
the whole procedure is purely algebraic.

\section{Large-$N$ limit from the eigenvalue distribution}
\label{sec:4}
We now discuss the Cartan algebra approach based on the integration over the matrix model eigenvalues in the large-$N$ limit.

A specific application of this approach to the $\cN=2$ \textbf{ABCDE} theories 
has already been presented in \cite{Fiol:2015mrp} 
(see also \cite{Baggio:2016skg}), where it is shown that the ratio $\nu$ defined in 
(\ref{defnu1}) is the unique relevant parameter at large $N$. 
The study in \cite{Fiol:2015mrp} focused mainly on the Wilson loop v.e.v. and its eventual extrapolation at strong coupling. Here, instead, we discuss how this method can be applied to 
the calculation of the two-point functions of chiral primaries 
and of their one-point functions in presence of a Wilson loop. 
In doing so, we also provide new results about the large-$N$ mixing of single-trace operators 
in terms of $\cN=2$ deformed orthogonal polynomials, setting in 
a broader perspective the findings of \cite{Rodriguez-Gomez:2016cem}. Our analysis 
also leads to an alternative and more compact derivation of the results 
obtained by the full Lie algebra methods described in the previous section which, 
at least in principle, may be a convenient starting point for a non-perturbative investigation.

\subsection{The large-$N$ universal integral equation}

For a generic model in the \textbf{ABCDE} series, the large-$N$ saddle-point equation
for the matrix model eigenvalues $m_u$ is obtained from the effective action (\ref{Sexpl}) 
and reads as follows
\begin{equation}
\begin{aligned}
\label{1.2}
0 &= \frac{8\pi^{2}}{\lambda}m_{u}-\frac{N_{F}}{2N}\,K(m_{u})
-\frac{N_{S}}{N}\,K(2m_{u})-\frac{N_{S}+N_{A}}{2N}\,\sum_{v\neq u}K(m_{u}+m_{u})\\
&~~~-\frac{1}{N}\sum_{v\neq u}\Big[\frac{1}{m_{u}-m_{v}}-K(m_{u}-m_{v})\Big]~.
\end{aligned}
\end{equation}
Here the function $K(x)$ is defined by
\begin{align}
K(x) &=  2\,(1+\gamma_\text{E})\,x-\frac{d}{dx}\log H(\ii x) = 
x\,\big[\psi(1+\ii x)+\psi(1-\ii x)+2\gamma_\text{E}\big]
\label{Kexp}
\end{align}
where $\psi(x)$ is the digamma function.
Notice that in $\cN=2$ superconformal theories, the linear part of $K(x)$ 
drops out in the saddle-point equation and thus it may be conveniently subtracted from the start\,%
\footnote{This property is related to the UV finiteness of the theories, as explained in \cite{Passerini:2011fe}.}. Moreover, exploiting the expansion (\ref{logHexp}), for small $x$ we have
\begin{align}
K(x) = -2\,\sum_{p=1}^{\infty}(-1)^{p}\,\zeta(2p+1)\,x^{2p+1}~.
\label{Kexp1}
\end{align}

The large-$N$ limit of (\ref{1.2}) is captured by an integral equation for the 
continuum limit of the discrete density 
\begin{equation}
\rho(m) = \frac{1}{N}\sum_{u=1}^{N}\delta\big(m-m_{u}\big)~.
\end{equation}
Assuming that for $N\to\infty$ the eigenvalues $m_{u}$
condense on a single\,%
\footnote{This is called a one-cut solution in resolvent language. This one-cut assumption is definitely correct in the perturbative framework, and it is
expected to hold at strong coupling too.}
segment $[\mu_{-},\mu_{+}]\subset \mathbb{R}$,
one finds that $\rho$ and $\mu_{\pm}$ are determined by the following
equation \cite{Fiol:2015mrp}:
\begin{align}
\label{1.4}
\int_{\mu_{-}}^{\mu_{+}} \!dy\,\rho(y)\,\Big[\frac{1}{x-y}-K(x-y)+(1-\nu)\,K(x+y)\Big] 
&= \frac{8\pi^{2}}{\lambda}\,x-\nu\,K(x)~,
\end{align}
where we have understood the prescription for taking the Cauchy principal value of the integral
and have introduced the parameter $\nu$ as in (\ref{defnu}), together with the
normalization condition
\begin{align}
\int_{\mu_{-}}^{\mu_{+}} \!dx\, \rho(x) &= 1~.
\end{align}
{From} the solution to (\ref{1.4}) we can readily compute the v.e.v. of single-trace operators. 
For example, the integral representation of the half-BPS Wilson loop v.e.v. is
\begin{equation}
\label{exaW}
\big\langle W_{C}\big\rangle = \int_{\mu_{-}}^{\mu_{+}}\!dx\,\rho(x)\,\rme^{2\pi x}~.
\end{equation}

The right hand side of (\ref{1.4}) is odd under $x\to -x$ and thus 
it is consistent to assume a symmetric density $\rho(x) = \rho(-x)$ supported on a symmetric 
cut  with $\mu_{+}=-\mu_{-}\,\equiv\,\mu$. 
With these assumptions,  we may replace $K(x+y)$ with $K(x-y)$ under integration and 
rewrite (\ref{1.4}) in the much simpler form 
\begin{equation}
\label{1.7}
\int_{-\mu}^{+\mu}
\!dy \,\rho(y)\,\Big[\frac{1}{x-y}-\nu\,K(x-y)\Big] = \frac{8\pi^{2}}{\lambda}\,x-\nu\,K(x)~.
\end{equation}
This implies that, whenever (\ref{1.7}) may be used, any $\nu=0$ model gives the same results as 
the $\cN=4$ SYM theory where the exact density is the Wigner semi-circle distribution
\begin{equation}
\label{muis}
\rho(x) = \frac{2}{\pi\mu^{2}}\,\sqrt{\mu^{2}-x^{2}}~~~\mbox{with}~~~ \mu = \frac{\sqrt\lambda}{2\pi}~.
\end{equation}
Using this distribution in (\ref{exaW}), one finds
\begin{equation}
\big\langle W_{C}\big\rangle = \frac{2}{\pi\mu^{2}}\,\int_{-\mu}^{+\mu} \!dx\,
\sqrt{\mu^{2}-x^{2}}\,\rme^{2\pi x} = \frac{2}{\sqrt\lambda}\,I_{1}(\sqrt\lambda)~,
\end{equation}
in agreement with (\ref{resw02}). Thus, for the $\nu=0$ theories we have $\Delta w=0$ in agreement
with (\ref{DeltaWCexpl}).

However, the symmetric one-cut assumption is too restrictive. 
Indeed, to compute observables that are more general than (\ref{exaW}), like for instance two-point functions or one-point functions in presence of a Wilson loop, 
we need to extend the action by suitable sources coupled to the operators 
that are inserted in the correlators. This leads to asymmetries in the cut when one considers 
``odd'' operators, {\emph{i.e.}} operators involving traces of odd powers of the matrix 
model  variable. Moreover, intermediate calculations require to consider the unrestricted 
integral equation (\ref{1.4}) that does not reproduce the $\cN=4$ results for $\nu=0$. 
In other words, one cannot expect that such ``odd'' observables in the $\nu=0$ theories are equal to those in the $\cN=4$ SYM theory.
Several examples of this claim will be illustrated and discussed later.

\subsection{Single-trace mixing at large $N$ and two-point functions}

As we remarked in the previous section, an important bottleneck is the operator mixing 
that seems to require an \textit{ad hoc} treatment depending on the operators under study. 
We now discuss how to deal in general with the mixing in the single-trace sector
starting from the integral equation (\ref{1.4}). Then, as an application, we 
recompute the two-point functions and the one-point functions in presence of the half-BPS Wilson loop. As a technical point, note that we shall work in the U$(N)$ matrix model which is expected to give the same results as the SU($N$) model in the large-$N$ limit, since the difference between enforcing or not the tracelessness condition turns out to be subleading at large $N$ \cite{Rossi:1996hs}; see also Appendix \ref{app:UN}. 
Indeed, we will perfectly reproduce the previous results obtained in the SU$(N)$ matrix model following the full Lie algebra approach.

To  evaluate the two-point functions of single-trace operators and eventually impose
the orthogonality conditions, we need to add a set of 
source terms to the effective action of the matrix model. We thus modify the action $S$ given in (\ref{Sexpl}) according to
\begin{align}
\label{Ssource}
S \to S+ N\,\sum_{n>2}\sigma_{n}\tr P_{n}~,
\end{align} 
where $\sigma_n$ is the source term for a monic polynomial
$P_{n}$ in the matrix $M$ of degree $n$:
\begin{align}
\label{PnM}
P_n = M^n + \ldots~.
\end{align}
These polynomials are determined by imposing the orthogonality condition
\begin{equation}
\label{ortmn}
\langle n, m\rangle = \widetilde G_n\, \delta_{n,m}~,
\end{equation}
with respect to the pairing
\begin{equation}
\label{pairingdef}
\langle n, m\rangle \equiv \langle \tr P_{n}\tr P_{m}\rangle -\langle \tr P_{n}\rangle\,\langle \tr P_{m} \rangle~. 
\end{equation}
In presence of the sources $\sigma_n$, the eigenvalue distribution satisfies a modified integral equation, that we will consider below.

Since the normal-ordered operators have to be orthogonal with respect to the identity, $P_{n}$ must also satisfy $\langle \tr P_{n}\rangle = 0$. This can be left to the end. This is because the
connected two-point function in the above pairing is unchanged 
if we replace $\tr P_{n}$ by $\tr P_{n} + c_{n}$.
As a result we can impose \eqref{ortmn} and determine $P_{n}$. The normal-ordered operator 
is then $\tr P_{n} - \langle \tr P_{n}\rangle$.

\subsubsection{The $\cN=4$ SYM theory}

Let us begin with the simplest case of the $\cN=4$ SYM theory. In previous sections, 
we have adopted the convention that quantities in the $\cN=4$ case are distinguished by a $(0)$ super/sub-script. Here, however, to avoid excessive clutter, we suppress this index where 
there is no risk of confusion.

We start by considering even sources $\sigma_{n}$ with $n\in 2\mathbb N$. They
modify the integral equation (\ref{1.7}) with $\nu=0$ as follows:
\begin{equation}
\label{2.1}
\int_{-\mu(\sigma)}^{+\mu(\sigma)}\!dy\, \frac{\rho(y;\sigma)}{x-y} = 
\frac{8\pi^{2}}{\lambda}\,x+\frac{1}{2}\sum_{n\ge 4}\sigma_{n}\,P_{n}'(x)~.
\end{equation}
Both the density and the cut edge now depend on $\sigma_{n}$.

In terms of this modified density, the pairing is expressed as\,%
\footnote{Note that the density integral provides a trace divided by a factor of $N$ which is compensated by differentiation with respect to $\sigma$ which inserts $N$ times the trace of the associated field.}
\begin{equation}
\label{2.2}
\langle n, m\rangle = 
-\frac{\partial}{\partial\sigma_{m}}\int_{-\mu(\sigma)}^{+\mu(\sigma)}\!dx\,\rho(x; \sigma)
\,P_{n}(x)\bigg|_{\sigma=0}~.
\end{equation}
Taking into account that the density vanishes at the source-dependent edges of the cut, this becomes
\begin{equation}
\label{2.3}
\langle n, m\rangle 
= -\int_{-\mu_0}^{+\mu_0}\!dx\,\rho'_m(x)\,P_{n}(x)
\end{equation}
where $\mu_{0}=\mu(0)$ and
\begin{align}	
\label{defrhop}
	\rho'_m(x) = 
	\left. \frac{\partial \rho(x, \sigma)}{\partial \sigma_{m}}
	\right|_{\sigma=0}~. 
\end{align}
To determine $\rho'_m$, we take a derivative of (\ref{2.1}) 
with respect to $\sigma_{m}$, obtaining
\begin{equation}
\label{2.4}
\int_{-\mu_{0}}^{+\mu_{0}} \!dy \,\frac{\rho'_m(y)}{x-y}
= \frac{1}{2}P_{m}'(x)~.
\end{equation}
Since the differentiated density is expected to be unbounded at the cut edges, 
the general solution to (\ref{2.4}) is \cite{tricomi1985integral}
\begin{equation}
\label{2.5}
\begin{aligned}
\rho'_m(x) &= \frac{C_{m}}{\sqrt{\mu_{0}^{2}-x^{2}}}
+\frac{1}{2\pi^{2}\sqrt{\mu_{0}^{2}-x^{2}}}\int_{-\mu_{0}}^{+\mu_{0}}\!
 dy\,\frac{\sqrt{\mu_{0}^{2}-y^{2}}}{y-x}\,P_{m}'(y)~,
 \\[1mm]
 &=\frac{C_{m}}{\sqrt{\mu_{0}^{2}-x^{2}}}-\frac{m}{2\pi\sqrt{\mu_{0}^{2}-x^{2}}}\,P_{m}(x)
\end{aligned}
\end{equation}
where $C_{m}$ is an arbitrary constant\,%
\footnote{Indeed, $(\mu_{0}^{2}-x^{2})^{-1/2}$ is a zero-mode of the Cauchy integral kernel
$1/(x-y)$ in (\ref{2.4}).}.
Inserting (\ref{2.5}) into (\ref{2.3}), we get 
\begin{equation}
\label{2.6}
\langle n,m \rangle=-C_m\int_{-\mu_0}^{+\mu_0}\!dx\,\frac{P_n(x)}{\sqrt{\mu_0^2-x^2}}
-\frac{1}{2\pi^{2}}\int_{-\mu_{0}}^{+\mu_{0}}\!dx\,\frac{P_{n}(x)}
{\sqrt{\mu_{0}^{2}-x^{2}}}\,\int_{-\mu_{0}}^{+\mu_{0}}
\!dy\,\frac{\sqrt{\mu_{0}^{2}-y^{2}}}{y-x}\,P_{m}'(y)~.
\end{equation}
The orthogonality condition $\langle n, m \rangle \propto \delta_{n,m}$ 
is realized if we take $P_{n}$ to be proportional to the Chebyshev polynomial of the second kind 
$T_{n}$. Indeed, in such a case 
the contribution proportional to $C_{m}$ is identically zero (so that we can safely set $C_m=0$)
and the double integral in (\ref{2.6}) vanishes for $n\not=m$.

To see this, let us observe that integrating first over $x$ gives a result proportional 
to the Chebyshev polynomial of the first kind $U_{n-1}(y)$ (see (\ref{A.2}) in Appendix~\ref{app:cheb}).
On the other hand, we have $T_{n}'(x) = n\,U_{n-1}(x)$, and thus (\ref{2.6}) reproduces
the orthogonality relation of polynomials $U_n$ with respect to the weight 
$\sqrt{\mu^{2}_{0}-y^{2}}$.

In conclusion, in the $\cN=4$ SYM theory, taking into account the normalization choice in
(\ref{PnM}), we have
\begin{equation}
\label{2.8}
P_{n}(x) = \frac{\mu_{0}^{n}}{2^{n-1}}~T_{n}\Big(\frac{x}{\mu_{0}}\Big)
= 2\,\Big(\frac{\lambda}{16\pi^2}\Big)^{\frac{n}{2}}\,T_n\Big(\frac{2\pi}{\sqrt{\lambda}}\,x\Big)
\end{equation}
where in the second step we used (\ref{muis}). Plugging this into (\ref{2.5}), we get
\begin{equation}
\label{rhopo}
\rho'_m(x) =-\frac{m}{2\pi\sqrt{\mu_{0}^{2}-x^{2}}}\,\frac{\mu_{0}^{m}}{2^{m-1}}~T_{m}\Big(\frac{x}{\mu_{0}}\Big)~,
\end{equation}
while the pairing becomes
\begin{equation}
\langle n,m\rangle =\widetilde G^{(0)}_{n} \,\delta_{n,m}
\end{equation}
with
\begin{align}
\label{2.9}
\widetilde G^{(0)}_{n} &= -\int_{-\mu_{0}}^{+\mu_{0}}\!dx \, P_{n}(x)\, 
\rho'_n(x)
= \frac{n}{2\pi}\int_{-\mu_{0}}^{+\mu_{0}}\!dx\,\frac{P_{n}(x)^{2}}{\sqrt{\mu_{0}^{2}-x^{2}}}
= n\,\Big(\frac{\lambda}{16\pi^{2}}\Big)^{n}
\end{align}
where the last step follows from (\ref{2.8}) and the orthogonality properties of the Chebyshev polynomials.

This result is in full agreement with \cite{Rodriguez-Gomez:2016ijh} 
where it is shown that (\ref{2.8}) and (\ref{2.9}) hold also for odd $n$.
It is also in complete agreement with the results of the full Lie Algebra approach given 
in Section~\ref{subsec:resN4}. Indeed, comparing 
(\ref{2.8}) with (\ref{omega0isC2}), and taking into account the rescaling (\ref{atoM}) between the matrices $M$ and $a$, we see that for $n\ge 2$
\begin{align}
\label{Pisp}
P_n(M) = \Big(\frac{g^2}{8\pi^2}\Big)^{\frac n2}\, p_n(a)~~~\mbox{and}~~~
	\tr P_n(M) = \Big(\frac{g^2}{8\pi^2}\Big)^{\frac n2}\, O_n^{(0)}(a)~.
\end{align} 
In a perfectly consistent way, the two-point functions (\ref{2.9}) and (\ref{G0}) are related as follows:
\begin{align}
	\label{tGG}
		\widetilde G^{(0)}_n = \Big(\frac{g^2}{8\pi^2}\Big)^n \, G^{(0)}_n~.
\end{align}

\subsubsection{The ABCDE theories}

In a $\cN=2$ theory of the \textbf{ABCDE} series 
with parameter $\nu$ and generic (even or odd) sources, 
the modified integral equation that determines the eigenvalue density reads
\begin{equation}
\label{2.10}
\begin{aligned}
\int_{-\mu(\sigma)+c(\sigma)}^{+\mu(\sigma)+c(\sigma)}\!dy \,\Big[\frac{1}{x-y}-K(x-y)
+(1-\nu)\,&K(x+y)\Big]\,\rho(y; \sigma) \\
&= 
\frac{8\pi^{2}}{\lambda}\,x-\nu K(x)+
\frac{1}{2}\sum_{n}\sigma_{n}P_{n}'(x)
\end{aligned}
\end{equation}
where we have allowed for a non zero cut center $c(\sigma)$. Of course $c(\sigma)=0$ if all 
sources are even.
The orthogonality condition has the same form as in (\ref{ortmn}) and the pairing is expressed in terms of the eigenvalue distribution by the analogue of (\ref{2.2}). Thus we must find polynomials $P_n$ such that
\begin{equation}
\label{2.11}
\langle n, m\rangle  =
-\int_{-\mu_{0}}^{+\mu_{0}}\!dx\,\rho'_m(x) \,P_{n}(x)
= \widetilde G_{n}\,\delta_{n,m}~,
\end{equation}
where again $\mu_{0}\equiv \mu(0)$.
Taking a derivative of (\ref{2.10}) with respect to $\sigma_{m}$ yields
\begin{equation}
\label{2.12}
\int_{-\mu_{0}}^{+\mu_{0}}\!dy \,\Big[\frac{1}{x-y}-K(x-y)+(1-\nu)K(x+y)\Big]\,
\rho'_{m}(y) = \frac{1}{2} P_{m}'(x)~.
\end{equation}
To solve this equation together with the orthogonality condition (\ref{2.11}), we make an Ansatz 
that corresponds to deforming\,%
\footnote{It would be interesting to understand more deeply such a deformation, as it 
happened in a different context for the hypergeometric polynomials \cite{Beccaria:2009rw}.}
the $\cN=4$ polynomials (\ref{2.8}) into   
\begin{align}
\label{2.13}
P_{n}(x) = \frac{\mu_0^n}{2^{n-1}}\sum_{k\le n}b_{k}^{(n)}\,T_{k}\Big(\frac{x}{\mu_{0}}\Big) 
\end{align}
with $b^{(n)}_{n}=1$, and the differentiated $\cN=4$ density (\ref{rhopo}) into 
\begin{align}
	\label{rhopo2}
		\rho_{m}'(x) = -\frac{m}{2\pi\sqrt{\mu_{0}^{2}-x^{2}}}
		\frac{\mu_0^m}{2^{m-1}}\,\sum_{k\ge m}
		a_{k}^{(m)}\,T_{k}\Big(\frac{x}{\mu_{0}}\Big)
\end{align}
with $a^{(m)}_{m}=1$ at tree level. Substituting this Ansatz in (\ref{2.12}) and in the orthogonality conditions, one realizes that in $\rho_m'$, actually, only a finite number of terms are needed in 
the sum over $k$.

The coefficients $b_k^{(n)}$ in (\ref{2.13}) may be easily computed once we keep only a certain (arbitrarily fixed) number of terms in the expansion (\ref{Kexp1})\,%
\footnote{Convolutions are conveniently evaluated by 
\begin{equation*}
\int_{-\mu_{0}}^{\mu_{0}}  dy \frac{1}{x-y}\,\rho'_{n}(y) =
-\frac{n}{2\pi}\int_{-\mu_{0}}^{\mu_{0}} dy \frac{1}{x-y}\,
\frac{1}{\sqrt{\mu_{0}^{2}-y^{2}}}\,\sum_{k}
a_{k}^{(n)}\,T_{k}\left(\frac{y}{\mu_{0}}\right) = 
\frac{n}{2\mu_{0}}\sum_{k} a_{k}^{(n)}\,U_{k-1}\left(\frac{x}{\mu_{0}}\right)~.
\end{equation*}
}.
Then, from
\begin{equation}
\label{2.15}
\begin{aligned}
\langle n, m\rangle  &= \frac{\mu_0^{n+m}}{2^{n+m-2}} \,\frac{m}{2\pi}
\int_{-\mu_{0}}^{+\mu_{0}}\!dx\,
\frac{1}{\sqrt{\mu_{0}^{2}-x^{2}}}\,\sum_{k}
a_{k}^{(m)}\,T_{k}\Big(\frac{x}{\mu_{0}}\Big)
\,\sum_{k'}b_{k'}^{(n)}\,T_{k'}\Big(\frac{x}{\mu_{0}}\Big) 
\\[1mm]
& = m\ \Big(\frac{\mu_{0}}{2}\Big)^{m+n}\,\sum_{m \le k \le n} a_{k}^{(m)}\,b_{k}^{(n)}~,
\end{aligned}
\end{equation}
we obtain, due to the restrictions in (\ref{2.13}), the diagonal values
\begin{equation}
\label{defnn}
		\widetilde G_n = n \Big(\frac{\mu_0}{2}\Big)^{2n} \,a^{(n)}_{n}~.
\end{equation}
The coefficients $a^{(n)}_{n}$ are rational functions of $\mu_{0}$ and may be Taylor expanded in powers of $\mu_{0}$, which, in turn, is equivalent to the weak-coupling expansion.
Keeping in (\ref{Kexp1}) the terms up to $\zeta(11)$, we find
\begin{align}
\label{2.17}
a^{(2)}_{2} &= 1-\frac{3}{4}\,\zeta(3)\,\nu\,\mu_{0}^{4}+\frac{5}{2}\,\zeta(5)\,\nu\,\mu_{0}^{6}
-\Big(\frac{245}{32}\,\zeta(7)\,\nu-\frac{9}{16}\,\zeta(3)^2\,\nu^{2}
\Big)\,\mu_{0}^{8}\notag\\[1mm]
&\quad+\Big(\frac{189}{8}\,\zeta(9)\,\nu-\frac{15}{4}\,\zeta(3)\,\zeta(5)\,\nu^{2}
\Big)\,\mu_{0}^{10}-\Big(\frac{38115}{512}\,\zeta(11)\,\nu-\frac{735}{64}\,\zeta(3)\,\zeta(7)\,\nu^{2}\notag\\[1mm]
&\qquad~~-\frac{825}{128}\,\zeta(5)^2\,\nu^{2}+\frac{27}{64}\,\zeta(3)^3\nu^3\Big)
\,\mu_{0}^{12}+\cdots~,\notag \\[2mm]
a^{(3)}_{3} &= 1+\frac{5}{8}\,\zeta(5)\,(\nu-2)\,\mu_{0}^{6}-\frac{105}{32}\,\zeta(7)\,(\nu-2)\,\mu_{0}^{8}+\frac{1701}{128}\,\zeta(9)\,(\nu-2)\,\mu_{0}^{10}\notag\\[1mm]
&\quad-\Big(\frac{12705}{256}\,\zeta(11)\,(\nu-2)-\frac{25}{64}\,\zeta(3)\,\zeta(7)\,(\nu-2)^2
\Big)\,\mu_{0}^{12}+\cdots~,
\notag \\[2mm]
a_{4}^{(4)} &= 1-\frac{35}{64}\, \zeta(7)\,\nu\,\mu _0^8   +\frac{63}{16}\, \zeta(9)\,\nu\, 
\mu _0^{10}  -\frac{2541}{128}\,\zeta(11)\, \nu
\,\mu _0^{12}    + \cdots ~,\\[2mm]
a_{5}^{(5)} &= 1+\frac{63}{128}\, \zeta(9)\,(\nu -2)\, \mu _0^{10}-\frac{1155}{256}\, \zeta(11)
\,(\nu -2)\, \mu _0^{12} +\cdots~,  \notag\\[2mm]
a_{6}^{(6)} &= 1-\frac{231}{512}\, \zeta(11)\,\nu\,\mu _0^{12}+\cdots~, \notag \\[2mm]
a_{n}^{(n)} &= 1 +\mc O\left(\mu _0^{2n+2}\right) \quad \text{for} \  n > 6~.\notag
\end{align}
The next step is finding the expression of $\mu_0$.

\paragraph{Determination of $\mu_{0}$:}

To obtain explicit expansions in powers of $\lambda$ we need to determine the expression of
the cut endpoint $\mu_{0}(\lambda)$. To this aim, we 
restart from (\ref{2.10}) without sources, {\em {i.e.}} from (\ref{1.7}) with $\mu\to \mu_{0}$.
Since we are interested in the contributions coming from a finite set of terms in the expansion (\ref{Kexp1}), we can conveniently represent the density, which is an even function, in the form
\begin{equation}
	\label{2.19}
		\rho(x) = \sqrt{\mu_{0}^{2}-x^{2}}\,\sum_{k}a_{2k}\,U_{2k}\Big(\frac{x}{\mu_{0}}\Big)~, 
\end{equation}
where the sum is over a finite number of terms and $a_0$ is fixed by the normalization condition
\begin{align}	
\label{norma0}
		\int_{-\mu_{0}}^{+\mu_{0}}\!dx\,\rho(x) = \frac{\pi}{2}\,\mu^{2}_{0}\,a_{0}=1~.
\end{align} 
Then, we can use the formula
\begin{align}
	\int_{-\mu_{0}}^{+\mu_{0}} \!dy\, \frac{\rho(y)}{x-y} =
	\pi\mu_{0}\,\sum_{k} a_{2k}\,T_{2k+1}\Big(\frac{x}{\mu_{0}}\Big)~,
\end{align}
which follows from the second relation in (\ref{TUort}). 
In this way, replacing (\ref{2.19}) in (\ref{1.7}) gives an algebraic equation for $\mu_{0}$ that 
can be easily expanded to any desired order (see
Appendix~(\ref{app:edge}) for an efficient algorithm and details). With this procedure
we obtain the following expansion in terms of the rescaled 't Hooft coupling $\widehat{\lambda}$:
\begin{equation}
\label{2.21}
\begin{aligned}
\mu_{0} &= \sqrt{2\widehat\lambda}\,\,\bigg[1-\frac{3}{2} \,\zeta (3)\, \nu\,
\widehat\lambda^{\,2}+10\,\zeta (5)\, \nu\, \widehat\lambda^{\,3}
-\Big(\frac{455}{8}\zeta (7)\,\nu-\frac{63}{8}\, \zeta (3)^2\,\nu ^2\Big)\,\widehat\lambda^{\,4} 
\\[1mm]
&\qquad\quad+\Big(\frac{2583}{8}\,\zeta (9)\,\nu-\frac{255}{2}\,\zeta (3)\,\zeta (5)\,\nu ^2\Big)
\, \widehat\lambda^{\,5}
-\Big(\frac{30261}{16}\,\zeta (11)\,\nu-\frac{1975}{4}\,\zeta (5)^2\,\nu ^2\\[1mm]
&\quad\qquad\qquad
-\frac{13965}{16}\,\zeta (3)\,\zeta(7)\,\nu ^2+\frac{891}{16}\, \zeta (3)^3\,\nu ^3\Big)
\,\widehat\lambda^{\,6}+\cdots\bigg]~.
\end{aligned}
\end{equation}

\paragraph{Two-point functions:}

We are now in the position of evaluating the coefficients $\widetilde{G}_n$ in the two-point functions. Inserting (\ref{2.21}) into (\ref{2.17}) and (\ref{defnn}), we obtain
\begin{align}
\label{Ggamma}
	\widetilde G_{n} = \widetilde G^{(0)}_{n}\,\gamma_n~,
\end{align}
where $\gamma_n$ encodes the deviation of the $\cN=2$ result from the $\cN=4$ one
given in (\ref{2.9}). For the first few values of $n$ we explicitly find
\begin{align}
\gamma_2 &= 1 - \nu\,
				\bigg[9\,\zeta(3)\,\widehat{\lambda}^{\,2}  - 60\,\zeta(5)\,\widehat{\lambda}^{\,3}
				+ \big(350 \,\zeta(7) -90\,\zeta(3)^2\,\nu
				\big)\widehat{\lambda}^{\,4} \notag\\[1mm]
				&\qquad\quad-
				\Big(\frac{4095}{2}\,\zeta(9)-1350\,\zeta(3)\,\zeta(5)\,\nu\Big)\widehat{\lambda}^{\,5}
                +\Big(\frac{98637}{8}\,\zeta(11)-8820
				\,\zeta(3)\,\zeta(7)\,\nu				
				\notag\\[1mm]
				&\qquad\qquad~~-\frac{9975}{2}\,\zeta(5)^2\,\nu+
				945\,\zeta(3)^3\,\nu^2\Big)\widehat{\lambda}^{\,6}+\cdots\bigg]~,
				\label{datag2}\\[3mm]
\gamma_3&= 1 - \bigg[9\,\zeta(3)\,\nu\widehat{\lambda}^{\,2} -
				5\,\zeta(5) (13\nu -2)\widehat{\lambda}^{\,3} 
				+\Big( \frac{105}{4}\,\zeta(7) (15\nu-4)-81\,\zeta(3)^2\,\nu^2\Big)
				\widehat{\lambda}^{\,4}\notag\\[1mm]
				&\qquad\quad-\Big(\frac{189}{2}\,\zeta(9)\,(25\nu-9)-45\,\zeta(3)\,\zeta(5)\,(29\nu^2-4\nu)\Big)\widehat{\lambda}^{\,5}\notag\\[1mm]
				&\qquad\quad+\Big(\frac{231}{8}\,\zeta(11)\,(503\nu-220) 
				-\frac{315}{4}\,\zeta(3)\,\zeta(7)\,(113\nu^2-28\nu)
				\notag\\[1mm]
				&\qquad\qquad~~-\frac{25}{2}\,\zeta(5)^2\,(407\nu^2-104\nu+8)
				+	756\,\zeta(3)^3\,\nu^3\Big)\widehat{\lambda}^{\,6}+\cdots\bigg]~,
				\label{datag3}\\[3mm]
\gamma_4&=1 - \nu
		\,\bigg[12\,\zeta(3)\,\widehat{\lambda}^{\,2}  - 80\,\zeta(5)\,\widehat{\lambda}^{\,3}
		+ \Big(\frac{1855}{4} \,\zeta(7)-126\,\zeta(3)^2\,\nu\Big)\widehat{\lambda}^{\,4} 
		\notag\\[1mm]				
		&\qquad\quad
		-\Big(2709\,\zeta(9)-1860\,\zeta(3)\,\zeta(5)\,\nu\Big)\widehat{\lambda}^{\,5}
		+\Big(16401\,\zeta(11)-11970\,\zeta(3)\,\zeta(7)\,\nu				
				\notag\\[1mm]
				&\qquad\qquad~~-6750\,\zeta(5)^2\,\nu+
				1296\,\zeta(3)^3\,\nu^2\Big)\widehat{\lambda}^{\,6}+\cdots\bigg]~.
				\label{datag4}
\end{align}
These expressions generalize to higher orders 
our previous results in (\ref{gamma2is}), (\ref{gamma3is}) and (\ref{gamma4is}).
Additional data for $\gamma_n$ with $n=5,\ldots,11$ are collected in Appendix~\ref{app:data} .

\subsection{One-point functions in presence of the Wilson loop}
The previous analysis provides us with all necessary ingredients to compute also 
the v.e.v. of the half-BPS Wilson loop and the one-point functions defined in (\ref{1pointWL}).

The v.e.v. of the Wilson loop may be obtained from the density in (\ref{2.19}) and reads
\begin{equation}
\begin{aligned}
w &\equiv\,\big\langle W_C\big\rangle= \int_{-\mu_{0}}^{+\mu_{0}}\!dx\,
\sqrt{\mu_{0}^{2}-x^{2}}\,\sum_{k}a_{k}\,U_{k}\Big(\frac{x}{\mu_{0}}\Big)\,\rme^{2\pi x} 	\\
&=\mu_{0}^{2} \,\sum_{k}a_{k}\,\int_{-1}^{+1}\!dx\,
\sqrt{1-x^{2}}\,U_{k}(x)\,\rme^{2\pi \mu_{0}x}\\
&=\frac{\mu_{0}}{2}\, \sum_{k}a_{k}\,(k+1)\,I_{k+1}\big(2\pi\mu_{0}\big)
\end{aligned}
\label{wgen}
\end{equation}
where the last step follows form the second equation in (\ref{A.3}).

When a chiral operator is inserted we can proceed as for the two-point functions in (\ref{2.2}) and 
find  
\begin{equation}
\widetilde w_{n} \equiv \big\langle \tr P_n\, W_C\big\rangle 
=-\frac{1}{N}\int_{-\mu_{0}}^{+\mu_{0}}\!dx\, 
\rho'_n(x)\,\rme^{2\pi x}
\end{equation}
where the differentiated density is given in (\ref{rhopo2}). 
Exploiting the first relation in (\ref{A.3}), it is easy to obtain 
\begin{equation}
\label{4.2}
\widetilde w_{n} = \frac{n}{N}\,\Big(\frac{\mu_{0}}{2}\Big)^{n}\,\sum_{k\ge n} a_{k}^{(n)}\,
I_{k}\big(2\pi \mu_{0}\big)~.
\end{equation}
This result is perfectly consistent with what we obtained within the full Lie algebra approach
in Section~\ref{sec:3} (see in particular (\ref{1pointWL})). In fact, the two results 
differ just by an overall factor due to the rescaling factor (\ref{atoM}), namely
\begin{equation}
\widetilde w_{n} = \Big(\frac{g^2}{8\pi^{2}}\Big)^{\frac{n}{2}}\,w_{n}~.
\end{equation}
We now give some explicit results.

\subsubsection{The $\cN=4$ SYM theory}

In $\cN=4$ SYM we just take $\mu_{0}=\frac{\sqrt\lambda}{2\pi}$, $a_k=0$ for $k\not=0$ and
$a_0$ given by the normalization condition (\ref{norma0}). With this input, it is immediate to find from (\ref{wgen}) the well-known $\cN=4$ result (\ref{resw02}).
On the other hand, from (\ref{4.2}) using $a^{(n)}_{n}=1$ for $n>1$, we get
\begin{equation}
\widetilde w_{n}^{(0)} = \frac{n}{N}\,\Big(\frac{\lambda}{16\pi^{2}}\Big)^{\frac{n}{2}}
\,I_{n}\big(\sqrt\lambda\big)~,
\label{wtn4}
\end{equation}
which is consistent with (\ref{wnN}) and the results of \cite{Semenoff:2001xp}.

\subsubsection{The ABCDE theories}

When $\mu_{0}$ is non-trivial, the result (\ref{4.2}) may be formally expanded in the Riemann 
$\zeta$-values and each term captures the exact dependence in $\lambda$. Proceeding in
this way, we find that the difference of the Wilson loop v.e.v. with respect to the $\cN=4$ result exactly matches the expression in (\ref{DeltaWCexpl}).

In the case of insertions, the above procedure yields
\begin{equation}
\Delta \widetilde{w}_n =\Big(\frac{g^2}{8\pi^{2}}\Big)^{\frac{n}{2}}\,\Delta w_{n}
\label{Deltatildewn}
\end{equation}
where for $n=2,3$ the quantities $\Delta w_2$ and $\Delta w_3$ precisely match the
differences (\ref{DW2is}) and (\ref{DW3is}) computed in the full Lie algebra approach.
Additional data for $n=4,5,6,7$ may be found  in Appendix~\ref{app:data}.

\subsection{Asymptotic correlators of operators with large dimension}

Inspection of the explicit expressions for $\gamma_{n}$ and $\Delta w_{n}$ shows that each 
transcendentality structure depends on $n$ in a simple way when $n$ is large enough. 
This is because beyond some point, {\em {i.e.}} for $n$ greater than a certain $n(\zeta)$ 
depending on the specific Riemann $\zeta$-values,
the coefficients $a^{(n)}_{k}$ become simply Kronecker deltas, $\delta^{n}_{k}$, 
and the $\cN=2$ expressions are reproduced by replacing
$\mu_{0}$ with the expansion (\ref{2.21}) in the corresponding $\cN=4$ formulas.
For the correction factors $\gamma_n$ in the two-point functions, 
this gives\,%
\footnote{The arrow emphasizes that for each $\zeta$-monomial we have to 
take $n>n(\zeta)$. The terms in the expansions written in (\ref{gamma_n_generic}) 
and (\ref{Deltaw_n_generic}) are correct for $n\ge 6$.}
\begin{align}
    \label{gamma_n_generic}
    \gamma_{n} \to & \ 1 -n\,\nu\bigg[3\,\zeta(3)\,\widehat{\lambda}^{\,2}-20\,\zeta(5)\,\widehat{\lambda}^{\,3}+\Big(\frac{455}{4}\,\zeta(7)-\frac{9}{2}\,\zeta(3)^2\,(n+3)\,\nu\Big)
    \,\widehat{\lambda}^{\,4} \notag \\[1mm]
    &\qquad\quad- \Big(\frac{2583}{4}\,\zeta (9)-15\,\zeta(3)\,\zeta(5)(4n+15)\,\nu\Big)
    \widehat{\lambda}^{\,5}\notag \\[1mm]
    &\qquad\quad+\Big(\frac{30261}{8} \,\zeta (11)-\frac{105}{4}\,\zeta (3)\,\zeta (7)\,
      (13n+60)\,\nu-\frac{25}{2}\,\zeta (5)^2\,(16n+71)\,\nu\notag\\[1mm]
&\qquad\qquad~~+\frac{9}{2}\,\zeta (3)^3\,(n^2+9n+20)\,\nu^{2}\Big)
\,\widehat{\lambda}^{\,6} + \cdots\bigg]~
\end{align}
while for the shifts $\Delta w_n$ in the one-point functions, it yields
\begin{align}
\label{Deltaw_n_generic}
      \Delta w_{n} &\to  -\frac{n\,\nu}{16}\,(2 \pi  N)^{n-2}\,\bigg[12\,\zeta (3) \,\mathbf{I}_{n-1}\,\widehat{\lambda}^{\,\frac{n}{2}+1}-80\,\zeta (5)\,\mathbf{I}_{n-1}\,
      \widehat{\lambda}^{\,\frac{n}{2}+2}\notag\\[1mm]
       &\qquad\qquad+\Big(455\,\zeta (7) \,\mathbf{I}_{n-1}-9\,\zeta (3)^2 
       \,\nu\,(\mathbf{I}_{n-2}+8 \,\mathbf{I}_{n-1})\Big)\,\widehat{\lambda}^{\,\frac{n}{2}+3}+\cdots\bigg]~.
   \end{align}

The detailed analysis presented in this section shows the complete equivalence of the methods based
on the use of the distribution of the matrix eigenvalues in the large-$N$ limit and those of the full Lie algebra approach based on the recursion relations satisfied by the v.e.v.'s of multitrace operators, 
that we discussed in Section~\ref{sec:3}.

\part*{Part II}
\label{part:2}
In the second part of this paper we study in detail the \textbf{D} and \textbf{E} superconformal theories in the large-$N$ limit. These models, which both have $\nu=0$, are known to have a holographic dual description in terms of an orbifold of $\text{AdS}_5\times S^5$ \cite{Ennes:2000fu}, and in some sense they can be regarded as the next-to-simplest theories 
after the $\cN=4$ SYM theory. It would therefore be extremely interesting to be able to extrapolate the various observables which so far we have discussed in a weak-coupling perturbative approach
and see what one can say in the planar limit at finite or strong coupling. 
As we shall see in the following, remarkable simplifications occur when $\nu=0$ and some resummation methods can be applied to our perturbative expansion.

\section{The full Lie algebra approach for the $\nu=0$ theories}
\label{sec:nu0-fla}

As discussed in Section~\ref{sec:3}, when $\nu=0$ the interaction action of the matrix model 
in the large-$N$ limit simply reduces to the odd part $S_{\mathrm{odd}}(a)$ given in (\ref{Soddis}). Moreover, the v.e.v. of the Wilson loop and the two-point functions of even single-trace operators are not corrected with respect to their $\cN=4$ values. We concentrate therefore on the odd single trace operators
and analyze their two-point functions and their one-point functions in presence of the Wilson loop. The key property we will exploit is the Wick-like factorization (\ref{tracewick}) of the expectation values of such odd operators in the Gaussian model at large $N$. 
The linear relation
(\ref{omega0isC}) ensures that this property also applies to correlators of the operators 
$O^{(0)}_{2i+1}(a)$, which are normal-ordered with respect to the Gaussian measure. Their 
correlators are therefore diagonal, as one can see from (\ref{G0}). 

Let us rescale them by setting
\begin{align}
	\label{defome}
		O^{(0)}_{2i+1}(a) = \sqrt{G^{(0)}_{2i+1}} ~ \omega_i(a)~,
\end{align}
where $G^{(0)}_{2i+1} = (2i+1) (N/2)^{2i+1}$ (see (\ref{G0})). 
At large $N$, the operators $\omega_i(a)$ have a canonical two-point function
\begin{align}
	\label{corrome}
		\big\langle \omega_i(a)\, \omega_j(a)\big\rangle_{(0)} = \delta_{ij}
\end{align}
and the correlators of many of them are again computed using Wick's theorem. 
In other words, we can regard the matrix operators $\omega_i(a)$ as a set of real variables 
$\omega_i$ normally distributed. Indeed we can write
\begin{align}
	\label{corrome1}
		\big\langle \omega_{i_1}(a)\, \omega_{i_2}(a) \ldots \omega_{i_n}(a)\big\rangle_{(0)} 
		= \int \!D\omega\, \,\omega_{i_1}\,\omega_{i_2}\ldots \omega_{i_n} \, 
		\rme^{-\frac 12 \,\omega^T \,\omega}~,
\end{align}		
where we have denoted by $\omega$ the (infinite) column vector of components $\omega_i$ and have defined
\begin{align}
	\label{defDome}
		D\omega = \prod_{i=1}^\infty \!\frac{d \omega_i}{\sqrt{2\pi}}~.
\end{align}
The rules (\ref{corrome}) and (\ref{corrome1}) exponentiate, so that
for any constant matrix $A$ we have
\begin{align}
	\label{expxMx}
\big\langle \rme^{\,\omega^T(a)\, A\, \omega(a)}\big\rangle_{(0)} 
= \int\! D\omega\,\,  \rme^{-\frac 12\, \omega^T (\mathbb{1} - 2 A)\, \omega}
=\mathrm{det}^{-\frac{1}{2}}\big(\mathbb{1}-2A\big).
\end{align} 

In order to efficiently compute the observables involving the odd single-trace operators, it is convenient to rewrite the interaction action $S_{\mathrm{odd}}(a)$ in terms of the quantities $\omega_i(a)$ we have just introduced. 
To this aim, we first invert the relation (\ref{omega0isC}), which for odd operators leads to
\begin{align}
\label{invnorm}
\Omega_{2i+1}(a) = \sum_{k=0}^{n-1} \Big(\frac{N}{2}\Big)^k \,\binom{2i+1}{k}
\, O^{(0)}_{2n-2i+1}(a)~;
\end{align}  
we then rescale the operators according to (\ref{defome}) and plug everything in the
action (\ref{Soddis}).
After carrying out the algebra, the resulting expression takes the form
\begin{align}
	\label{Soddome}
		S_{\mathrm{odd}}(a)= - \frac{1}{2}\, \omega^T(a)\,\mathsf{X}\,\, \omega(a)
\end{align}
where $\mathsf{X}$ is an infinite numerical matrix whose entries are given by
\begin{align}
	\label{Xij}
		\mathsf{X}_{ij} = - 8 \sqrt{(2i+1)(2j+1)} \,
		\sum_{p=0}^\infty (-1)^p\, c_{i,j,p}\, \,\zeta(2i+2j+2p+1) \,\Big(\frac{\widehat{\lambda}}{2}
		\Big)^{i+j+p+1}~,
\end{align}
with
\begin{align}
	\label{defcijp}
		c_{i,j,p} =\sum_{m=0}^p \frac{(2i+2j+2p+1)! }{m!\,(2i+m+1)!\,(p-m)!\,(2j+p-m+1)!}~.
\end{align}
It is interesting to observe that the matrix elements $\mathsf{X}_{ij}$ can be given an integral representation 
in terms of the Bessel functions of the first kind $J_{\ell}(x)$; indeed one can show that
\begin{align}
	\label{Xalb}
	\mathsf{X}_{ij} = -8 (-1)^{i+j} \sqrt{(2i+1)(2j+1)} \int_0^\infty \!\frac{dt}{t}\, 
		\frac{\rme^t}{(\rme^t-1)^2}\,
		J_{2i+1}\Big(t \sqrt{2\widehat{\lambda}}\Big)\, 
		J_{2j+1}\Big(t \sqrt{2\widehat{\lambda}}\Big)~.	
\end{align}
This expression resums
the perturbative expansion in $\widehat{\lambda}$ given (\ref{Xij}), which in turn can be recovered
by Taylor expanding the Bessel functions and performing the resulting $t$-integral using
\begin{align}
	\label{tint}
		\int_0^\infty\! dt\, \frac{\rme^t}{(\rme^t-1)^2}\, t^{2p+1} = (2p+1)! \,\zeta(2p+1)~.
\end{align}	

\paragraph{Odd observables:}
Let us now consider the v.e.v. of an operator that can be constructed with the odd single-traces of the matrix $a$, and thus can be written entirely in terms of the operators $\omega_i(a)$. From the definition (\ref{vevmat}) of the v.e.v. in the $\cN=2$ matrix model and the form (\ref{Soddome}) 
of the interaction action for the $\nu=0$ theories, we have
\begin{align}
	\label{vevEome}
		\big\langle f\big(\omega(a)\big)\big\rangle 
		= \frac{
		\big\langle f\big(\omega(a)\big)\,\rme^{-S_{\mathrm{odd}}(a)}\big\rangle_{(0)}}{
		\big\langle \rme^{-S_{\mathrm{odd}}(a)}\big\rangle_{(0)}}
		= \frac{
		\big\langle f\big(\omega(a)\big)\,\rme^{\,\frac{1}{2}\, \omega^T(a)\, \mathsf{X}\, \,\omega(a)}\big
		\rangle_{(0)}}{
		\big\langle \rme^{\,\frac{1}{2}\,\omega^T(a)\, \mathsf{X}\,\, \omega(a)}\big\rangle_{(0)}}~.
\end{align}	 
The definition (\ref{corrome}) and the property (\ref{expxMx}) allow us to rewrite this 
expression in terms of ordinary gaussian integrals as follows:
\begin{align}
	\label{vevEome2}
       \big\langle f\big(\omega(a)\big)\big\rangle 
		= \frac{1}{\cZ}
		\int\!D\omega\, f(\omega)\, \rme^{-\frac{1}{2} \,\omega^T\, (\mathbb{1} - \mathsf{X})\, \omega}~,
\end{align} 
where

\begin{equation}
\label{pfome}
\cZ=\int\! D\omega\, \rme^{-\frac{1}{2}\,\omega^T\,(\mathbb{1}-\mathsf{X})\, \omega}
=\mathrm{det}^{-\frac{1}{2}}\big(\mathbb{1}-\mathsf{X}\big)~.
\end{equation}

\paragraph{Two-point functions:} Using this free-field formalism, we now reconsider the 
computation of the observables $\gamma_{2i+1}$ that parametrize, according to (\ref{defGN2}), the two-point functions of the normal ordered odd operators $O_{2i+1}(a)$. 
The starting point is finding the correlators of the operators $O^{(0)}_{2i+1}$ in the 
$\cN=2$ theories with $\nu = 0$, to which we have to apply the Gram-Schmidt procedure. 
Let us thus define the matrix $\widehat{G}$ with elements
\begin{align}
	\label{defGtilde}
		\widehat G_{ij} = \big\langle O^{(0)}_{2i+1}(a)\, O^{(0)}_{2j+1}(a)\big\rangle 
		= \sqrt{G^{(0)}_{2i+1}\,G^{(0)}_{2j+1}}~ \big\langle \omega_i(a)\,\omega_j(a)\big\rangle~.
\end{align}  
The two-point function in the right hand side is immediately computed using the free variable description given in (\ref{vevEome2}), leading to
\begin{align}
	\label{omeome}
	\big\langle \omega_i(a)\,\omega_j(a) \big\rangle
	= \Big[(\mathbb{1} - \mathsf{X})^{-1}\Big]_{ij}~.
\end{align}
Applying the Gram-Schmidt procedure one can express the observables $\gamma_{2i+1}$ in 
terms of the correlators $\widehat{G}_{ij}$ by means of the analogue of (\ref{ratiodetG0}). 
Taking into account (\ref{defGN2}), we find
\begin{align} 
	\label{noGome}
		G_{2i+1,2j+1} = G^{(0)}_{2i+1}\, \gamma_{2i+1}\,\delta_{ij} 
	= \frac{\det{\widehat G}_{(i+1)}}{\det{\widehat G}_{(i)}}\,\delta_{ij} ~.
\end{align}
Here $\widehat G_{(i+1)}$ is the submatrix of $\widehat G$ comprising its first $i$ rows and
columns, namely the matrix of correlators of the operators $O^{(0)}_{2k+1}$ with $k\leq i$.
It then follows that 
\begin{align}
	\label{gammaome}
		\gamma_{2i+1} =  \frac{\det  \Big[(\mathbb{1} - \mathsf{X})^{-1}\Big]_{(i+1)}}{\det  \Big[
		(\mathbb{1} - \mathsf{X})^{-1}\Big]_{(i)}}~. 
\end{align} 		
This ratio of determinants can be rewritten in a different way taking into account the expansion
\begin{align}
	\label{1mX}
		(\mathbb{1} - \mathsf{X})^{-1} = \mathbb{1} + \mathsf{X} + \mathsf{X}^2 + \mathsf{X}^3 + \ldots~
\end{align}
and introducing the infinite matrices
\begin{align}
	\label{defYi}
		\overbar{\mathsf{X}}_{(i)} \equiv \mathsf{X}~\text{with the first $i-1$ rows and columns deleted.}
\end{align}
Of course, $\overbar{\mathsf{X}}_{(1)} = \mathsf{X}$. Then one has
\begin{align}
	\label{idgsa}
		\gamma_{2i+1} 
	= \Big[\mathbb{1} + \overbar{\mathsf{X}}_{(i)} +   \overbar{\mathsf{X}}^2_{(i)} + \ldots \Big]_{1,1} 
	= \Big[(\mathbb{1} - \overbar{\mathsf{X}}_{(i)})^{-1}\Big]_{1,1}~.
\end{align}
This expression, together with the form of the matrix $\mathsf{X}$ given in (\ref{Xalb}), is very powerful. In fact, it readily reproduces the formul\ae\, in (\ref{gamma3is}), (\ref{gamma5is}) and (\ref{datag5})--(\ref{s4g11}) for $\gamma_3$ up to $\gamma_{11}$ for $\nu=0$, and also it yields 
in a quite straightforward way the expansions to very high orders in $\widehat{\lambda}$ for any desired value of the index $i$ and any power of Riemann $\zeta$-values.
In the next section we will show how one can obtain equivalent expressions using the eigenvalue approach discussed in Section~\ref{sec:4} and provide also a few explicit examples of 
expansions to high orders.

We conclude by observing that it is also possible to exploit the Gaussian variables $\omega_i$ to express in a very efficient way the observables $w_{2i+1}$ defined (\ref{1pointWL}), which are related to the one-point functions of odd operators in presence of the BPS Wilson loop. 
To avoid redundancy, however we will discuss them only in the eigenvalue approach in the next section.

\section{The eigenvalue distribution approach for the $\nu=0$ theories}
\label{sec:nu0-ca}

In the previous section \ref{sec:nu0-fla}, we have discussed the $\nu=0$ theories $\textbf{D}$ and $\textbf{E}$ in the ``full Lie algebra'' approach. Here, 
consider them in the ``Cartan algebra approach''. We discuss in full details both the 
calculation of the two-point unctions and that of the one-point functions in presence of the Wilson loop.

\paragraph{Two-point functions:}

In the \textbf{D} and \textbf{E} models, the cut edge $\mu_{0}$ given in (\ref{2.21}) simply
reduces to 
$\mu_{0}=\sqrt{2\widehat{\lambda}}=\frac{\sqrt\lambda}{2\pi}$. 
The two-point functions (\ref{defnn}) read therefore 
\begin{equation}
\label{3.1}
\widetilde G_{n} = n\,\Big(\frac{\lambda}{16\pi^{2}}\Big)^{n}\,a^{(n)}_{n}~.
\end{equation}
Comparing with (\ref{2.9}) and (\ref{Ggamma}), we see that the coefficients $a^{(n)}_n$ yield in this case directly the quantities $\gamma_n$ which encode the deviation of the $\cN=2$ result from
to the $\cN=4$ one:
\begin{align}
	\label{aisgamma}
		\gamma_n = a^{(n)}_n~.
\end{align}
Let us recall that the coefficients $a^{(n)}_{m}$ appear in the differentiated density (\ref{2.13}),
which in the present case is determined by the integral equation (\ref{2.12}) with $\nu=0$, namely 
\begin{equation}
\label{3.2}
\int_{-\mu_0}^{+\mu_0}\! dy\, \Big[\frac{1}{x-y}-K(x-y)+K(x+y)\Big]\,
\rho'_{n}(y) = \frac{1}{2} P_{n}'(x)~.
\end{equation}
Since we are interested in odd chiral primaries with $n=2k+1$, according to (\ref{rhopo2}), 
the differentiated density $\rho_n'$ has an expansion in Chebyshev $T$-polynomials of odd degree. After inserting the Ansatz (\ref{rhopo2}) in the left hand side of (\ref{3.2}),
the part of the integral operator that depends on the function $K$ can be written as
\begin{align}
	\label{expKf}
		- \frac{\mu_0^{2k+1}}{2^{2k}} \sum_{j\geq k} a_{2j+1}^{(2k+1)}\, f_{2j+1}(x)~, 
\end{align}
where
\begin{equation}
\begin{aligned}
	\label{fis}
		f_{2j+1}(x) &= \int_{-\mu_0}^{+\mu_0}\!dy\,
		\Big[K(x+y)-K(x-y)\Big]\,\frac{T_{2j+1}(y/\mu_0)}{\sqrt{\mu_0^{2}-y^{2}}}\\
		&=2\int_{-\mu_0}^{+\mu_0}\! dy\,K(x+y)\,\frac{T_{2j+1}(y/\mu_0)}{\sqrt{\mu_0^{2}-y^{2}}}~.
\end{aligned}
\end{equation}
We can now use the following integral representation for $K(x)$ valid in the strip 
$|\operatorname{Im}(x)|<1$, given by
\begin{equation}
\label{0.3}
K(x) = 2\,x\,\int_{0}^{\infty}\!dt\,\frac{1-\cos(t x)}{\rme^{t}-1}~,
\end{equation}
and obtain
\begin{align}
\label{3.3}
f_{2j+1}(x) & = 4\int_{0}^{\infty}\,\frac{dt}{\rme^{t}-1}
\int_{-\mu_0}^{+\mu_0}\! dy\,(x+y)\,\Big[1-\cos(t (x+y))\Big]\,\frac{T_{2j+1}(y/\mu_0)}{\sqrt{\mu_0^{2}-y^{2}}}~.
\end{align}
This expression can be expanded in the even Chebyshev $U$-polynomials as follows
\begin{equation}
\label{ftoU}
f_{2j+1}(x) =\frac{\pi}{\mu_0} \,\sum_{i}\mathsf{Y}_{i, j}\,U_{2i}(x/\mu_0)
\end{equation}
where
\begin{align}
	\label{YtoH}
		\mathsf{Y}_{i,j} = \frac{8}{\mu_0\pi^{2}}\, \int_{0}^{\infty}\!dt\, \frac{H_{i,j}(t)}{\rme^{t}-1}~,
\end{align}
with
\begin{equation}
H_{i,j}(t) = \int_{-\mu_0}^{+\mu_0}\!dx\,U_{2i}(x/\mu_0)\,\sqrt{\mu_0^2-x^{2}}\, 
\int_{-\mu_0}^{+\mu_0}\!dy\,\frac{T_{2j+1}(y/\mu_0)}{\sqrt{\mu_0^2-y^{2}}}\,(x+y)\,
\Big[1-\cos\big(t(x+y)\big)\Big]~.
\end{equation}
With some work, it is possible to evaluate this expression in closed form using the
Bessel functions of the first kind and to show that
\begin{align}
\label{Yis}
\mathsf{Y}_{i,j} &= -8(-1)^{i+j}\,(2i+1)\,\int_{0}^{\infty}\!\frac{dt}{t}\,\frac{\rme^{t}}
{(\rme^{t}-1)^{2}}\,J_{2i+1}(\mu_0\,t)\,J_{2j+1}(\mu_0 \,t)~.
\end{align}
The matrix $\mathsf{Y}$ is related to the matrix $\mathsf{X}$ 
introduced in (\ref{Xalb}) in a simple way:
\begin{equation}
\mathsf{Y}_{i,j} = \sqrt{\frac{2i+1}{2j+1}}\,\,\mathsf{X}_{i,j}~.
\end{equation}
Inserting this result into (\ref{ftoU}) and (\ref{expKf}) we can solve the integral equation (\ref{3.2}) for the coefficients $a_{2j+1}^{(2k+1)}$, taking into account the ansatz (\ref{2.13}) for the polynomials $P_{2k+1}$.
After some straightforward manipulations, we find that the 
vector $\mathsf{A}_{(k)} =\big\{a^{(2k+1)}_{2k+1}, a^{(2k+1)}_{2k+3}, \cdots\big\}$ 
is given by
\begin{equation}
\label{3.17}
\mathsf{A}_{(k)} = \big(\mathbb{1}-\overbar{\mathsf{Y}}_{(k)}\big)^{-1}\,\mathsf{C}~,
\end{equation}
where $\mathsf{C}=(1,0,0,\dots)$ and $\overbar{\mathsf{Y}}_{(k)}$ is the matrix obtained from
$\mathsf{Y}$ by removing its first $k-1$ rows and columns. 
Using this into (\ref{aisgamma}), we conclude that the two-point coefficients 
$\gamma_{2i+1}$ are given by
\begin{equation}
\label{resca}
\gamma_{2i+1} = 
\Big[(\mathbb{1} - \overbar{\mathsf{Y}}_{(i)})^{-1}\Big]_{1,1}~.
\end{equation}
This result is fully equivalent to (\ref{idgsa}). For the first few values of $i$ we get explicitly
\begin{subequations}
\label{oddresults}
\begin{align}
 \gamma_{3} &= 1-10\,\zeta(5)\,\widehat{\lambda}^{\,3}
 +105\,\zeta(7)\,\widehat{\,\lambda}^4
 -\frac{1701}{2}\,\zeta(9)\,\widehat{\lambda}^{\,5}+
 \Big(\frac{12705}{2}\,\zeta(11)+100\,\zeta(5)^2\Big) \widehat{\lambda}^{\,6}\notag \\[1mm]
    &\qquad-\Big(\frac{184041}{4}\,\zeta(13)+2100\,\zeta(5)\,\zeta (7)\Big) 
    \widehat{\lambda}^{\,7}
    +\Big(\frac{5270265}{16}\,\zeta(15)+17010\,\zeta(5)\,\zeta(9)\notag\\[1mm]
    &\qquad+
    \frac{44835}{4}\,\zeta (7)^2\Big)\widehat{\lambda}^{\,8}
    -\Big(\frac{18803785}{8}\,\zeta(17)+127050\,\zeta (5)\,\zeta (11)\notag\\[1mm]
    &\qquad+
    \frac{368235}{2}\,\zeta (7)\,\zeta (9)+
    1000\,\zeta (5)^3 \Big)\widehat{\lambda}^{\,9} + \cdots ~, \label{gamma3expl} \\[2mm]    
  \gamma_{5} &= 1-\frac{63}{2}\,\zeta (9)\,\widehat{\lambda}^{\,5}
  +\frac{1155}{2}\,\zeta (11)\,\widehat{\lambda}^{\,6}
  -\frac{27885}{4}\,\zeta (13)\,\widehat{\lambda}^{\,7}
  +\frac{1126125}{16}\,\zeta (15)\,\widehat{\lambda}^{\,8}\notag\\[1mm]
  &\qquad  -\frac{5165875}{8}\,\zeta (17)\,\widehat{\lambda}^{\,9} + \cdots ~,\label{gamma5expl} 
  \\[2mm]
   \gamma_{7} &= 1-\frac{429}{4}\,\zeta (13)\,\widehat{\lambda}^{\,7}+
   \frac{45045}{16}\,\zeta (15)\,\widehat{\lambda}^{\,8}
   -\frac{1446445}{32}\,\zeta (17)\,\widehat{\lambda}^{\,9} +\cdots ~.\label{gamma7expl}
\end{align}%
\end{subequations}
We have reported here only the first few terms of the expansions but, as we will discuss in 
Section~\ref{sec:resumm}, using (\ref{resca}) we can push the expansions to very high orders with minor effort.  

\paragraph{One-point functions in presence of the Wilson loop:}
The one-point functions of the chiral odd operators in presence of the circular Wilson loop, given in (\ref{4.2}), can be written for the $\nu=0$ theories as follows (``t'' denotes matrix transposition)
\begin{align}
	\label{wY}
		\widetilde w_{2i+1} &= \frac{2i+1}{N}\,\left(\frac{\lambda}{16\pi^2}\right)^{\frac{2i+1}{2}}\,
		\sum_{k=1}^{\infty}\Big[\mathbb{1}-\overbar{\mathsf{Y}}^{\,t}_{(i)}\Big]^{-1}_{1k}
		\,\mathsf B_{k}~,
\end{align}
where $\mathsf B_{k}$ is the $k$-th component of the vector
\begin{align}
	\label{Bvec}
		\mathsf{B} &=  \Big(I_{2i+1}(\sqrt\lambda), I_{2i+3}(\sqrt\lambda), \dots\Big)
\end{align}
with $I_\ell$ being the modified Bessel functions of the first kind.
Taking the difference with respect to the $\cN=4$ results given in (\ref{wtn4}), we obtain
the corrections $\Delta\widetilde w_{2i+1}$ which are related to quantities 
$\Delta w_{2i+1}$ computed in the full Lie algebra approach by a simple rescaling 
due to the relation (\ref{atoM}), namely
\begin{align}
	\label{rescDw}
		\Delta \widetilde w_{2i+1} = \Big(\frac{g^2}{8\pi^2}\Big)^{\frac{2i+1}{2}}
		\Delta w_{2i+1}~.
\end{align}	
We report below the explicit expressions of $\Delta w_{2i+1}$ we obtain from the above procedure
for the first few values of $i$:
\begin{subequations}
\begin{align}
    \Delta w_{3} &= -3\pi  \sqrt{N}\, \bigg[10\,\zeta (5)\,\mathbf{I}_3\,\widehat{\lambda}^{\,\frac{7}{2}}-\frac{35}{2}\,\zeta (7)\,(7 \,\mathbf{I}_3-8 \,\mathbf{I}_4)\,\widehat{\lambda}^{\,\frac{9}{2}}+126\,\zeta (9)\,(9 \,\mathbf{I}_3-20 \,\mathbf{I}_4+20 \,\mathbf{I}_5)\,
    \widehat{\lambda}^{\,\frac{11}{2}}\notag \\
    &\quad-\Big(\frac{3465}{4}\,\zeta (11)\,(11 \,\mathbf{I}_3-36 \,\mathbf{I}_4+72 \,\mathbf{I}_5-64 \,\mathbf{I}_6)+100\,\zeta (5)^2\,\mathbf{I}_3\Big)\,\widehat{\lambda}^{\,\frac{13}{2}}\notag \\
    &\quad+\Big(\frac{2165}{4}\,\zeta (13)\,(143 \,\mathbf{I}_3-616 \,\mathbf{I}_4+1848 \,\mathbf{I}_5-3360 \,\mathbf{I}_6+2688 \,\mathbf{I}_7)\notag\\
&\qquad~~+175\,\zeta (5)\,\zeta (7)\,(13 \,\mathbf{I}_3-8 \,\mathbf{I}_4)\Big)\,
\widehat{\lambda}^{\,\frac{15}{2}}+\cdots\bigg]~, \\[1mm]
  \Delta w_{5} &=-5\big(\pi\sqrt{N}\big)^3\,\bigg[126\,\zeta (9)\,\mathbf{I}_5
  \,\widehat{\lambda}^{\,\frac{13}{2}}-231\,\zeta (11)\,(11 \,\mathbf{I}_5-12 \,\mathbf{I}_6)\,\widehat{\lambda}^{\,\frac{15}{2}}    \notag \\
    &\qquad +2574\,\zeta (13)\,(13 \,\mathbf{I}_5-28 \,\mathbf{I}_6+28 \,\mathbf{I}_7)\,
    \widehat{\lambda}^{\,\frac{17}{2}}+\cdots\bigg]~,\\[1mm]
  \Delta w_{7} &= -7\,\big(\pi  \sqrt{N}\big)^5\,\bigg[1716\,\zeta (13) \,\mathbf{I}_7
  \, \widehat{\lambda}^{\,\frac{19}{2}}+\cdots\bigg]
\end{align}
\end{subequations}
where $\mathbf{I}_\ell$ are the rescaled Bessel functions (\ref{bI}). These expressions agree with
(\ref{DW3is}), (\ref{DW5is}) and (\ref{DW7is}) for $\nu=0$ and generalize them to higher perturbative
orders.

\section{Diagrammatic analysis for the $\n=0$ theories}
\label{sec:diagrams}
In this section we perform a diagrammatic analysis of the two-point correlator (\ref{twopointdef}) 
for the $\cN=2$ SCFTs in flat space with $\nu=0$. As we shall see, these models are simple enough to push the perturbative analysis very far. Our main goal is to understand the field-theory difference between correlators with even and odd operators and in particular to trace the diagrammatic origin of the fact that the 
even correlators do not receive corrections in the large-$N$ limit with respect to the $\cN=4$ case, while the odd correlators do. 
We mainly concentrate on the two-point function \eqref{twopointdef} of single-trace operators 
$O_n (x) = \tr \varphi^n(x)$, namely
\begin{equation}
	\label{TwopointFT}
		\big\langle O_m(x) \,\overbar{O}_n(0)\big\rangle 
		= \frac{G_{n}(g,N)}{(4\pi^2 x^2)^{2n}\phantom{\big|}}\,\delta_{m,n}~.
\end{equation}
Superconformal symmetry guarantees that the space-time dependence $1/(4\pi^2 x^2)^{2n}$ is preserved at the quantum level, so we have simply to compute the coefficient $G_n$ and compare it with the matrix model results. In particular our achievements are twofold:
\begin{itemize}
\item
We provide a systematic diagrammatic analysis up to three loops for the two-point function coefficients $G_2$, $G_3$, $G_4$ and $G_{5}$.
\item
We infer the general contribution for the first deviation from the $\cN=4$ theory of the 
coefficient $G_{n}$ for generic $n$.
\end{itemize}
We use the same tools which are frequently used in supersymmetric contexts \cite{Pomoni:2013poa,Mitev:2014yba,Billo:2017glv,Billo:2018oog,Billo:2019job, Billo:2019fbi}, namely 
the $\cN=1$ superspace formalism and the diagrammatic difference between $\cN=2$ and $\cN=4$ theories. We refer to \cite{Billo:2019fbi} for a detailed account of the Lagrangian and the Feynman rules that are needed for the computations.

\subsection{Field theory calculations up to 3-loops}
\begin{figure}[H]
\begin{center}
\includegraphics[scale=0.7]{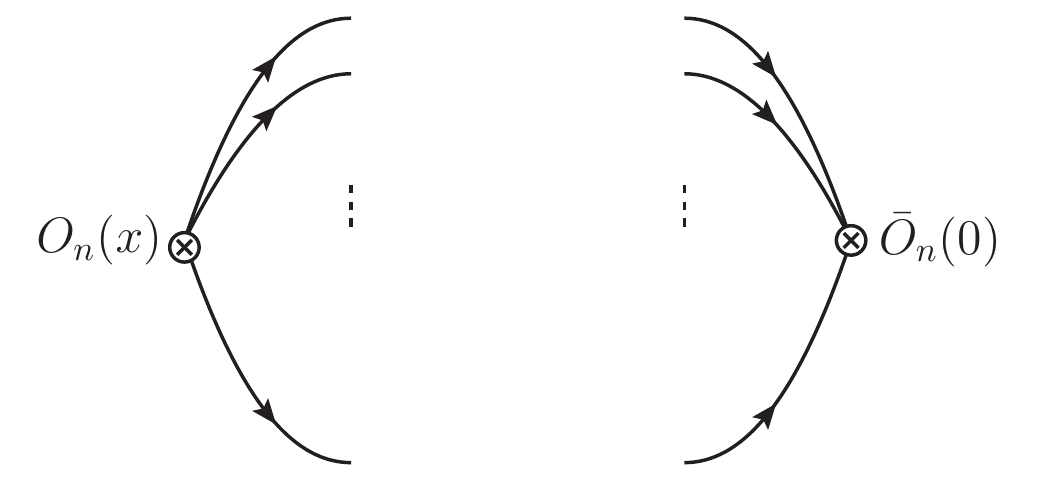}
\end{center}
\caption{The diagram representing the insertion of a chiral operator $O_n=\tr \phi^n$ in $x$ 
and of the corresponding anti-chiral operator $\overbar{O}_n=\tr \overbar{\phi}^n$ in the origin.}
\label{Fig::tree}
\end{figure}

\begin{figure}[H]
\includegraphics[scale=0.47]{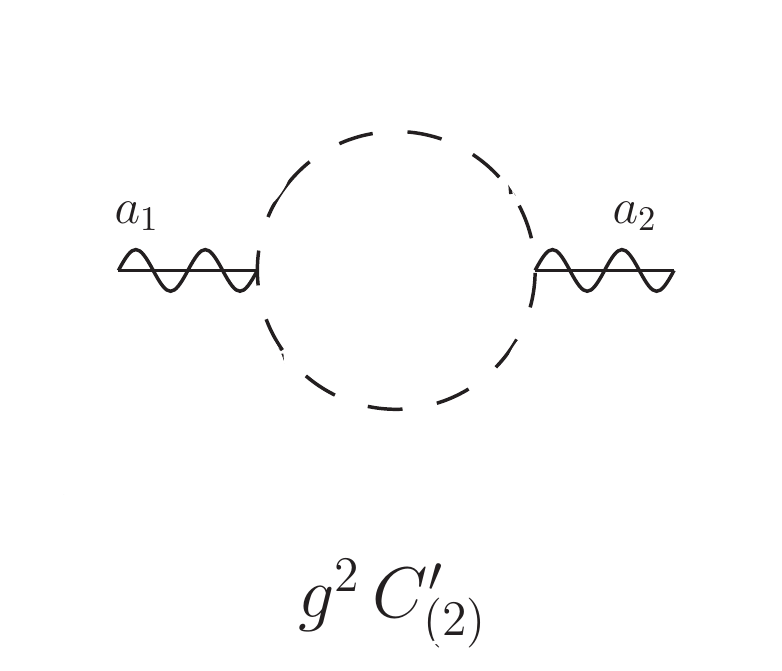}\hspace{-0.6cm}
\includegraphics[scale=0.47]{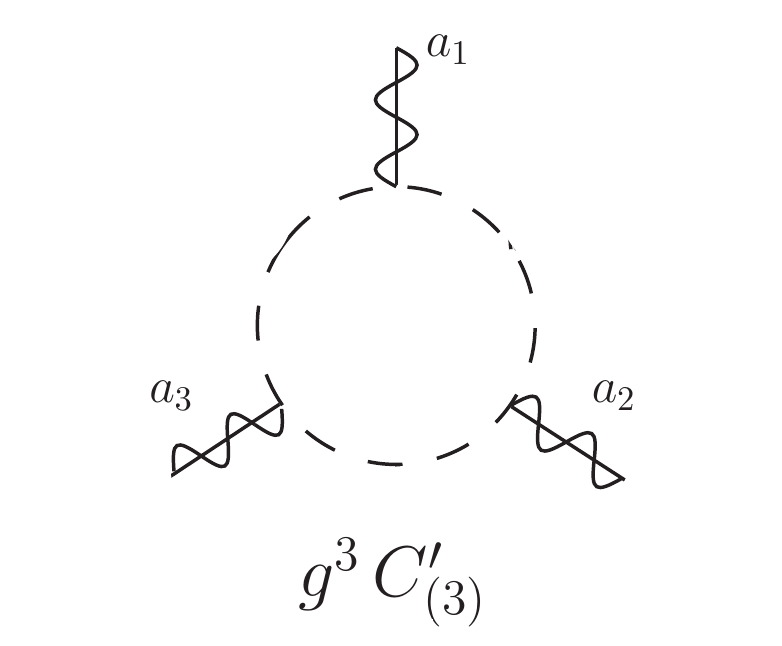}\hspace{-0.8cm}
\includegraphics[scale=0.47]{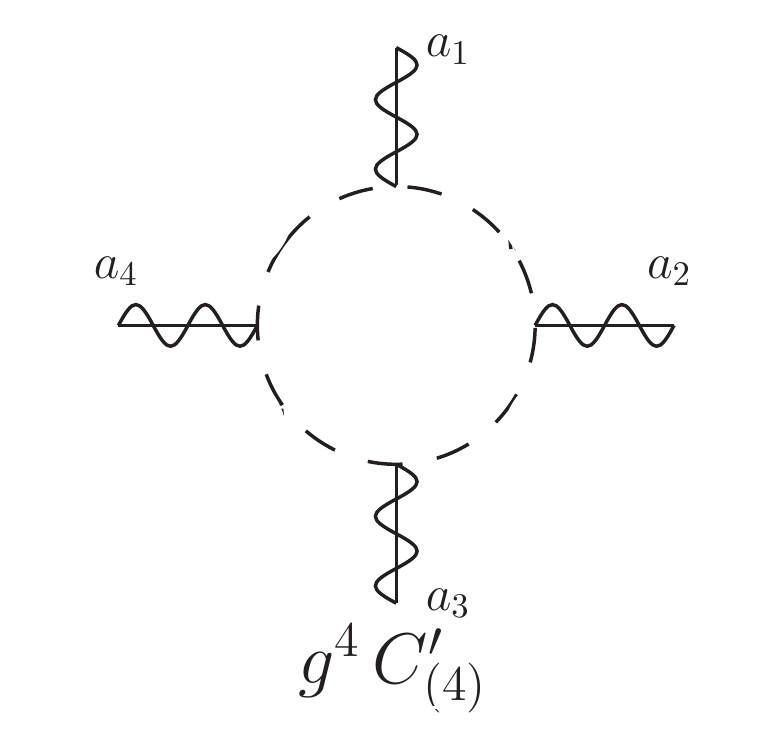}\hspace{-.2cm}
\includegraphics[scale=0.47]{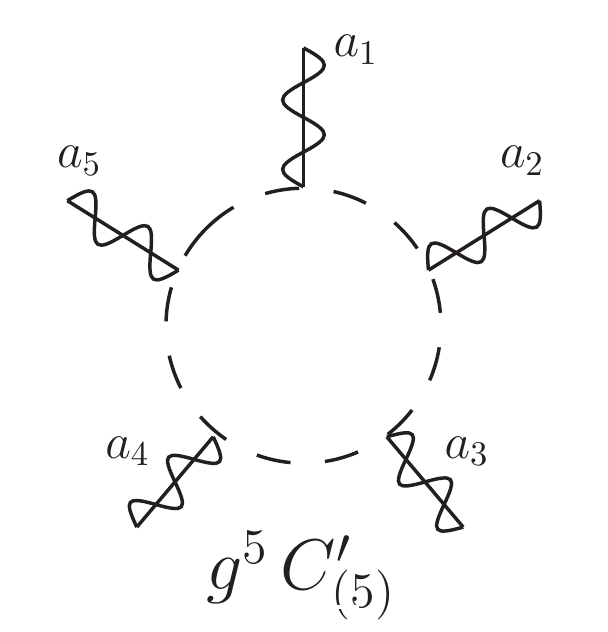}\hspace{-.2cm}
\includegraphics[scale=0.47]{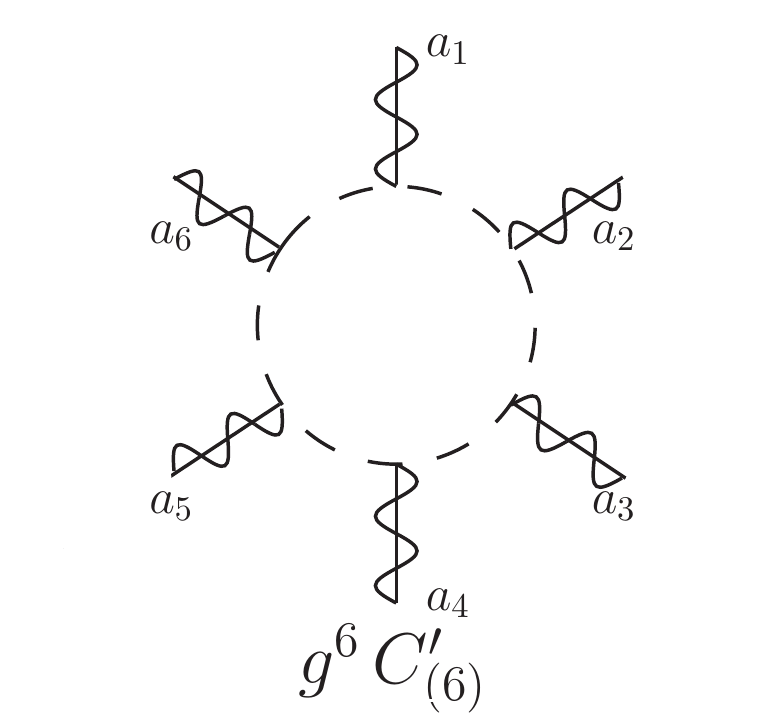}
\caption{Building blocks up to $g^6$ order. The dashed lines stand for hypermultiplet loops, the wavy/solid lines stand for fields belonging to $\cN=2$ vector multiplet and the coefficients 
$C'_{(n)}$ are color factors.}
\label{Fig:BBlocks}
\end{figure}

We start from the operator insertions represented in Figure~\ref{Fig::tree}. The tree-level contribution
to the two-point function is obtained by contracting the legs in this diagram in all possible ways 
using the tree-level propagators. Doing so, we obtain
\begin{equation}
\label{Gtree}
G_n^{(0)} = n\,\Big(\frac{N}{2}\Big)^n
\end{equation}
In the $\cN=4$ SYM theory all higher-loop corrections cancel and the tree-level result (\ref{Gtree})
represents the full answer to the correlator in the large-$N$ limit, matching the matrix model result \eqref{G0}.
The $\cN=2$ theories, instead, contain a full tower of quantum corrections. Here we follow 
the same notation introduced in (\ref{Gnis}) and factorize the $\cN=4$ result \eqref{Gtree}, so that any $\cN=2$ two-point function coefficient
\begin{equation}\label{GnN2}
G_n =n \, \Big( \frac{N}{2}\Big)^n\,\gamma_n
\end{equation}
is uniquely described by $\gamma_n$ which is a function of the 't Hooft coupling.
The loop corrections of $\cN=2$ theories can be effectively worked out in the diagrammatic difference with $\cN=4$. This procedure allows us to discard all diagrams which only contain fields from the $\cN=2$ vector multiplet, since they are in common with the $\cN=4$ SYM theory, 
and to isolate the genuine $\cN=2$ contributions, corresponding to the diagrams where hypermultiplet are present. As we see from Figure \ref{Fig::tree}, the operator insertions only contain fields belonging to the vector multiplet, hence the hypermultiplets appear only inside loops. The diagrams surviving in the difference can then be organized in terms of some building blocks, as displayed in Figure \ref{Fig:BBlocks}.

This procedure reduces the number of Feynman diagrams to be computed and allows a rapid comparison with the matrix model. Indeed each building block carries a color factor $C'_{(n)}$, where $n$ counts the number of adjoint fields attached to hypermultiplet loops. In the difference theory this color factor is given by the combination
\begin{equation}
C'_{a_1\dots a_n}= (\Tr_{\cR} - \Tr_{\mathrm{adj}})\, T_{a_1}\dots T_{a_n}
\end{equation}
which exactly reproduces what we found in the matrix model (see (\ref{deftrp})). 
We now systematically classify the Feynman diagrams by looking at their color factors. Many of them vanish because of the matter content $\cR$ of the $\n=0$ theories or can be discarded 
because they are subleading in the large $N$ limit.

\paragraph{One loop:} At order $g^2$ there is a unique way to insert a hypermultiplet loop inside Figure~\ref{Fig::tree}, namely by using the one-loop correction to the scalar propagator. This diagram corresponds to the first building block of Figure \ref{Fig:BBlocks}, but its color factor $C'_{(2)}$ vanishes because of conformal invariance due to the condition \eqref{trpa2}.

\begin{figure}[H]
\hspace{2cm}\includegraphics[scale=0.5]{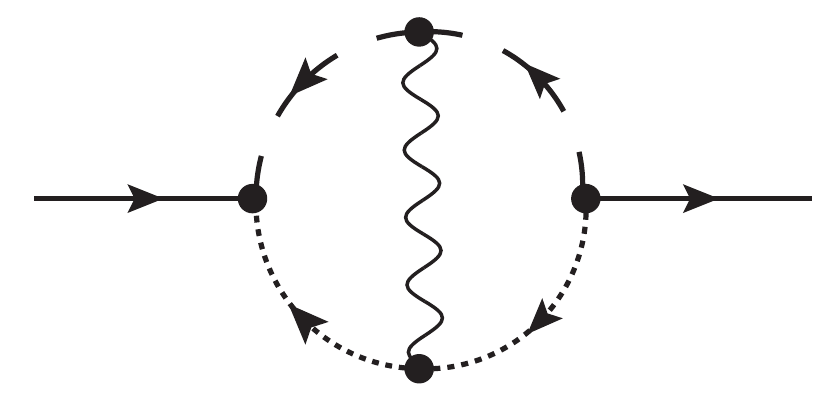}\hspace{1.5cm}
\includegraphics[scale=0.5]{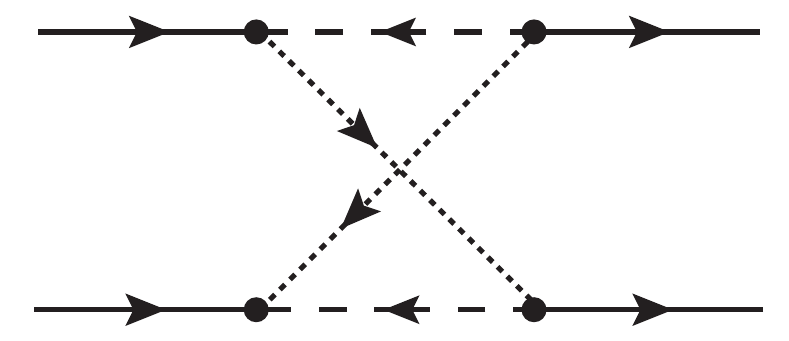}
\caption{Diagrams contributing at $g^4$ order in $\cN=1$ superspace formalism. The dashed and dotted lines stand for the $\cN=1$ chiral fields $Q,\tilde Q$ belonging to the $\cN=2$ hypermultiplet, the wavy line stand for the $\cN=1$ vector multiplet, the straight line for the adjoint chiral multiplet}
\label{Fig:2loops}
\end{figure}

\paragraph{Two loops:}
Since $C'_{(2)}$ always vanishes, the diagrams at order $g^4$ can only be built out of the $C'_{(3)}$ and $C'_{(4)}$ building blocks. Any diagram proportional to $C'_{(3)}$ is again proportional to the  coefficient $\beta_0$ of the $\beta$-function, and so it vanishes in any superconformal theory. There are only two diagrams which can be built out of the third diagram of Figure \ref{Fig:BBlocks} with a $C'_{(4)}$ color factor. They have been fully analyzed in the conformal SQCD, which is theory \textbf{A}, in \cite{Billo:2017glv,Billo:2018oog} and are represented in Figure \ref{Fig:2loops}.

These diagrams do give a non-trivial contribution to correlation functions for $\n\neq 0$ theories, and actually account for the terms linear in $\zeta(3)$ in $\gamma_n$. However, if we analyze the color factor $C'_{(4)}$ in the \textbf{D} and \textbf{E} theories by expanding the trace combination 
in \eqref{trpa2k} and using FormTrace \cite{Cyrol:2016zqb}, we obtain that $C'_{(4)}$ is always subleading in the large-N limit.
This fact explains why the $\n=0$ theories have no two-loop order term proportional to 
$\zeta(3)$ in the large-N limit. 

\begin{figure}[H]
\begin{center}
\includegraphics[scale=0.23]{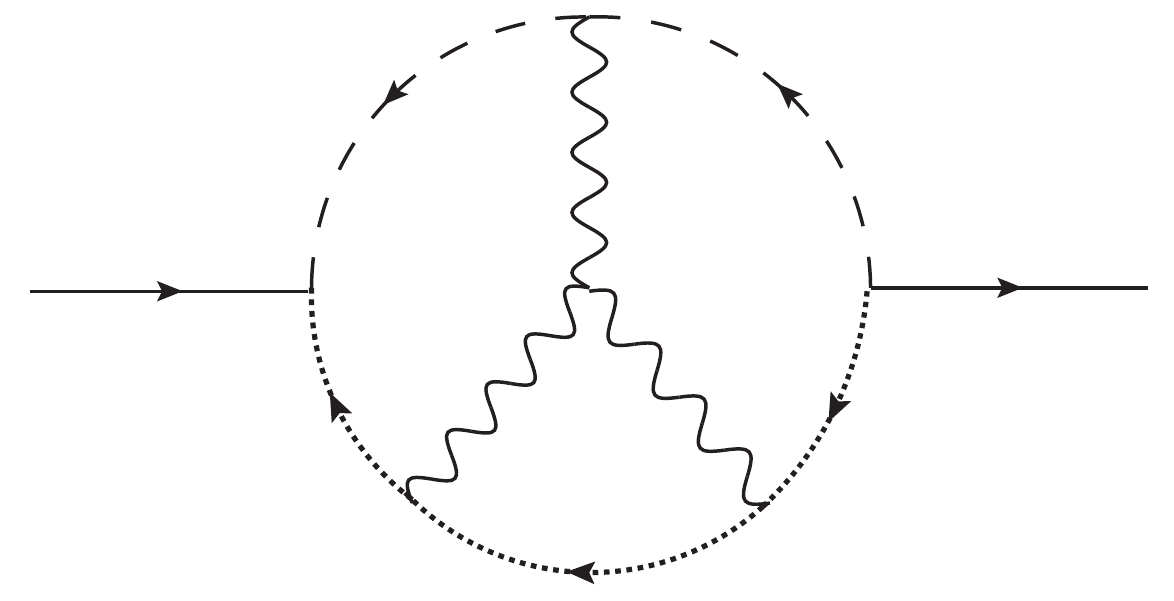}\hspace{.2cm}
\includegraphics[scale=0.23]{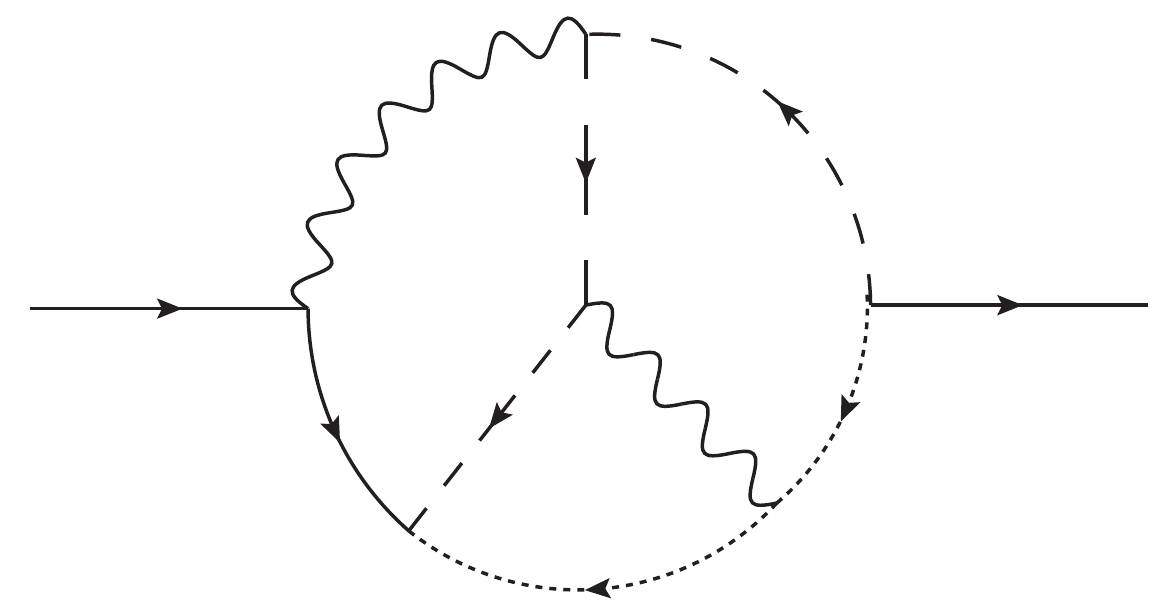}\hspace{.2cm}
\includegraphics[scale=0.23]{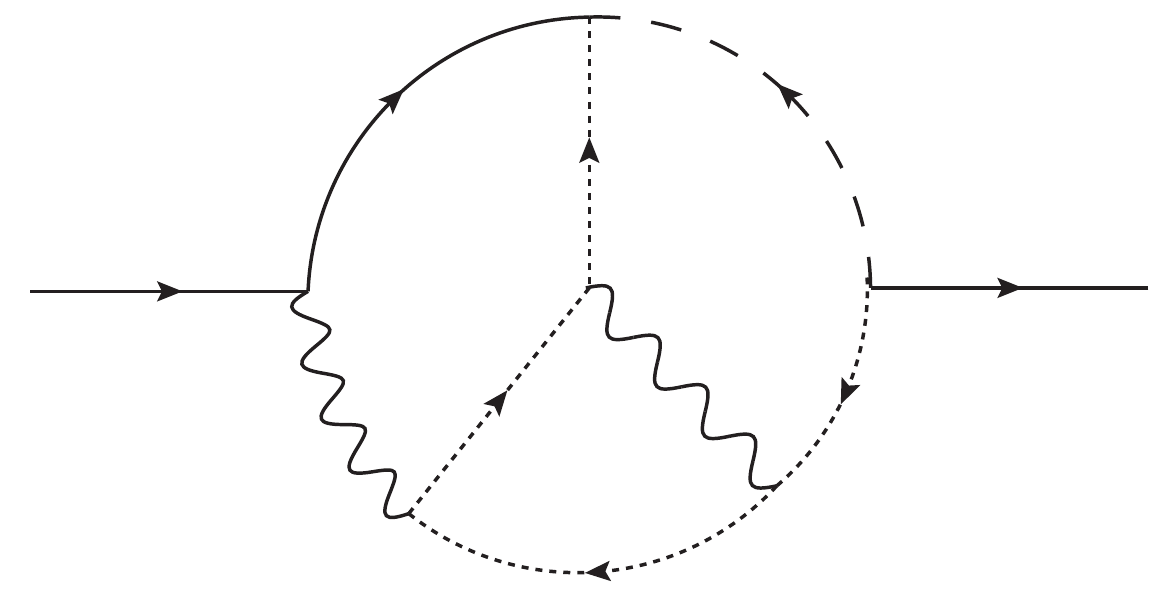}\hspace{.2cm}
\includegraphics[scale=0.23]{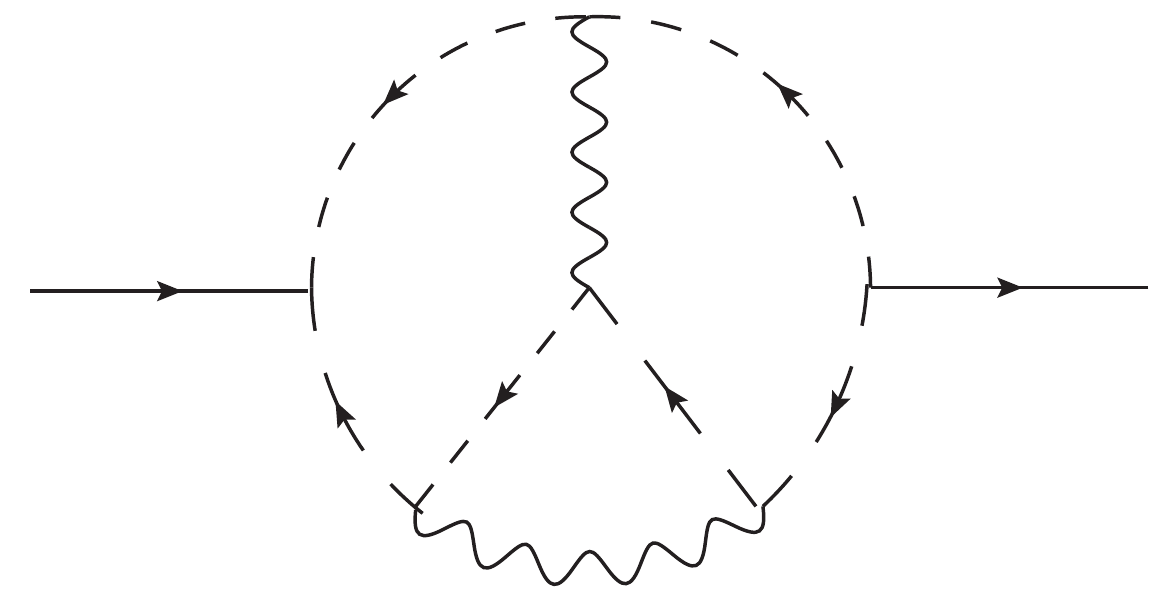}\hspace{.2cm}
\includegraphics[scale=0.23]{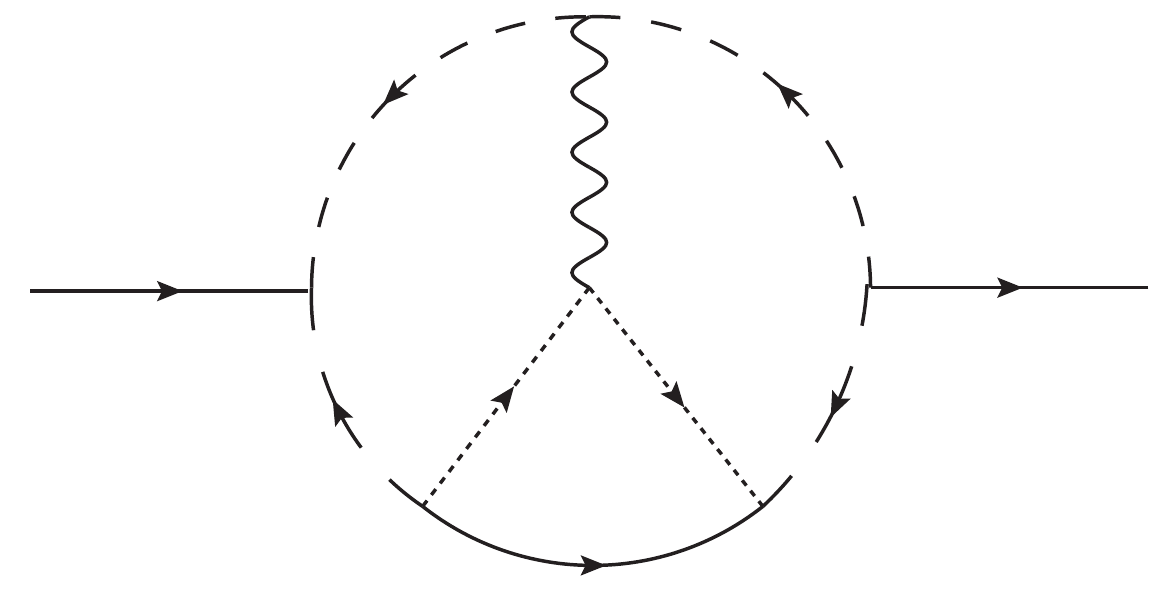}\\
\includegraphics[scale=0.23]{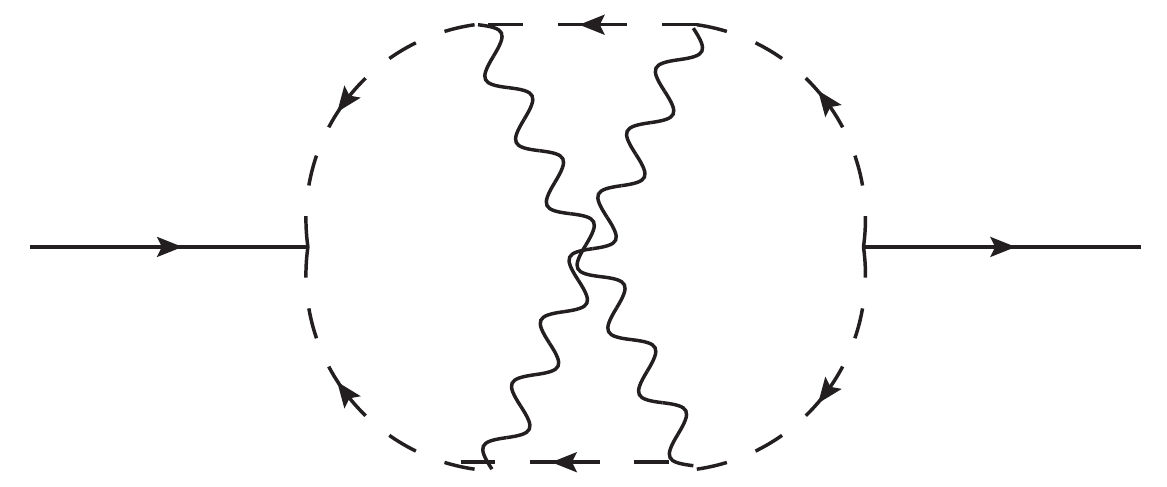}\hspace{.2cm}
\includegraphics[scale=0.23]{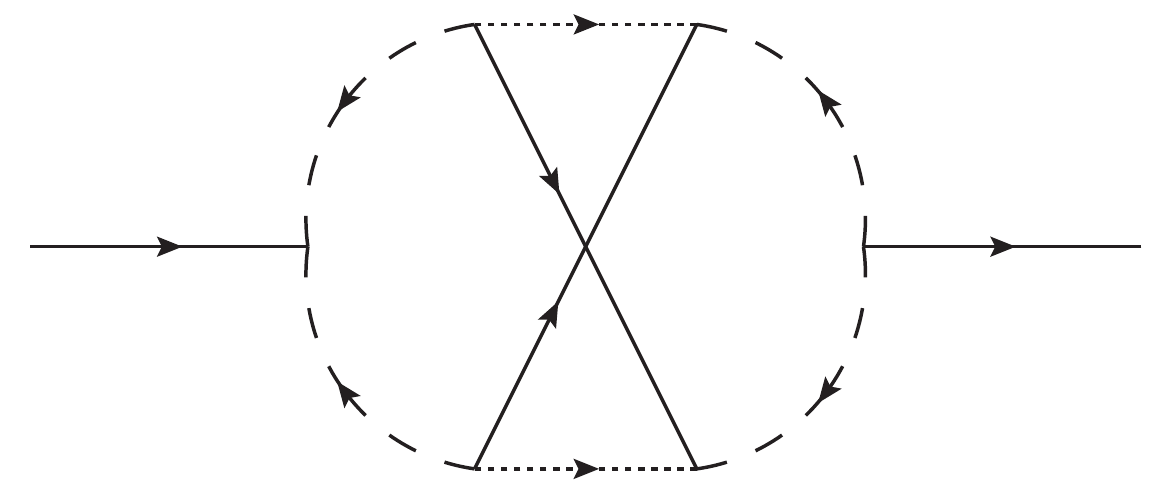}\hspace{.2cm}
\includegraphics[scale=0.23]{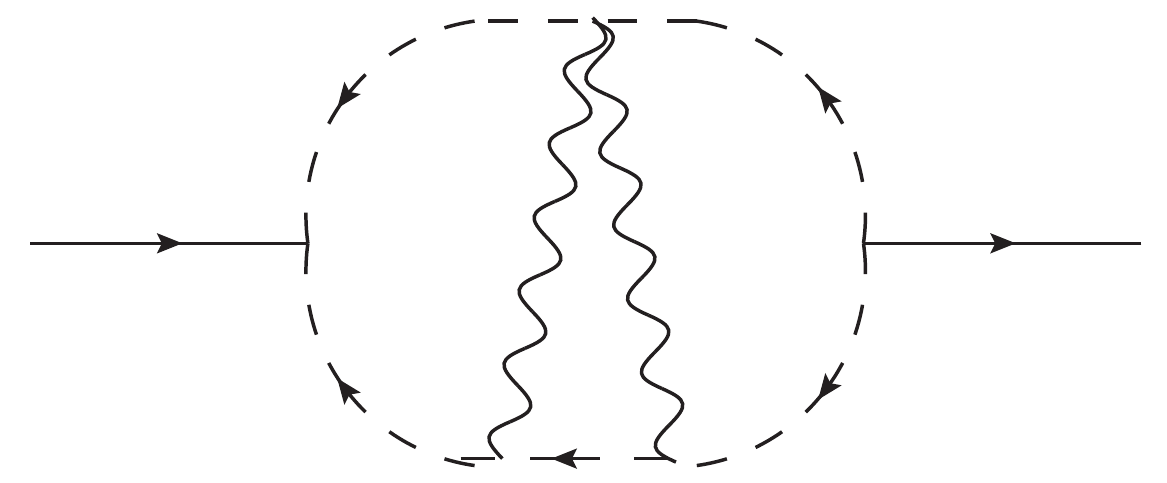}\hspace{.2cm}
\includegraphics[scale=0.23]{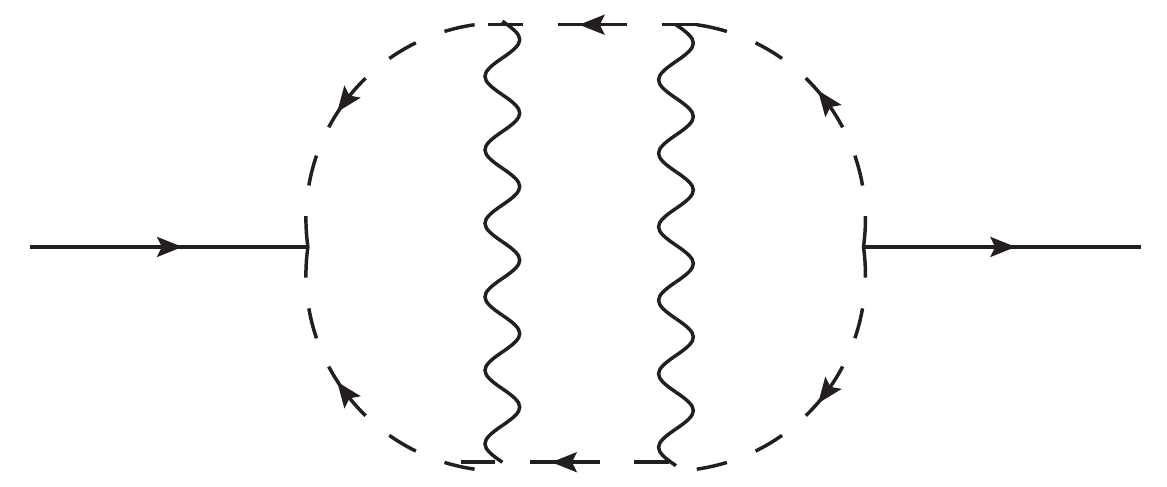}\hspace{.2cm}
\includegraphics[scale=0.23]{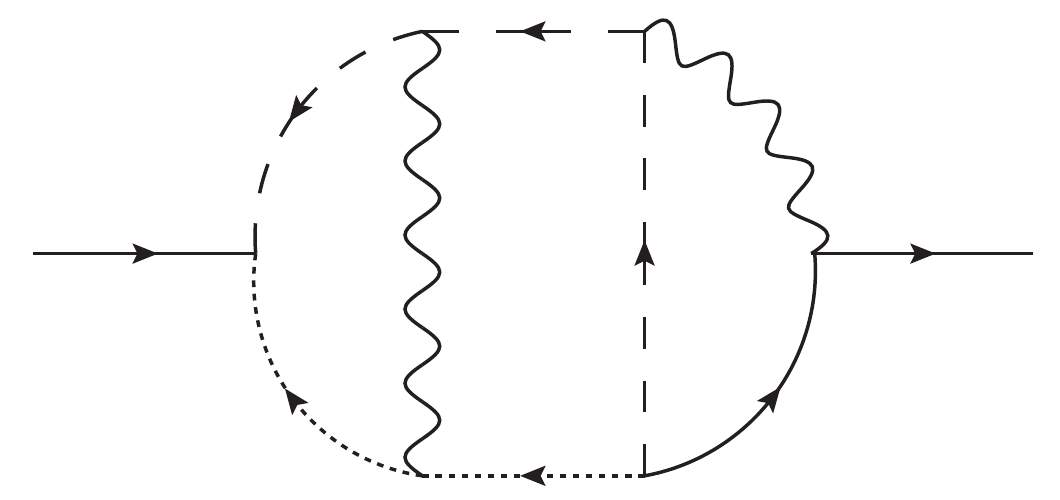}
\end{center}
\caption{Three-loop corrections to the scalar propagator}
\label{Fig::all1}
\end{figure}

\begin{figure}[H]
\begin{center}
\includegraphics[scale=0.25]{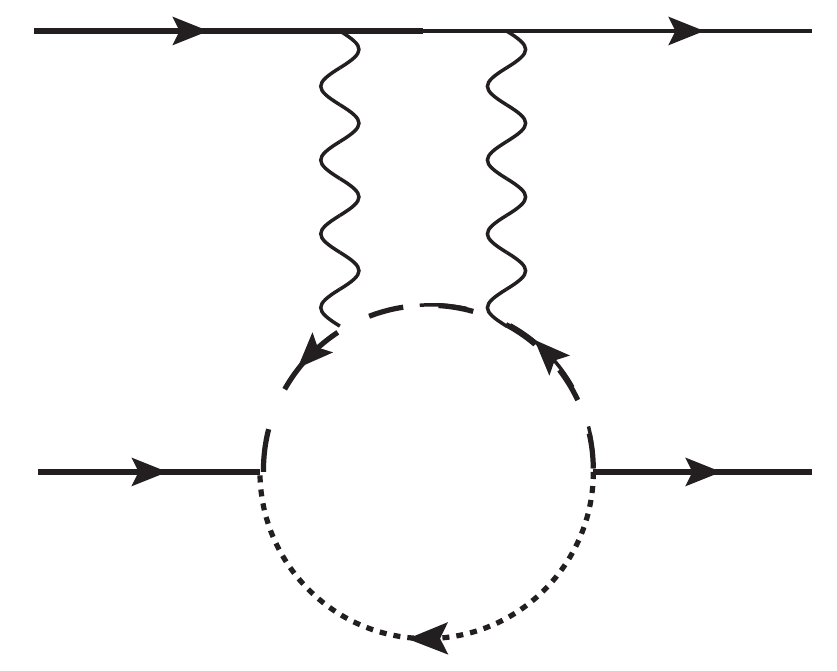}\hspace{.2cm}
\includegraphics[scale=0.25]{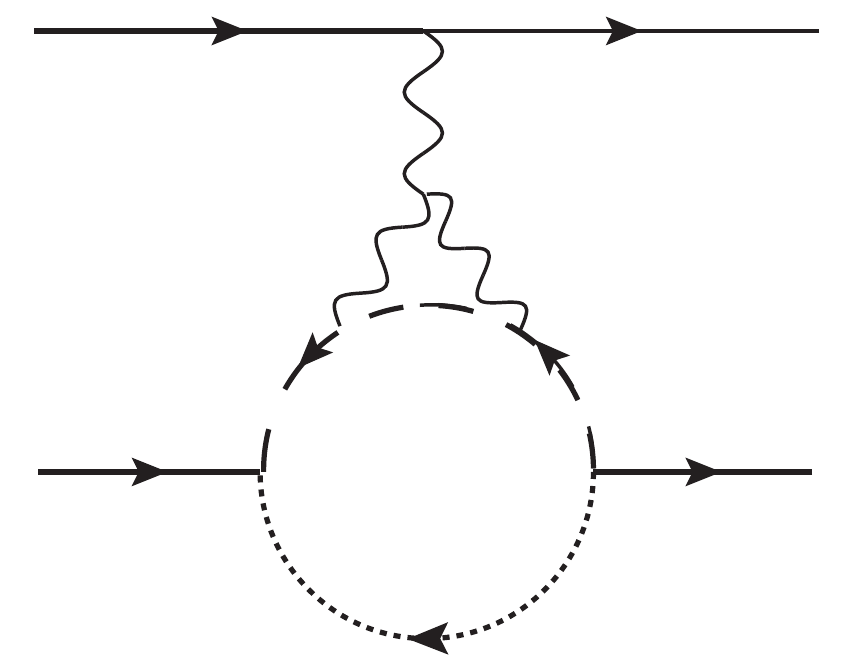}\hspace{.2cm}
\includegraphics[scale=0.25]{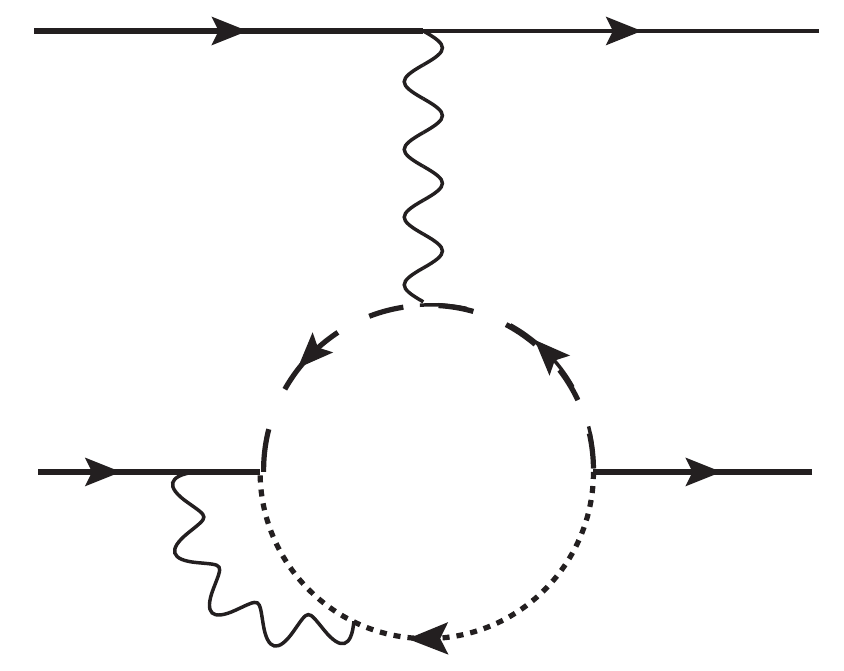}\hspace{.2cm}
\includegraphics[scale=0.25]{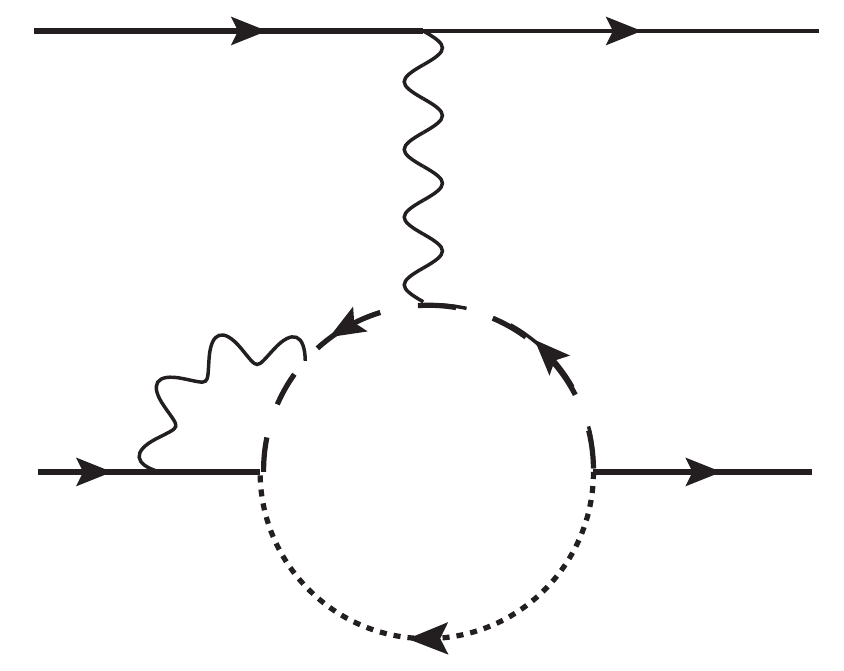}\hspace{.2cm}
\includegraphics[scale=0.25]{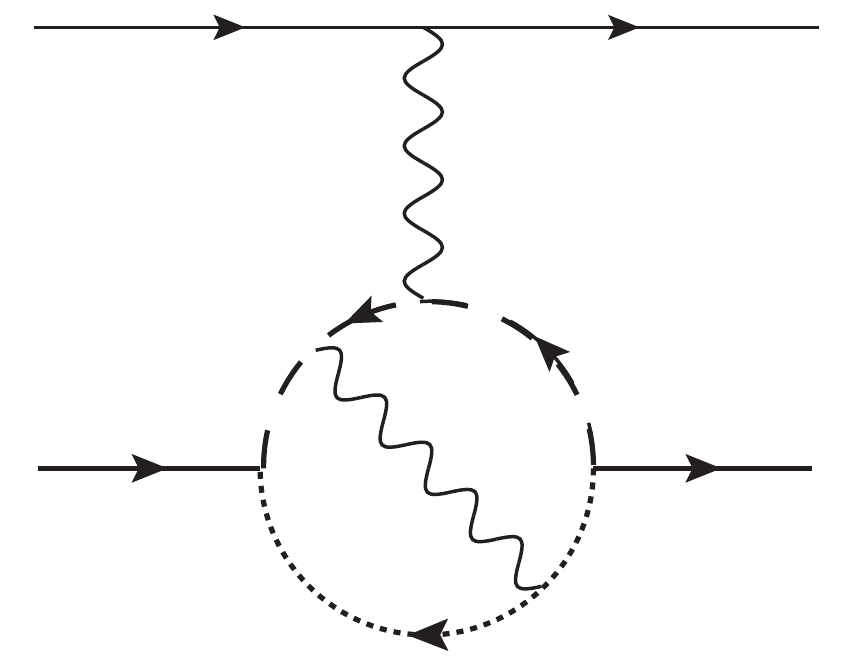}\hspace{.2cm}
\includegraphics[scale=0.25]{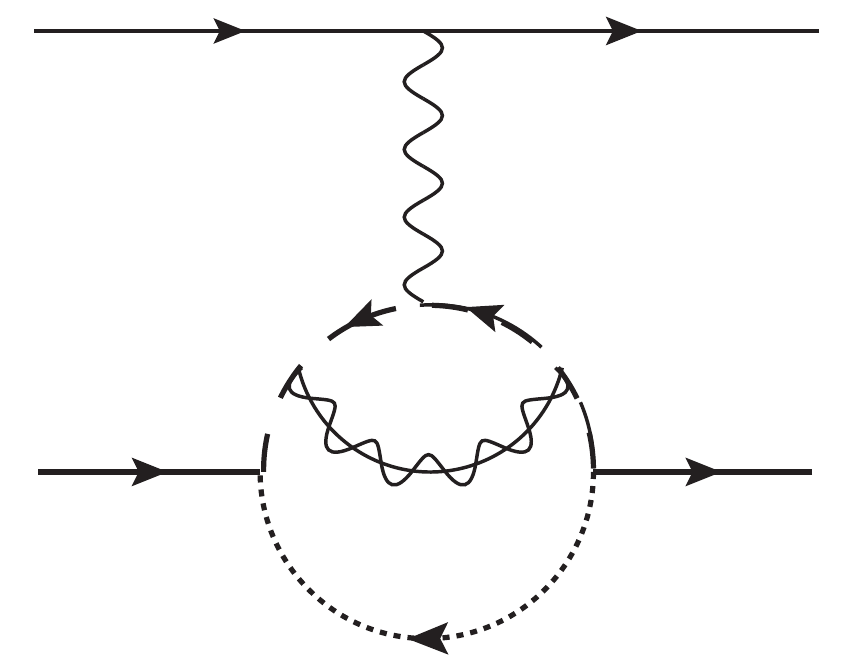}\\
\includegraphics[scale=0.3]{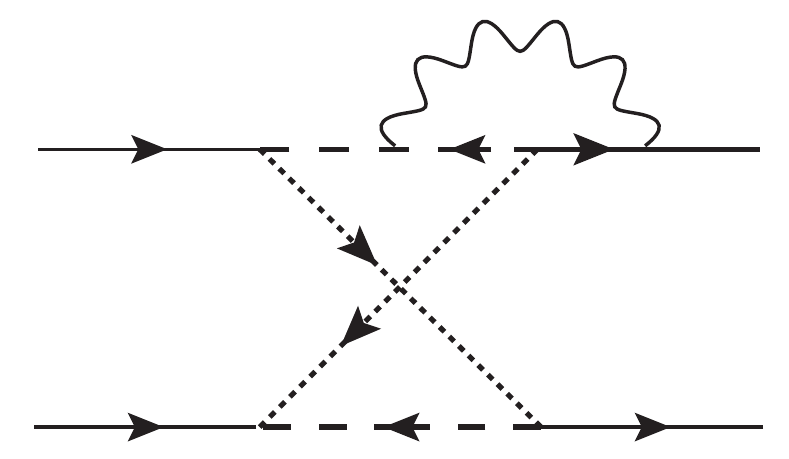}\hspace{.2cm}
\includegraphics[scale=0.3]{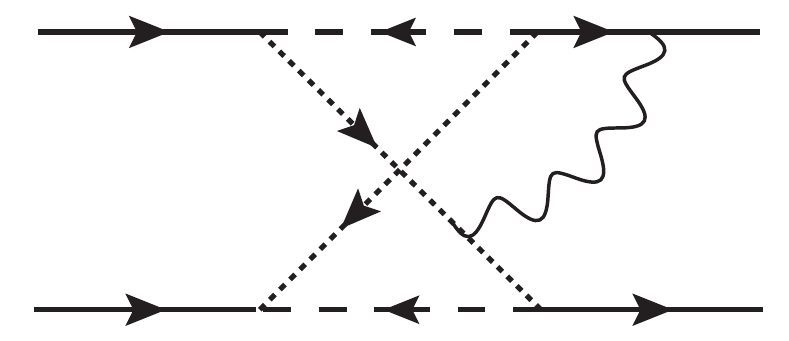}\hspace{.2cm}
\includegraphics[scale=0.3]{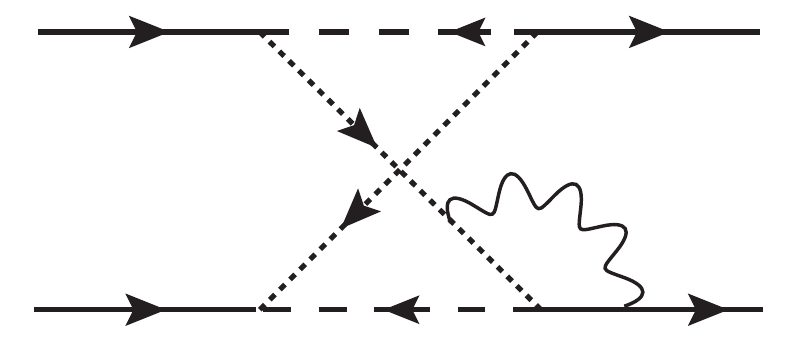}\hspace{.2cm}
\includegraphics[scale=0.3]{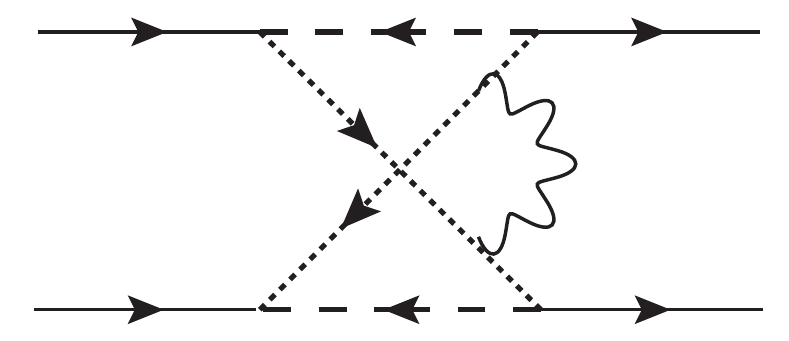}\\[0.2cm]
\includegraphics[scale=0.3]{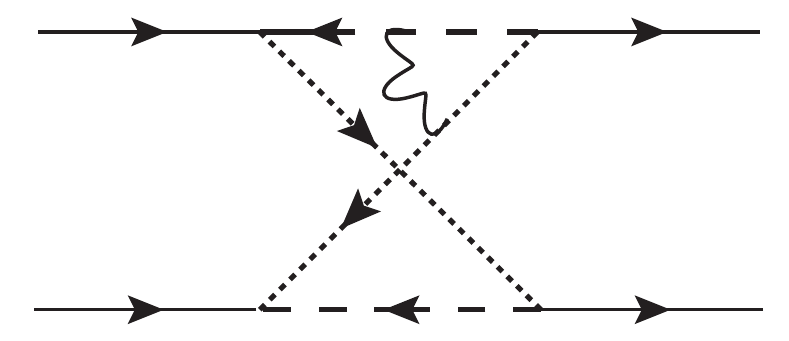}\hspace{.2cm}
\includegraphics[scale=0.3]{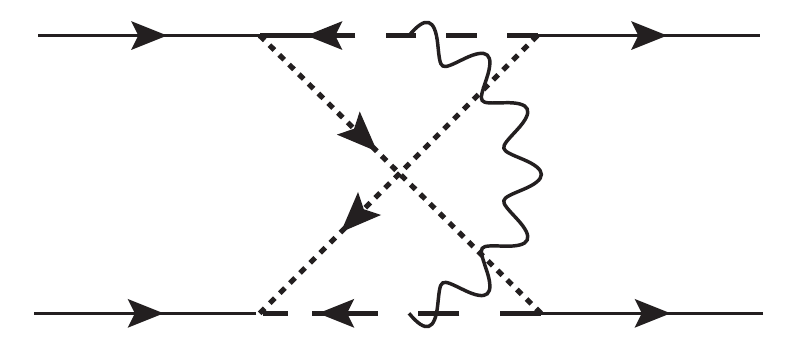}\hspace{.2cm}
\includegraphics[scale=0.3]{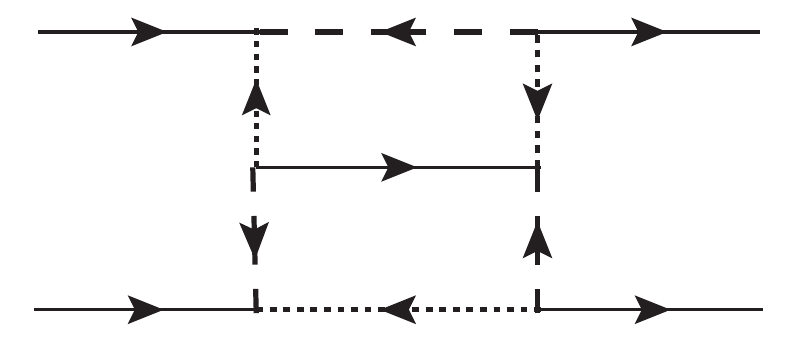}

\end{center}
\caption{Diagrams inserted inside two legs of Figure \ref{Fig::tree}}
\label{Fig::all2}
\end{figure}

\begin{figure}[H]
\begin{center}
\includegraphics[scale=0.3]{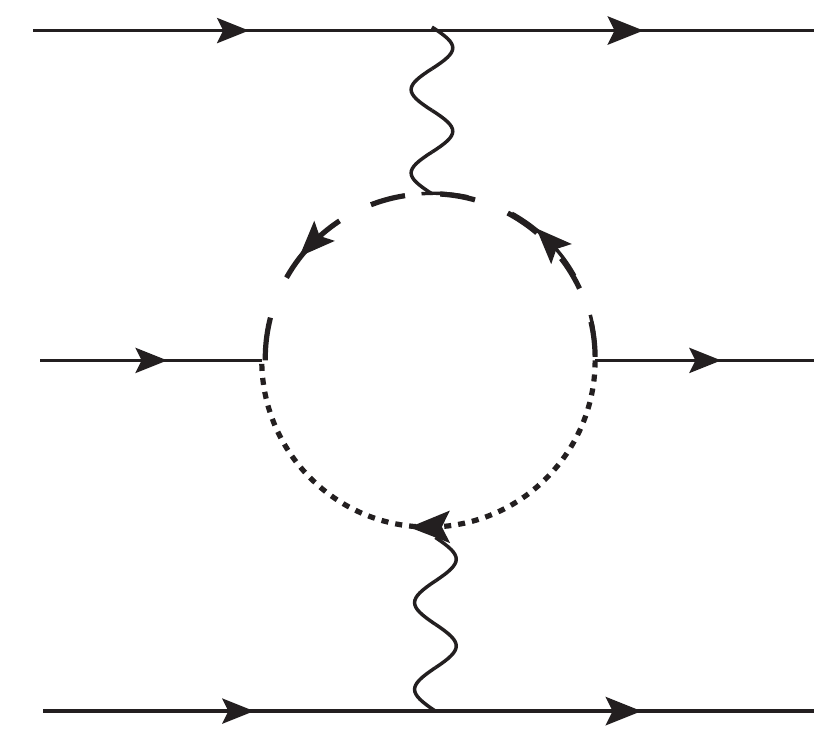}\hspace{.2cm}
\includegraphics[scale=0.3]{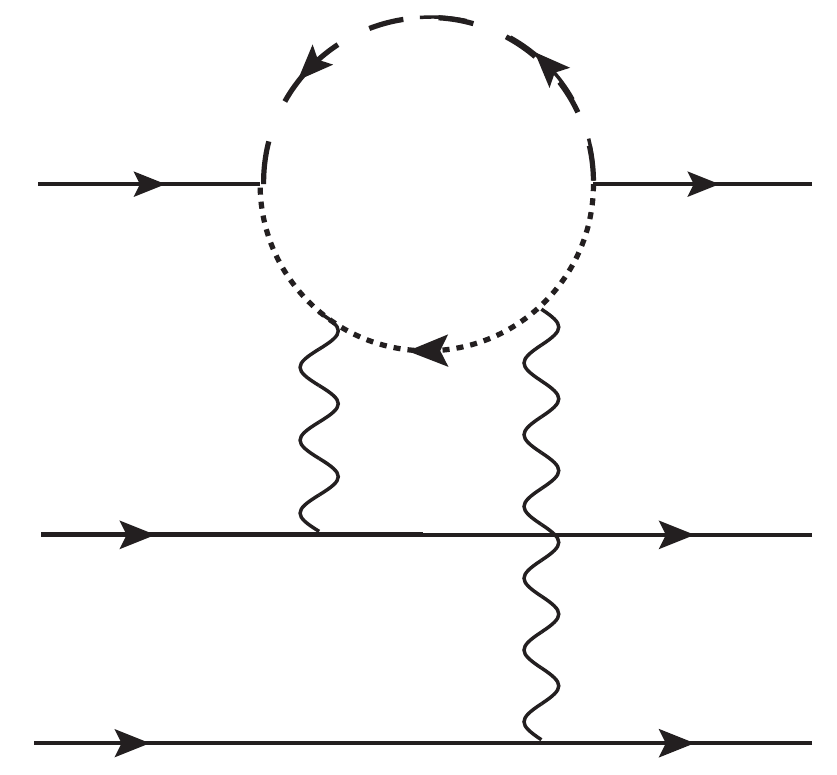}\hspace{.2cm}
\includegraphics[scale=0.3]{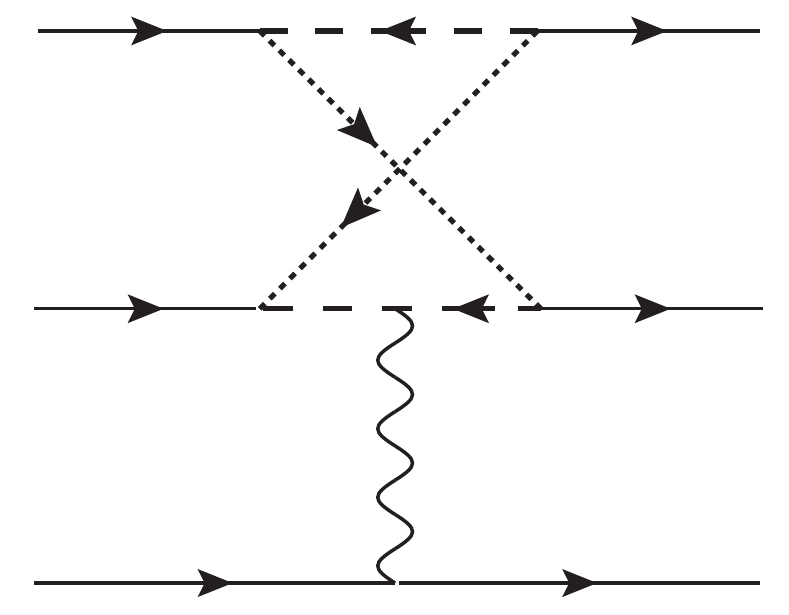}\hspace{.2cm}
\includegraphics[scale=0.3]{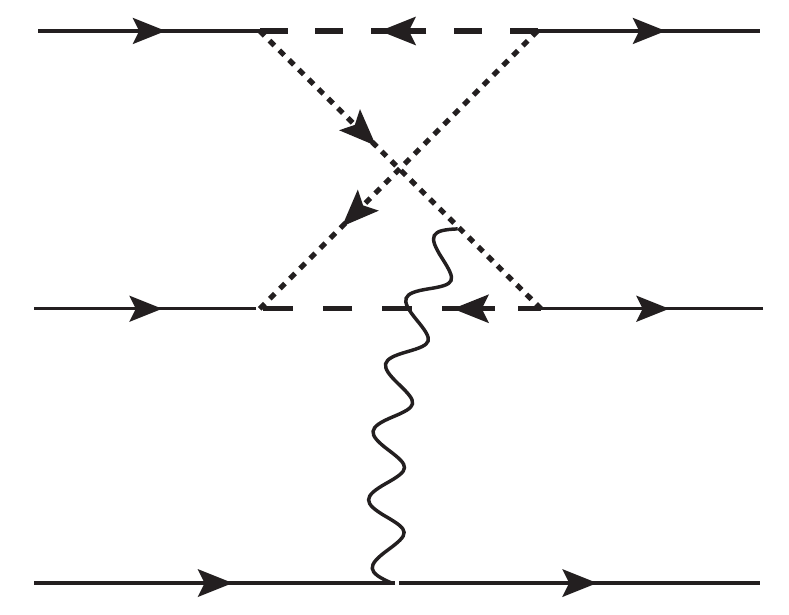}
\end{center}
\caption{Diagrams inserted inside three legs of Figure \ref{Fig::tree}}
\label{Fig::all3}
\end{figure}

\paragraph{Three loops:} 
At order $g^6$ we analyze all diagrams that could provide a $\zeta(5)$ term (we
do not consider diagrams made of $\zeta(3)$-subdiagrams). They can arise only from 
the $C'_{(4)}$, $C'_{(5)}$ and $C'_{(6)}$ building blocks. Such a list is rather long as one can see from  Figures \ref{Fig::all1}, \ref{Fig::all2} and \ref{Fig::all3}. However, in the $\n=0$ theories all these diagrams have a color factor which is either zero or subleading in $N$.

\begin{figure}[H]
\begin{center}
\includegraphics[scale=0.45]{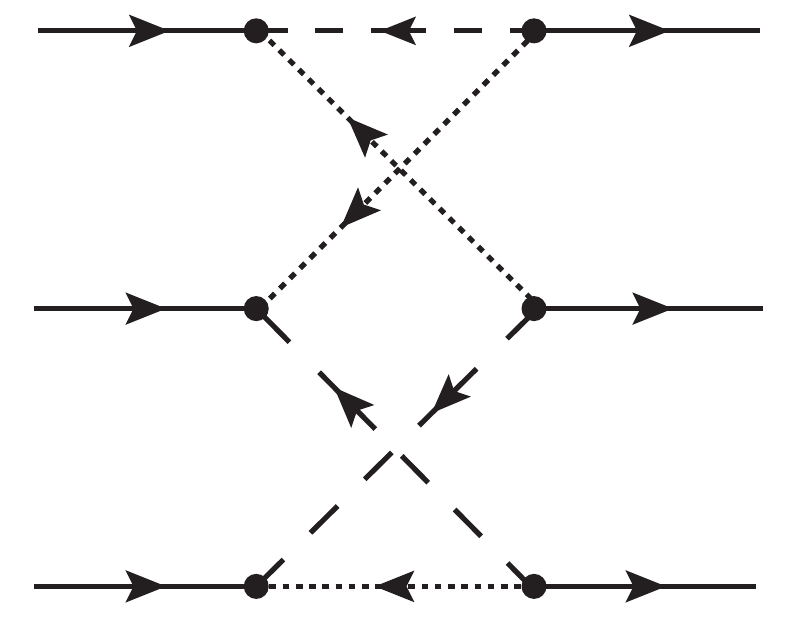}
\end{center}
\caption{Diagram to be inserted inside Figure \ref{Fig::tree}. Its space-time loop integral is denoted by $W_6(x)$ and does not depend on $n$. When we insert this diagram inside $G_n$ it yields a 
prefactor denoted by $K_n$.}
\label{Fig:exagon}
\end{figure}

The unique special diagram that deserves to be analyzed in detail is the one depicted in Figure \ref{Fig:exagon}, which we call the ``hexagon diagram'' since it gives rise to an hexagon when it is unfolded. We analyze its contribution in the following subsection.

\subsection{Hexagon diagram in even and odd correlators}
First of all, it is obvious that 
the hexagon diagram cannot be inserted inside $G_2$. This fact confirms that for the $\nu=0$ theories
\begin{equation}
\gamma_2= 1+\cO(\lambda^4)
\end{equation}
in agreement with the matrix model result.

We now evaluate the contribution of the hexagon diagram to $G_3$ and $G_4$. 
Remembering that each vertex brings a power of $\sqrt{2} g$ and computing the
corresponding color factor in the large-$N$ limit, we obtain the following
prefactors
\begin{align}\label{colorExagon}
K_{3} \simeq 2 \lambda^3~, \hspace{1cm} K_{4} \simeq 0~.
\end{align}
This shows that the hexagon diagram is planar only inside $G_3$
whereas it is subleading, and hence non-planar, inside $G_4$. Therefore we conclude that for 
$\nu=0$ the factor $\gamma_3$ receives a three-loop correction, whereas $\gamma_4$
remains 1, at least up to four loops, namely
\begin{equation}
\gamma_4=1+\cO(\lambda^4)~.
\end{equation}
This diagrammatic analysis highlights the difference between even and odd correlators. For the
two-point functions of even operators we expect the same cancellations in the planar limit also at higher loops, so that the total result for any even correlator in the $\n=0$ theories is equal to 
the $\cN=4$ one, confirming the matrix model predictions.

Let us now return to $G_3$ and $\gamma_3$. The relevant space-time integral $W_6(x)$
contributing to the two-point function of $O_3$ has been computed in appendix C 
of \cite{Billo:2017glv} and is
\begin{equation}
\label{exagonintegral}
W_6(x) = \Big(\frac{-1}{16\pi^2}\Big)^3\,\times\,\frac{20 \zeta(5)}{3}\,\times\,\Big(\frac{1}{4\pi^2 x^2}\Big)^3~.
\end{equation}
The space-time dependence is the expected one because of conformal invariance as prescribed by \eqref{TwopointFT}, and so we can just focus on the remaining terms.
Multiplying by $K_3$ given in \eqref{colorExagon} and taking into account a symmetry factor of $3!$ corresponding to all possible ways of inserting the hexagon inside the correlator, we get:
\begin{equation}\label{G3}
\gamma_3=1-10\,\zeta(5)\,\widehat{\lambda}^{\,3} + \cO(\widehat{\lambda}^{\,4})~.
\end{equation} 
This result confirms the matrix model prediction \eqref{gamma3expl} up to three loops.

Notice that if we had inserted the hexagon diagram inside correlators with odd operators with higher order, we would have obtained a non-planar result. This implies that the coefficient $G_{2k+1}$ with $k>1$ starts deviating from the $\cN=4$ result at higher-loop orders.

\subsection{Higher loops for higher order correlators}
We are able to generalize the previous reasoning to higher correlators. In particular we show that for each correlator in $\n=0$ theories the first deviation from $\cN=4$ theory is given by the insertion of the following diagram:
\begin{align}
\label{Davydichev}
\parbox[c]{.2\textwidth}{\includegraphics[width = .2\textwidth]{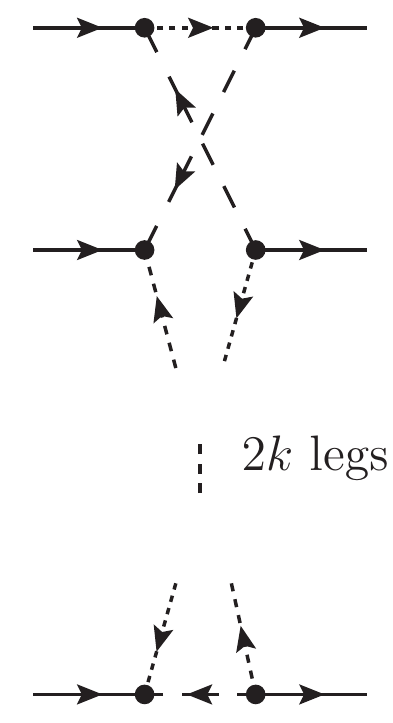}} \quad
=~ \Big(\frac{-1}{16\pi^2}\Big)^k \,\times\,\binom{2k}{k} \,\frac{\zeta(2k-1)}{k} \,\times \,
\Big(\frac{1}{4\pi^2 x^2}\Big)^k~.
\end{align}
This is a hypermultiplet loop with $2k$ adjoint scalar legs, and represents a generalization of the hexagon diagram. The result (\ref{Davydichev}) is obtained by exploiting a map with the $k$-loop ladder diagrams contributing to the four-point function of a $\phi^3$-scalar theory, which was computed in \cite{Usyukina:1993ch}.
Following the reasoning of Figure \ref{Fig:BBlocks}, the color factor arising from the insertion of the diagram with $2k$ legs is the totally symmetric tensor
\begin{equation}
\label{colorDiag}
C^\prime_{(a_1 \dots a_{2k})} = \big(\Tr _\cR-\Tr_{\mathrm{adj}}\big) T_{(a_1}\dots T_{a_{2k})}~.
\end{equation}
We again evaluate this color factor using FormTrace \cite{Cyrol:2016zqb}, and find that at large $N$
it is given by
\begin{align}\label{oddeven}
C^\prime_{(a_1 \dots a_{2k})} \simeq \begin{cases} -\nu\, \tr T_{(a_1}\dots T_{a_{k}}\,\tr T_{a_{k+1}}\dots T_{a_{2k})} ~~~~~ k ~\mathrm{even}\\ ~\,2\,\, \tr T_{(a_1}\dots T_{a_{k}}\,\tr T_{a_{k+1}}\dots T_{a_{2k})} ~~~~~ k~ \mathrm{odd} \end{cases}~.
\end{align}
This shows that this $2k$-leg diagram can contribute to correlation functions in $\n=0$ theories only for odd $k$.
 
Proceeding as before, we see that for each $G_k$ with $k=2i+1$, the first non-trivial deviation from $\cN=4$ appears at $\cO(\lambda^{2i+1})$, and its full contribution is given by the insertion of the diagram \eqref{Davydichev} with $(4i+2)$ legs. We have explicitly computed the $i=2$ and $i=3$ cases corresponding to $k=5$ and $k=7$ respectively, finding
\begin{align}\label{G5}
\gamma_5&= 1-\frac{63}{2}\,\zeta(9) \,\widehat{\lambda}^{\,5} +\cO(\widehat{\lambda}^{\,6})~, \notag \\
\gamma_7&= 1-\frac{429}{4}\,\zeta(13) \,\widehat{\lambda}^{\,7} +\cO(\widehat{\lambda}^{\,8})~.
\end{align}
These perfectly match the first perturbative corrections in the matrix model
results (\ref{oddresults}). Then it is quite easy to infer the general behavior for the $\nu=0$ theories:
\begin{equation}
\gamma_{2i+1}= 1-\frac{1}{2^{2i-1}}\binom{4i+2}{2i+1}\,\zeta(4i+1)
\, \widehat{\lambda}^{\,2i+1} +\cO(\widehat{\lambda}^{\,2i+2})~.
\end{equation}
For each correlator we recognize the contribution coming from the integral
\eqref{Davydichev} with $4i+2$ legs, while the coefficient $\frac{1}{2^{2i-1}}$ is due to the multiplicity and color factors.

This diagrammatic analysis can be readily generalized to the one-point functions of chiral primaries
in presence of the Wilson loop, and also in this case one nicely recovers the first perturbative
terms in perfect agreement with the matrix model results.

\section{Resummation in the $\nu=0$ theories}
\label{sec:resumm}

The computational tools that we have developed in the previous sections are particularly efficient
for the $\mathbf{D}$, $\mathbf{E}$ models and allow us to generate perturbative expansions to
a very high order without too much effort. In this section we try to analyze these long 
perturbative expansions in order to get some preliminary non-perturbative information.
This is clearly a very important issue given the difficulties that are notoriously
encountered in the strong coupling analysis of these models, at both numerical 
and analytical level
\cite{Passerini:2011fe,Baggio:2016skg,Fiol:2015mrp,Rodriguez-Gomez:2016ijh}.

We begin with a short recap of the available results. The strong-coupling scaling of the two-point 
function $G_{2}$ for SQCD ({\it{i.e.}} theory \textbf{A} with $\nu = 1$) was considered in \cite{Rodriguez-Gomez:2016ijh}. The same scaling was reproduced in \cite{Baggio:2016skg} 
which extended the analysis to $G_{4}$. For the correction factors $\gamma_{2}$ and 
$\gamma_{4}$ the results are that for large $\lambda\to\infty$,
\begin{equation}
\begin{aligned}
\gamma_2\big|_{\nu=1}&\,\sim \,\Big(\frac{\log \lambda}{\lambda}\Big)^{2}~,\\
\gamma_4\big|_{\nu=1}&\,\sim\, \lambda^{2}\,\Big(\frac{\log \lambda }{\lambda}\Big)^{6}~.
\end{aligned}
\end{equation}
Similarly the Wilson loop scaling was first considered in \cite{Passerini:2011fe} for SQCD and the analysis was extended to general $\nu$ in \cite{Fiol:2015mrp}, which obtained for large $\lambda$
the following behaviors\,%
\footnote{
The Wilson loop v.e.v. $w$ in the $\nu=0$ models is equal to the one in $\mathcal{N}=4$ SYM. 
The power correction, which is well-known from the matrix model solution, has been 
a hard test of AdS/CFT correspondence and has been recovered only quite recently in \cite{Medina-Rincon:2018wjs}.}
\begin{equation}
\begin{aligned}
w\big|_{\nu = 0} &\,\sim\, \lambda^{-3/4}\,\,\rme^{\sqrt{\lambda}} ~,\\ 
w\big|_{\nu = \frac{1}{2}} &\,\sim\, \lambda^{5} ~,\\
w\big|_{\nu = 1} & \,\sim\, \lambda^{3} ~. 
\end{aligned}
\end{equation}
Here, we point out two shortcomings of these results:
\begin{itemize}
    \item
        Apart from the case of the Wilson loop without insertions, it has not been possible to figure out the coefficient that should go in front of these scaling factors, which have been obtained primarily using the Wiener-Hopf method which is quite hard to control, see \textit{e.g.} Appendix D of \cite{Baggio:2016skg}.

    \item
        All these results are for the even sector of observables. This means in particular that they don't shed any light on the difference between the strong coupling dynamics of the \textbf{D} and 
\textbf{E} models compared to the $\cN=4$ SYM theory. 
\end{itemize}
Although the first point is beyond the reach of the numerical analysis of the long perturbative expansions we employ, we are going to explore the structure that can be seen from such an approach. 
We also take a first step addressing the second point, presenting strong numerical evidence that the
two-point correction $\gamma_3$ and one-point correction $\Delta w_{3}$ have a 
power-law growth. 

\subsection{The two-point function correction factor $\gamma_{3}$}

We begin our analysis with the discussion of the correction coefficient $\gamma_{3}$. 
In order to compute a long expansion, we use the method described in Sections~\ref{sec:nu0-fla} 
and \ref{sec:nu0-ca}, and write (see for instance (\ref{idgsa}) for $i=1$)
\begin{equation}
\label{gamma3exp}
\gamma_{3}=\big(\mathbb{1}+\mathsf X+\mathsf X^{2}+\mathsf X^{3}+\cdots\big)_{1,1}~,
\end{equation}
where the matrix $\mathsf{X}$ is defined in (\ref{Xalb}).
The powers $\mathsf{X}^k$ can be computed using the sum rule
\begin{equation}
\begin{aligned}
\cG(t,t') &= \cG(t',t) =8\,\sum_{m=1}^{\infty}(2m+1)\,J_{2m+1}(t)\,J_{2m+1}(t')\\
&= \frac{4 t t' }{t^{2}-t'^{2}}\,\Big[t \,J_1(t) \, J_0(t')-t' \, J_0(t)\,   J_1(t')\Big]-8 J_1(t) J_1(t')~.
\end{aligned}
\end{equation}
This relation gives the following pattern of iterated integrals
\begin{equation}
\begin{aligned}
\mathsf{X}_{1,1} &= -24\,\int_{0}^{\infty}\! Dt\,J_{3}(t z)^{2}~, \\[1mm]
(\mathsf{X}^{2})_{1,1} &= +24\,\int_{0}^{\infty}\!Dt\,Dt'\,
J_{3}(tz)\,\cG(tz, t'z)\,J_{3}(t'z)~,\\[1mm]
(\mathsf X^{3})_{1,1} &= -24\,\int_{0}^{\infty}\! Dt\,Dt'\,Dt''\,
J_{3}(tz)\,\cG(tz, t'z)\,\cG(t'z, t''z)\, J_{3}(t''z)~,
\end{aligned}
\end{equation}
and so on, where for convenience we have set
\begin{equation}
Dt=\frac{dt}{t}\,\frac{\rme^{t}}{(\rme^{t}-1)^{2}}
\quad\mbox{and}\quad z= \frac{\sqrt\lambda}{2\pi} = \sqrt{2\widehat\lambda}~.
\label{Dtz}
\end{equation}
After expansion in powers of $z$, the $t, t', \dots$ integrals are trivial using (\ref{tint}).
The first cases are 
\begin{align}
\mathsf{X}_{1,1} &=  -\frac{5}{4}\,\zeta(5)\, z^6+\frac{105}{16}\,\zeta(7)\, z^8
-\frac{1701}{64}\,\zeta(9)\, z^{10}+\frac{12705}{128}\,\zeta(11) \,
z^{12}-\frac{184041}{512}\,\zeta(13) \,z^{14}\notag \\
&\quad+\frac{5270265}{4096}\,\zeta(15)\,z^{16}
-\frac{18803785}{4096}\, \zeta(17) \,z^{18}+\cdots~,\notag \\[2mm]
(\mathsf{X}^{2})_{1,1} &=\frac{25}{16}\,\zeta(5)^2\, z^{12}-\frac{525}{32}\,\zeta(5)\,\zeta(7)
\,z^{14}
+\Big(\frac{44835}{1024}\,\zeta(7)^2+\frac{8505 }{128}\,\zeta(5)\,\zeta(9)\Big) z^{16}
\label{Xn}\\
&\quad-\Big(\frac{368235}{1024}\,\zeta(7)\,\zeta(9)
+\frac{63525}{256}\,\zeta(5)\,\zeta(11)\Big) z^{18}+\cdots~, \notag \\[2mm]
(\mathsf{X}^{3})_{1,1} &= -\frac{125}{64}\,\zeta(5)^3\, z^{18}+\cdots~.\notag
\end{align}
The algorithm can be easily coded and pushed to large order. Here, we discuss the analysis of the perturbative expansion of $\gamma_{3}$ up to order $\lambda^{100}$. 

\begin{figure}[H]
\begin{center}
\includegraphics[scale=0.52]{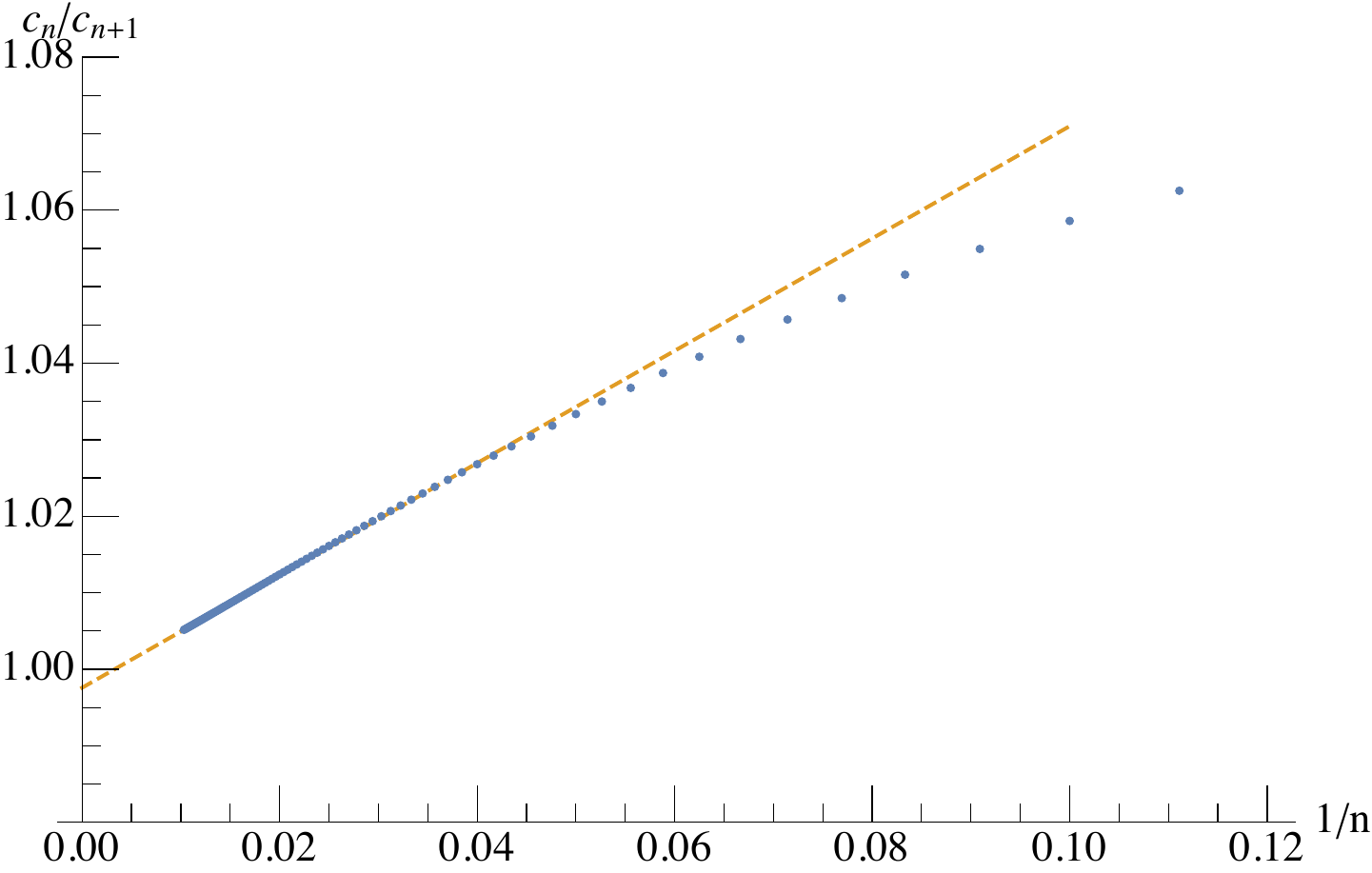}\hspace{0.2cm}
\includegraphics[scale=0.52]{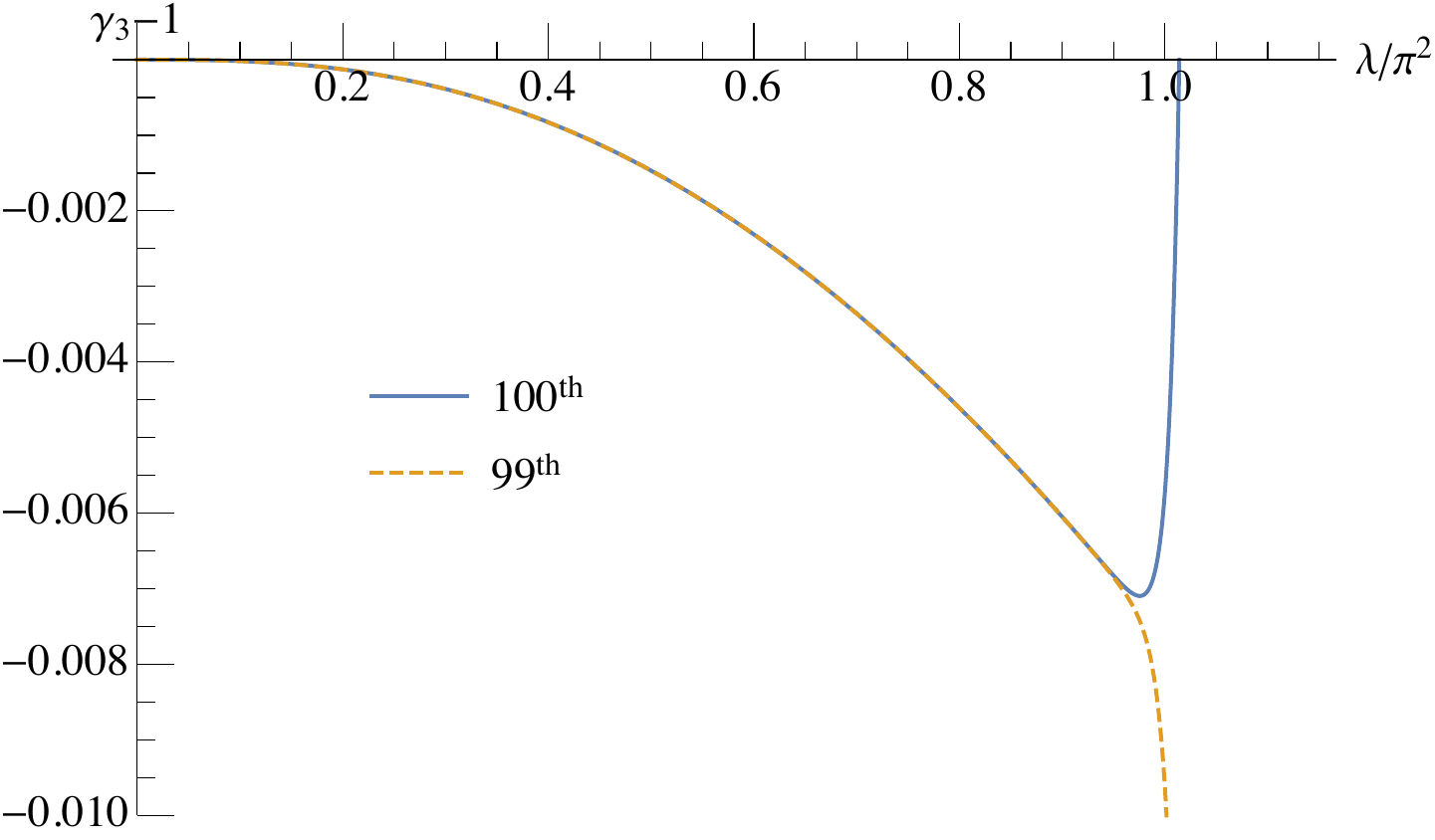}
\end{center}
\caption{(Left) The Domb-Sykes plot to estimate of the radius of convergence $R$ 
by ratio test $c_{n}/c_{n+1}$. The intercept of the linear asymptote is 0.997, very close to 1. (Right) The $99^\text{th}$ and $100^\text{th}$-order truncated Taylor polynomials of $\Delta\gamma_{3}$.}
\label{Fig::Radius}
\end{figure}

Plugging (\ref{Xn}) in (\ref{gamma3exp}), we obtain an explicit expansion of the form 
\begin{equation}
\Delta \gamma_{3}\,\equiv\,
\gamma_3-1= \sum_{n=3}^{\infty}c_{n}\,\Big(\frac{\lambda}{\pi^{2}}\Big)^{n}~.
\end{equation}
As a first step, we can estimate the radius of convergence $R$ (in terms of $\frac{\lambda}{\pi^{2}}$) by the ratio test, {\em i.e.} from 
\begin{equation}
R = \lim_{n\to\infty}\Big|\frac{c_{n}}{c_{n+1}}\Big|~.
\end{equation}
Using our data, the result for $R$ is shown in the left panel of Figure~\ref{Fig::Radius} where it appears that there is indeed a finite radius of convergence at $\lambda_{c}/\pi^{2}\simeq 1$.
This is confirmed if we plot the $99^\text{th}$ and $100^\text{th}$-order truncated series for 
$\Delta \gamma_{3}$, as shown in the right panel of Figure~\ref{Fig::Radius} where we observe 
the expected (alternating) numerical blow up near the estimated critical 
value of $\lambda_{c}/\pi^{2}\simeq 1$. 

\begin{figure}[H]
\begin{center}
\includegraphics[scale=0.52]{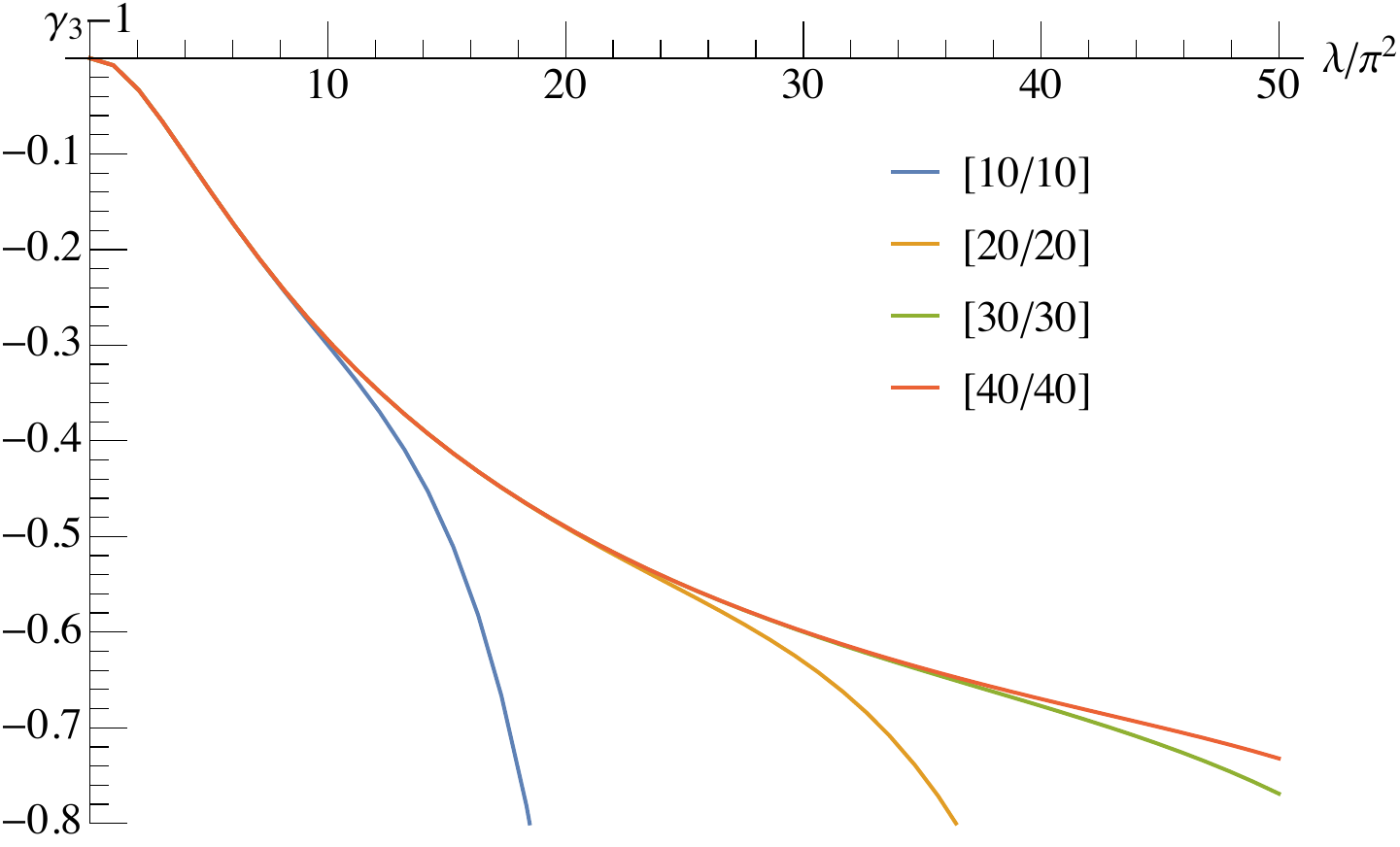} \hspace{0.2cm}
\includegraphics[scale=0.52]{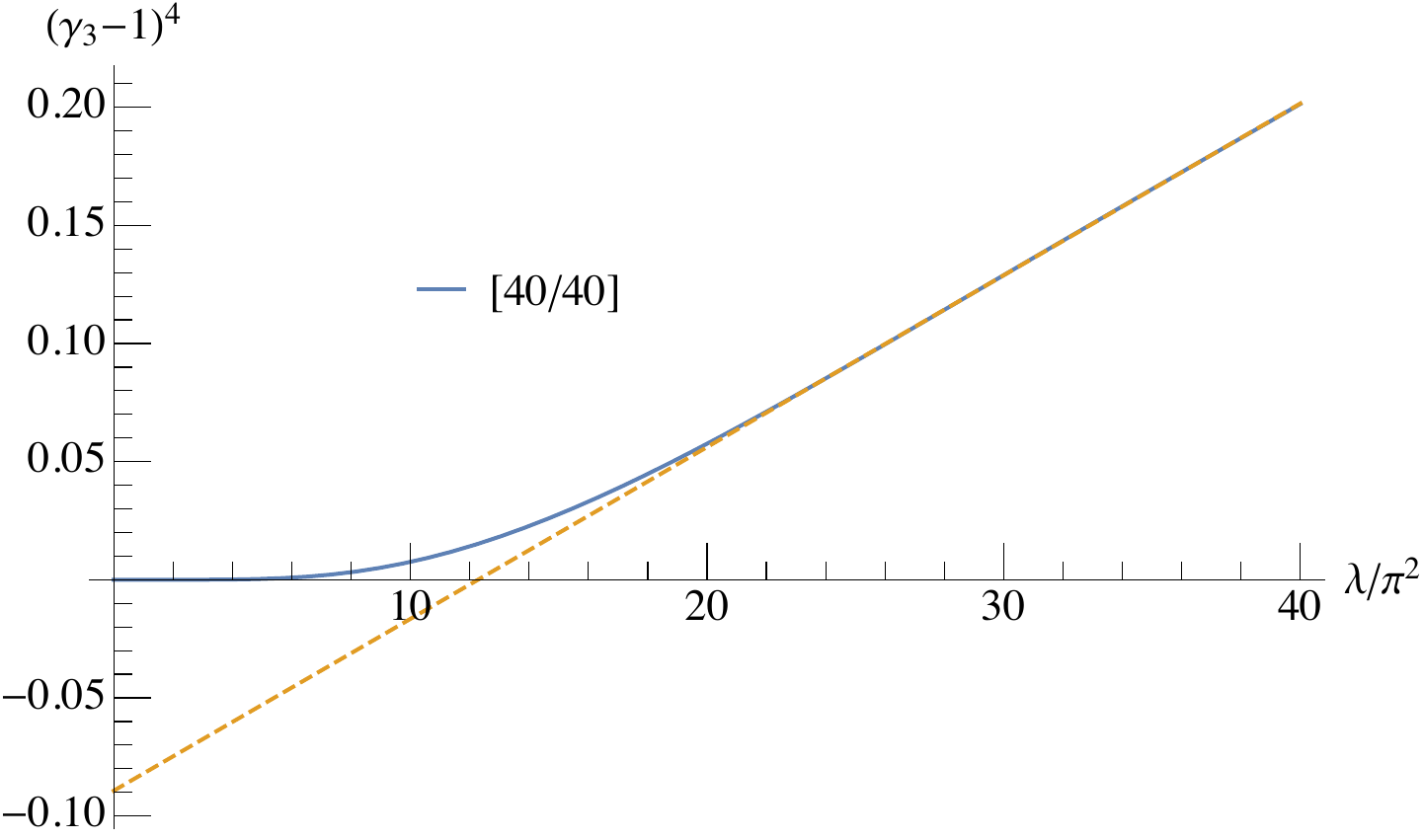} 
\end{center}
\caption{(Left) Pad\'e-Borel resummation of $\Delta\gamma_{3}$ using diagonal $[M/M]$ approximants with 
$M=10,20,30,40$. (Right) Linear fit to  $(\Delta\gamma_3)^{4}$ in the intermediate 
coupling region $\lambda/\pi^{2}\sim 30$}
\label{Fig::Borel}
\end{figure}

Further information can be gained by considering  the denominator of Pad\'e approximants to 
$\Delta\gamma_{3}/\lambda^{3}$ and looking at the zero nearest to $\lambda=0$.
Taking as an example the diagonal $[M/M]$ approximant we find the following table:
\begin{equation}
\def\arraystretch{1.3}
\begin{array}{cllll}
\toprule
\textsc{M} &  \qquad 10 & \qquad 15  & \qquad 20  & \qquad 25 \\
\midrule
\text{nearest zero}\  \frac{\lambda}{\pi^{2}} \quad& 
-1.00639 & 
-1.00191 & 
-1.00044 & 
-0.99983
 \\
\bottomrule
\end{array}\notag
\end{equation}
which strongly supports the exact result $\lambda_{c} = \pi^{2}$, due to a singularity on the negative real axis.
Unfortunately, we have no theoretical control on the properties of $\Delta\gamma_{3}$, 
like for example its large order behavior or its analyticity structure. 
This prevents us to perform any rigorous resummation of the perturbative expansion. Nevertheless, we can try to analytically continue beyond the convergence radius by
applying the Pad\'e-Borel resummation technique
\cite{janke1998resummation}. We evaluate a Pad\'e approximant $P_{[M/K]}(\lambda)$ of the 
Borel-improved series, namely
\begin{equation}
P_{[M/K]}(\lambda) = \bigg[\sum_{n=3}^{\infty}\frac{c_{n}}{(n-3)!}\,\Big(\frac{\lambda}{\pi^{2}}\Big)^{n}\bigg]_{[M/K]}~,
\end{equation}
and then transform back according to:
\begin{equation}
\Delta\gamma_{3, [M/K]}\,\equiv\, \lambda^{2}\,\int_{0}^{\infty}\!
dx\, P_{[M/K]}(x)\,\rme^{-x/\lambda}~.
\end{equation}
In the left panel of Figure~\ref{Fig::Borel} we show what is obtained by considering three diagonal 
Pad\'e approximations. 
In all cases, the reconstructed function coincides with the convergent perturbative sum when $\lambda<\lambda_{c}$. 
Beyond $\lambda_{c}$, the resummation appears to be well defined in the sense that its value at fixed $\lambda$ stabilizes
at increasing Pad\'e degree $M$.  Our data suggest that we can safely trust this reconstruction up to the rather large values 
$\frac{\lambda}{\pi^{2}}\simeq  35$ where the $M=30$ and $M=40$ curves are still indistinguishable. In this region, 
the asymptotic behavior appears to be $\Delta\gamma_{3}
\sim C\,\lambda^{1/4}$ for moderate $\lambda$, as shown in 
the right panel of Figure~\ref{Fig::Borel}. The exponent $1/4$ is just a qualitative estimate in the considered range of coupling, since one can expect also logarithmic corrections 
as discussed in \cite{Baggio:2016skg}.
It would be very interesting to match such numerical indications by an analytic strong-coupling calculation.

\subsection{The one-point function correction $\Delta w_{3}$}

\begin{figure}[H]
\begin{center}
\includegraphics[scale=0.37]{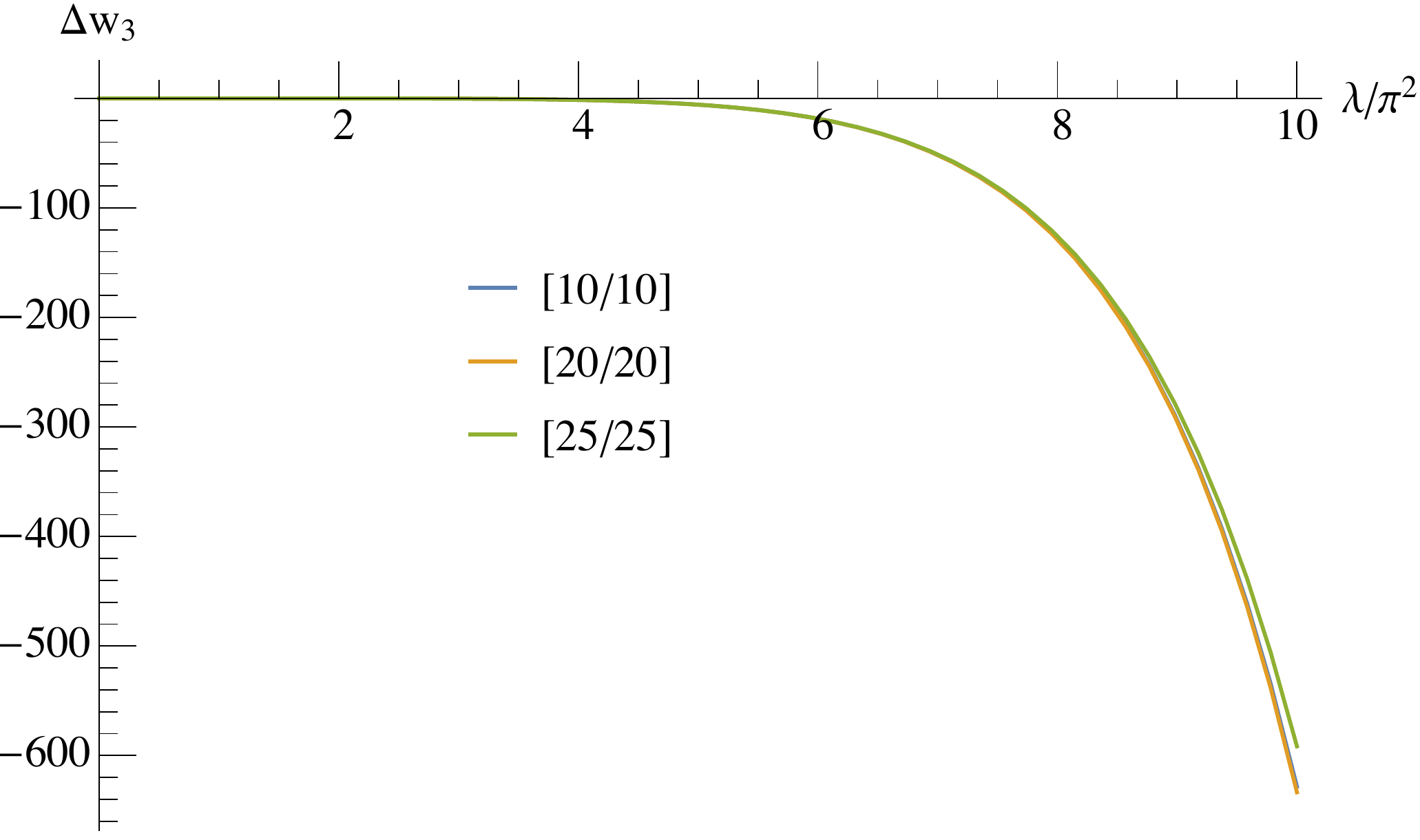}\hspace{0.5cm}
\includegraphics[scale=0.37]{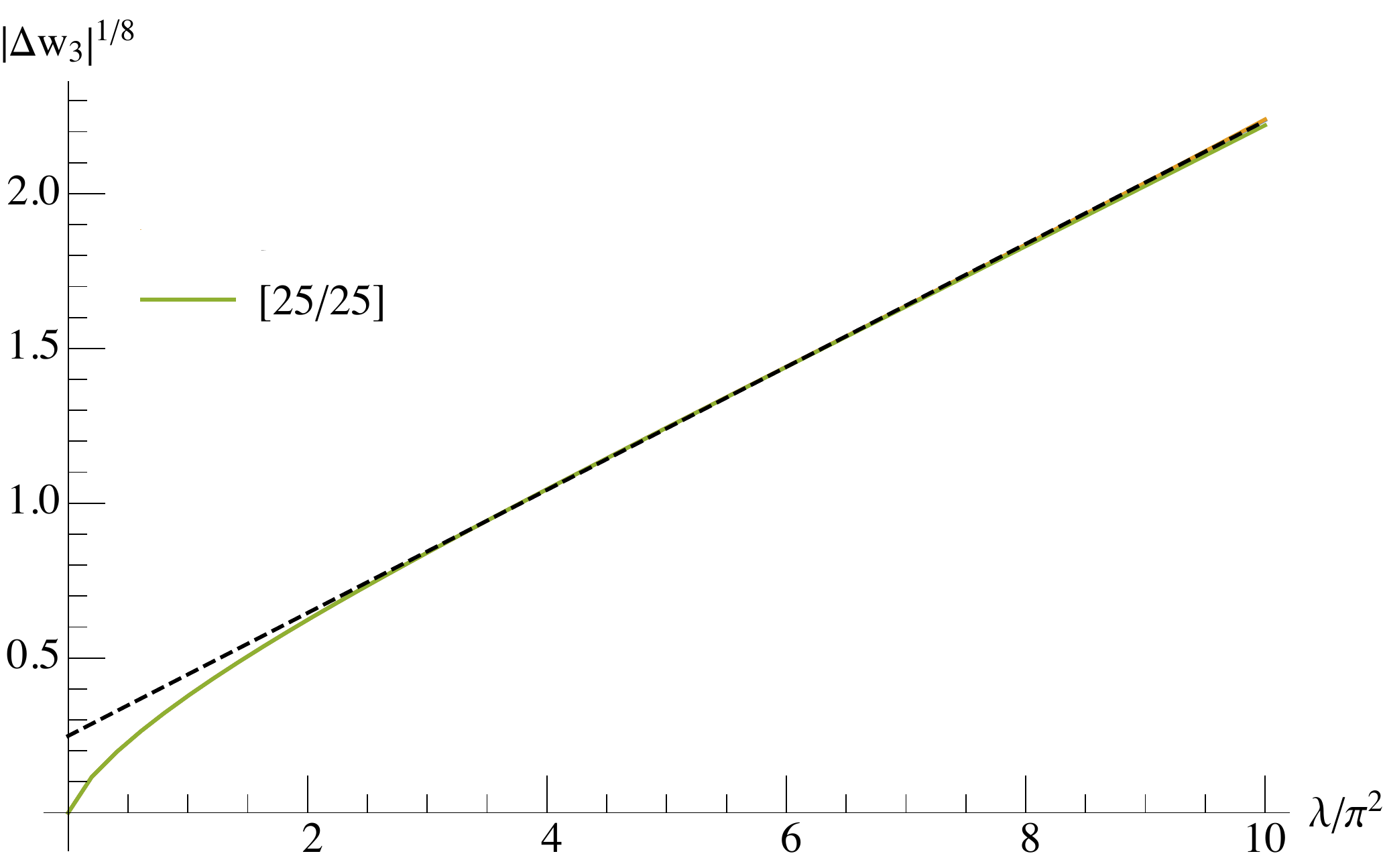}
\end{center}
\caption{(Left) Pad\'e-Borel resummation of $\Delta w_{3}$ 
with diagonal $[M/M]$ approximants. (Right) Qualitative (asymptotic) linear fit of 
$|\Delta w_3|^{1/8}$. The value $1/8$ is be taken as a simple rational approximation to the unknown exact 
exponent (up to possible logarithmic corrections).}
\label{Fig::BorelW3}
\end{figure}

The same numerical investigation can be worked out for $\Delta w_{3}$. From the
perturbative expansion up to order $\lambda^{60}$ we obtain a finite convergence 
radius consistent with that of $\Delta\gamma_{3}$ and a pattern which is very similar to that
of Figure~\ref{Fig::Radius}. The smallest singularity of the diagonal Pad\'e approximants is now
shown in the table
\begin{equation}
\def\arraystretch{1.3}
\begin{array}{cllll}
\toprule
\textsc{M} &   \qquad 10 & \qquad 15  & \qquad 20  & \qquad 25 \\
\midrule
\text{nearest zero}\  \frac{\lambda}{\pi^{2}} \quad & 
-1.01474 & 
-1.00265 & 
-1.00107 & 
-1.00006
 \\
\bottomrule
\end{array}\notag
\end{equation}
which  strongly suggests a singularity at the same position as in $\Delta\gamma_{3}(\lambda)$,
namely at $\lambda_c/\pi^{2}\simeq 1$. 
Finally, in Figure~\ref{Fig::BorelW3}, we present the results from the Pad\'e-Borel resummation. We emphasize again that this kind of resummation is just 
a numerical exploration, given the lack of theoretical control on the asymptotic properties of the perturbative expansion. 
Nevertheless, the analytic continuation of our numerical data turns out to be reliable at least up to $\lambda/\pi^{2}\simeq 10$ (left panel).  In this region, 
the asymptotic behavior appears to be $\Delta w_{3}\sim C\,\lambda^{8}$.
The specific value of the exponent should not be interpreted as an analytic prediction that 
we don't have. It is just an estimate valid in the range suggested by the numerical data. To guide
the eye, in the plot we have also included a dashed linear fit of $|\Delta w_3|^{1/8}$
to the rightmost part of the data. 
Again, we remind that such asymptotic representation may well be a crude approximation to the actual answer, due to possible logarithmic corrections that are natural in this context, but 
which are out of the reach of the current numerical analysis.

\section{Conclusions and perspectives}
\label{concl}
In this paper we have considered a set of Lagrangian $\cN=2$ conformal theories with SU($N$) gauge group. Following the localization procedure, we have reviewed the existing matrix model techniques for large values of the rank $N$. Developing both the full Lie algebra approach and the Cartan sub-algebra approach, we have computed a large set of observables in the planar limit.
We have generated efficient algorithms to produce perturbative series up to very high loop orders, without any limitations due to the conformal dimension, extending some results already present in the literature.

In the second part we have concentrated on $\cN=2$ theories whose fundamental matter content does not scale with $N$. We have shown several reasons why they can be considered as ``the next-to-simplest'' gauge theories. At the diagrammatic level it is immediate to reach the three-loops order in perturbation theory, and we can easily go even beyond for certain classes of observables. Moreover we have showed that many observables (like the Wilson loop v.e.v. and chiral correlators with even dimensions) are equivalent to those of the $\cN=4$ SYM theory in the large-$N$ limit. Only odd correlators feel the difference with $\cN=4$ and represent a set of interesting observables to explore the gravity dual of this special $\cN=2$ class of theories. 

From this perspective, a natural continuation of the present 
analysis of the field theory side could be the study of these special observables in 
the strong coupling regime $\lambda\gg 1$. To this aim, one should capture the evolution of the matrix
model eigenvalue
density by solving at large $\lambda$ the associated $\nu=0$ integral equation. As we remarked in the main text, this requires dealing with various deviations from the $\cN=4$ SYM case, 
like extra (odd) 
sources and the associated cut asymmetry, that play a role when computing the  odd observables.  
At the moment, it is unclear whether this can be done analitically, for instance by Wiener-Hopf methods.
The preliminary results presented in 
Section~\ref{sec:resumm} may be useful in this respect. 

To give an example, we presented strong support for a
finite convergence radius of the perturbative expansion at $|\lambda_{c}|/\pi^{2}=1$. 
This non-perturbative feature of the exact solution in the intermediate coupling range 
is not completely unexpected. Indeed, it resembles what happens in the $\cN=4$ SYM theory
where branch-point  
singularities are present 
on the negative real axis of the ’t Hooft coupling, with the leading one located 
at $\lambda = -\pi^{2}$ (see the discussion 
in \cite{Russo:2013sba}). The origin of this singularity is in the structure of the 
single-magnon dispersion relation, which was derived in the past
in perturbation theory \cite{Gross:2002su,Santambrogio:2002sb} and using
superconformal symmetry \cite{Beisert:2004hm,Beisert:2006ez}. 
Nowadays, it is well-understood 
in the more modern quantum algebraic curve treatment of the $\cN=4$ SYM
theory (see for instance \cite{Gromov:2017blm}).
The fact that the same singularity could also be present in the $\cN=2$ theories considered here
was already noticed in the study of mass deformations
of $\cN=4$ SYM, where $|\lambda_{c}|/\pi^{2}=1$ is indeed the radius of convergence of the free energy \cite{Russo:2014jma}.
This sort of universality could be ascribed to the similar structure of the combinatorics of planar diagrams; 
from a deeper perspective, it might be related to integrability structures yet to be fully clarified
\cite{Pomoni:2019oib}.

\vskip 1.5cm
\noindent {\large {\bf Acknowledgments}}
\vskip 0.2cm
We thank Gernot Akemann, Jorge Russo and Pierpaolo Vivo for many useful discussions.

\noindent
The work of A.L. is partially supported by ``Fondi Ricerca Locale dell'Universit\`a 
del Piemonte Orientale".
\vskip 1cm
\begin{appendix}
\section{Recursion relations in U$(N)$ theories}
\label{app:UN}

When the gauge group is U($N$) there are some modifications in the recursion formulas
described in Sections~\ref{sec:N2scft} and \ref{sec:3} which we are going to illustrate.

Let us consider a basis of $\mathfrak{u}(N)$ generators $T_b$, with $b=1,\ldots,N^2$,
normalized as in (\ref{normtrace}). Then one can show that the following fusion/fission identities hold:
\begin{equation}
\label{fusionfissionUN}
\begin{aligned}	
		\tr\big(T_b\, A\, T_b\, B\big) & = \frac{1}{2}\,\tr A\, \tr B~,\\
		\tr \big(T_b\, A \big) \,\tr\big(T_b \,B\big) & 
		= \frac{1}{2}\,\tr\big(A\,B\big)~,
\end{aligned}
\end{equation}
for any two $(N\times N)$ matrices $A$ and $B$. These are the U($N$) analogues of the
identities (\ref{fussion}) valid for SU($N$).

Given a matrix $a \in \mathfrak{u}(N)$, we consider the multitrace operators $\tr a^{n_1}\,\tr a^{n_2}\ldots$ and their v.e.v. in the Gaussian model
\begin{equation}
t_{n_1,n_2,\ldots} =\big\langle \tr a^{n_1}\,\tr a^{n_2}\ldots\big\rangle_{(0)}~.
\end{equation}
As in the SU($N$) case treated in the main text, we have $t_{2k+1}=0$
for $k=0,1,\ldots$, but differently from the SU($N$) case, now a v.e.v. $t_{n_1,n_2,\ldots}$
with an index $n_i=1$ is not any more vanishing, provided the total number of odd indices is even.

Using the previous definitions and the relations (\ref{fusionfissionUN}), it is easy to show that in
the large-$N$ limit the even single traces behave exactly like in the SU($N$) case, namely 
as in (\ref{t2kres1}). Also the even double traces at large $N$ satisfy the same factorization property
(\ref{t2k2k}) as in SU($N$) and their connected component is still given by (\ref{t2evenc}) which we
rewrite here for convenience
\begin{equation}
\begin{aligned}
t_{2k_1,2k_2}^{\mathrm{c}}=N^{k_1+k_2}\,
\frac{(2k_1-1)!!\,(2k_2-1)!!}{(k_1+k_2)\,(k_1-1)!(k_2-1)!}~.
\end{aligned}
\label{t2evencUN}
\end{equation}
On the other hand, the U($N$) odd double traces are different with respect to the SU($N$) ones in
the large-$N$ limit. Indeed, one finds
\begin{equation}
t_{2k_1+1,2k_2+1}=N^{k_1+k_2+1}\,\frac{(2k_1+1)!!\,(2k_2+1)!!}{2(k_1+k_2+1)\,k_1!\,k_2!}~,
\label{t2oddcUN}
\end{equation}
to be compared with (\ref{t2oddc}). We have verified in numerous examples that (\ref{t2evencUN})
and (\ref{t2oddcUN}) can be compactly written as
\begin{equation}
t_{n,m}^c=(2N)^{\frac{n+m}{2}}\,n\,m\,h_{n-1,m-1}
\label{tmnc}
\end{equation}
where
\begin{equation}
h_{n,m}=\frac{1}{2\pi^2}\int_{-1}^{+1}\!\int_{-1}^{+1}\!dx\,dy\,
\mathrm{arctanh}\Big(\frac{\sqrt{1-x^2}\sqrt{1-y^2}}{1-xy}\Big)\,x^n\,y^m
\label{hnm}
\end{equation}
for any $n$ and $m$. 

We finally remark that in the U($N$) matrix model the operator $\widehat{\Omega}_1(a)=\tr a$ 
is non-zero and that it mixes with all operators of odd dimensions. This fact implies that 
normal-ordered version of these operators always contains a component proportional to $\widehat{\Omega}_1(a)$.
For example, in the $\cN=4$ U($N$) SYM theory one finds that the normal-ordered single trace operator of dimension 3 at large $N$, instead of being simply $\widehat{\Omega}_3=\tr a^3$, is
\begin{equation}
O_3^{(0)}(a)=\widehat{\Omega}_3(a)-\frac{3N}{2}\,\widehat{\Omega}_1(a)~.
\end{equation}
For the \textbf{ABCDE} theories introduced in Section~\ref{sec:N2scft},
there is a modification of this result due to the interaction in the associated matrix model
and the normal-ordered operator of dimension 3 at large-$N $ is
\begin{equation}
O_3(a)=O_3^{(0)}(a)+\Delta O_3(a)
\label{O3UN}
\end{equation}
where
\begin{align}
\label{f31}
\Delta O_3(a) &=N\,\bigg[3\,\zeta(3)\,(2\nu-1) \,\widehat{\lambda}^{\,2}
-\frac{5}{2}\,\zeta(5)\,(11\nu-6)\,\widehat{\lambda}^{\,3}\notag\\
&\qquad+\Big(\frac{21}{10}\,\zeta(7)\,(93\nu-56)
+\frac{27}{2}\,\zeta(3)^2\,\nu(3\nu-2)\Big)\,\widehat{\lambda}^{\,4}
\notag\\
&\qquad
- \Big(\frac{189}{8}\,\zeta(9)\,(61\nu-40)-30\,
\zeta(3)\,\zeta_5\,(22\nu^2-16\nu+1)\Big)\,\widehat{\lambda}^{\,5}\\
&\qquad
+ \Big(\frac{99}{8}\,\zeta_{11}\,(706\nu-405)+\frac{225}{2}\,\zeta(5)^2\,(23\nu^2-18\nu+2)
\notag\\
&\qquad\qquad+\frac{315}{4}\,\zeta(3)\,\zeta(7)\,(58\nu^2-45\nu+4) 
\Big)\,\widehat{\lambda}^{\,6}
+\ldots\bigg]\,\widehat{\Omega}_1(a)~.\notag
\end{align}
It is interesting to remark that if we compute the two-point function 
$\big\langle O_3(a) O_3(a)\big\rangle$ in the large-$N$ limit we obtain the same result as in the
SU($N$) case for any $\nu$, namely the function $\gamma_3$ is still given by the expression given in (\ref{gamma3is}) or (\ref{datag3}).
This means that the mixing term (\ref{f31}) does not give any contribution in the planar limit. We have explicitly verified that the same thing occurs also for the mixing terms 
involving $\widehat{\Omega}_1(a)$ in the single trace operators with odd dimensions up to $n=7$.
These findings confirm the expectation that for the observables that exist in both theories,
the SU($N$) and U($N$) models are indistinguishable at large $N$.

\section{Chebyshev polynomials of the first and second kind}
\label{app:cheb}

The Chebyshev polynomials of the first and second kind $T_n(x)$ and $U_n(x)$ can be defined as
\begin{align}
T_{n}(x) = \cos(n\theta)\quad\mbox{and}\quad U_{n}(x) = \frac{\sin[(n+1)\theta]}{\sin\theta}~,
\end{align}
where $x= \cos\theta$, with $x \in[-1,1]$ and $\theta\in[0,\pi]$.
The two sets of polynomials are related by
\begin{align}
	\label{A.1.5}
		T_{n}'(x) = n\,U_{n-1}(x)~.
\end{align}
They obey the orthogonality relations
\begin{equation}
\begin{aligned}
&\int_{-1}^{+1}\!dx\,\frac{T_{n}(x)\,T_{m}(x)}{\sqrt{1-x^{2}}} =
\frac{\pi}{2}\,\big(\delta_{nm}+\delta_{n0}\,\delta_{m0}\big)~,\\[1mm]
&\int_{-1}^{+1}\!dx \,\sqrt{1-x^{2}}\,U_{n}(x)\,U_{m}(x) = \frac{\pi}{2}\,\delta_{nm}~.
\end{aligned}
\label{A.2}
\end{equation}
They also satisfy 
\begin{equation}
\begin{aligned}
	\label{TUort}
&	\int_{-1}^{+1}\!dy\, \frac{T_{n}(y)}{(x-y)\,\sqrt{1-y^{2}}} = -\pi\,U_{n-1}(x)~,\\[1mm]
& \int_{-1}^{+1}\!dy\,\sqrt{1-y^{2}}\,\frac{U_{n}(y)}{x-y} = \pi\,T_{n+1}(x)~.
\end{aligned}
\end{equation}
Other useful relations are
\begin{equation}
	\label{A.3}
	\begin{aligned}
		& \int_{-1}^{+1}\!dx\,\frac{T_{n}(x)}{\sqrt{1-x^{2}}}\,\rme^{a\,x} = \pi\,I_{n}(a)~,\\[1mm]
		&\int_{-1}^{+1}\!dx\,\sqrt{1-x^{2}}\,U_{n}(x)\,\rme^{a\,x} = \frac{\pi}{a}\,(n+1)\,I_{n+1}(a)
	\end{aligned}
\end{equation}
where $I_n$ is the modified Bessel function of the first kind. These relations 
are easily proved by changing variable to $x=\cos\theta$
and using the well-known integral representation of the Bessel function 
$ I_{n}(a) = \frac{1}{\pi}\int_{0}^{\pi}d\theta\, e^{a\cos\theta}\,\cos(n\theta)$.
\section{On the determination of the cut edge $\mu(\lambda)$}
\label{app:edge}
The solution of (\ref{1.7}) with the Ansatz (\ref{2.19}) cannot be given in closed form. Nevertheless,
there exists a simple iterative scheme that allows us to determine $\mu_{0}$ in an efficient way. 
After writing (\ref{1.7}) in the form 
\begin{equation}
	\label{B.1}
		\int_{-\mu_{0}}^{+\mu_{0}}\!dy\, \Big[\frac{1}{x-y}+\nu\,\big(K(x)-K(x-y)\big)
		\Big]\,\rho(y) = 
		\frac{8\pi^{2}}{\lambda}\,x,
\end{equation}
we insert the Ansatz (\ref{2.19}), and using (\ref{TUort}) we get
\begin{align}
\label{B.1.5}
		\pi \mu_0 \sum_k a_{2k} \,T_{2k+1}\Big(\frac{x}{\mu_0}\Big)
		+ \nu \sum_k a_{2k} &\!\int_{-\mu_0}^{+\mu_0}\!dy\,
		\frac{K(x)- K(x-y)}{x-y}\sqrt{\mu_0^2-y^2} \,
		U_{2k}\Big(\frac{y}{\mu_0}\Big)\notag\\[1mm]
		&		= \frac{8\pi^2\mu_0}{\lambda}\, T_1\Big(\frac{x}{\mu_0}\Big)~.
\end{align}
The second term in the left hand side above can be expanded in the Chebyshev polynomials of the 
$T$ type as follows
\begin{align}
	\label{defE}
		\int_{-\mu_0}^{+\mu_0}\!dy\,\frac{K(x)- K(x-y)}{x-y}\sqrt{\mu_0^2-y^2} \,
		U_{2k}\Big(\frac{y}{\mu_0}\Big) 
		= \pi\mu_0 \sum_{k'} E_{k',k}\,T_{2k'+1}\Big(\frac{x}{\mu_0}\Big)~.
\end{align}
Then the condition (\ref{B.1.5}) becomes
\begin{align}
	\label{condE}
		\sum_{k',k} T_{2k+1}\Big(\frac{x}{\mu_0}\Big) \big(\delta_{k',k} + \nu\, 
		E_{k,'k}\big) \,a_{2k} = \frac{8\pi}{\lambda}\, T_{1}\Big(\frac{x}{\mu_0}\Big)~.
\end{align}
Let us denote by $\mathsf{E}$ the matrix of elements $E_{k',k}$; notice that in the conventions we are using, the index labels start from $0$. The solution to (\ref{condE}) can then be written as
\begin{align}
	\label{solaE}
		a_{2k} = \frac{8\pi}{\lambda}\,
\Big[(\mathbb{1} +\nu\,\mathsf{E})^{-1}\Big]_{k,0}~.
\end{align}
In particular, 
\begin{align}
	\label{sola0E}
	a_{0} = \frac{8\pi}{\lambda}\,
\Big[(\mathbb{1} +\nu\,\mathsf{E})^{-1}\Big]_{0,0}~.
\end{align}

The coefficients $E_{k',k}$ can be determined using the orthogonality relation (\ref{A.2}) in
(\ref{defE}); they are given by
\begin{align}
	\label{Ekkis}
		E_{k',k}&= \frac{2}{\pi^2\mu_0}
		\int_{-\mu_{0}}^{+\mu_{0}}\!\!\frac{dx}{\sqrt{\mu_{0}^{2}-x^{2}}}
		\,T_{2k'+1}\Big(\frac{x}{\mu_{0}}\Big)\,
		\int_{-\mu_{0}}^{+\mu_{0}}\!dy\, \big(K(x)-K(x-y)\big)\,\sqrt{\mu_{0}^{2}-y^{2}}
		\,U_{2k}\Big(\frac{y}{\mu_{0}}\Big)~.
\end{align}
Repeating the analysis discussed after (\ref{3.3}), we can show that for $k>0$ we have 
\begin{align}
	\label{Ekn0}
		E_{k',k} &= 4\,(-1)^{k+k'}(2k+1)
		\int_{0}^{\infty}\!\frac{dt}{t}\,\frac{\rme^{t}}{(\rme^{t}-1)^{2}}\,
		J_{2k+1}(\mu_{0}\, t)\,J_{2k'+1}(\mu_{0}\, t)~,
\end{align}
while for $k=0$ we find instead
\begin{align}
	\label{Ek0}
		E_{k',0} = 2\,(-1)^{k'}\int_{0}^{\infty}\!\frac{dt}{t}\,\frac{\rme^{t}}{(\rme^{t}-1)^{2}}\,
		J_{2k'+1}(\mu_{0}\, t)\,\Big(2J_{1}(\mu_{0} \,t)-\mu_{0} \,t\Big)~.
\end{align}
These expression can easily be expanded in powers of $\mu_0$. If one is interested in the contributions from a finite set of $\z$-values, then the matrix $\sf E$ can be truncated to a finite dimensional matrix, and the integrals (\ref{Ekn0}) and (\ref{Ek0}) provide the generating function 
for all monomials built over that finite set in closed form. For example the first $(3\times 3)$
block of $\mathsf{E}$ reads (we denote $\zeta_{n}\equiv \zeta(n)$ for brevity)
{\small
\begin{equation}
\mathsf{E} = \left(
\begin{array}{lll}
-\frac{3}{4} \z_{3} \mu _0^4+\frac{5}{2}  \z_{5} \mu _0^6 -\frac{455}{64}  \z_{7} \mu _0^8+\cdots &
-\frac{3}{4} \z_{3} \mu _0^4  +\frac{45}{16} \z_{5} \mu _0^6 -\frac{147}{16}  \z_{7}\mu _0^8+\cdots&
\frac{5}{16}  \z_{5}\mu _0^6-\frac{35}{16}  \z_{7}\mu _0^8+\cdots \\[2mm]
\frac{5}{8} \z_{5}\mu _0^6 -\frac{175}{64}  \z_{7}\mu _0^8+\cdots & 
\frac{5}{8}  \z_{5}\mu _0^6-\frac{105}{32}  \z_{7}\mu _0^8+\cdots & 
-\frac{35}{64}  \z_{7}\mu _0^8+\cdots \\[2mm]
-\frac{21}{64} \z_{7} \mu _0^8 +\cdots& -\frac{21}{64} \z_{7}\mu _0^8 +\cdots& \cdots \\[2mm]
\end{array}
\right)
\\
\end{equation}
}
Inserting this explicit expression of $\mathsf{E}$ into (\ref{sola0E}) we obtain 
\begin{align}
\label{B.9}
a_{0} &= \frac{8 \pi}{\lambda }\, \bigg[
1
+\frac 34\,\zeta (3)\,\nu\, \mu_{0} ^4
-\frac 52 \,\zeta (5)\,\nu\, \mu_{0} ^6
+\left(\frac{9}{16} \nu ^2 \zeta (3)^2
+\frac{455}{64}\nu \zeta (7)\right) \mu_{0} ^{8} \notag\\[1mm]
&\qquad
- \Big(\frac{2583}{128}\,\zeta (9)\,\nu+\frac{135}{32}\,\zeta (3)\, \zeta (5)\,\nu^2\Big)\,
\mu_{0} ^{10} \\[1mm]
&\qquad+ \Big(\frac{30261}{512}\,\zeta (11)\,\nu 
+\frac{3255}{256}\,\zeta (3)\,\zeta (7)\,\nu ^2+\frac{1025}{128}\,\zeta (5)^2\,\nu ^2
+\frac{27}{64}\,\zeta(3)^3\,\nu ^3 \Big)\,\mu_{0} ^{12}
+\cdots\bigg]~.\notag
\end{align}
On the other hand, $a_0$ is fixed by the normalization condition (\ref{norma0}) to $a_0= 2/(\pi\mu_0^2)$. This turns the above relation into an equation for $\mu_0$ which can be solved perturbatively, obtaining the result reported in (\ref{2.21}) of the main text.

\section{Additional data for $\gamma_{n}$ and $\Delta w_{n}$}
\label{app:data}

Additional explicit expressions for the two-point correction factor 
$\gamma_{n}$, extending those in (\ref{datag2})-(\ref{datag4}) are:
\begin{align}
\label{datag5}
        \gamma_{5} &= 1-15\,\zeta(3)\,\nu\,\widehat{\lambda}^{\,2}+100\,\zeta(5)\,\nu\,\widehat{\lambda}^{\,3}-\nu\,\Big(\frac{2275}{4}\,\zeta(7)-180\,\zeta (3)^2\,\nu\Big)\,\widehat{\lambda}^{\,4}\notag \\
        &\qquad +\Big(\frac{63}{2}\,\zeta (9)\,(103\,\nu -1)-2625\,\zeta (3) \,\zeta (5) \,\nu ^2\Big)\,\widehat{\lambda}^{\,5} \\
        &\qquad-\Big(\frac{1155}{8}\,\zeta (11)\,(133\,\nu-4)
        -\frac{65625}{4}\,\zeta (3) \,\zeta (7)\,\nu ^2-\frac{18875}{2}\,\zeta (5)^2\,\nu ^2 
        \!+\!2025\,\zeta (3)^3\,\nu ^3 \Big)\,\widehat{\lambda}^{\,6} + \cdots~,\notag
\end{align}
\begin{align}
\label{datag6}
        \gamma_{6} &= 1-\nu\,\bigg[18\,\zeta(3)\,\widehat{\lambda}^{\,2}-120\,\zeta(5)
        \widehat{\lambda}^{\,3}+\Big(\frac{1365}{2}\,\zeta(7)-243\,\zeta (3)^2\,\nu\Big)\,\widehat{\lambda}^{\,4}\notag \\
        &\qquad -\Big(\frac{7749}{2}\,\zeta (9)-3510\,\zeta (3) \,\zeta (5) \,\nu\Big)\,\widehat{\lambda}^{\,5} \\
        &\qquad-\Big(\frac{181797}{8}\,\zeta (11)
        -21735\,\zeta (3) \,\zeta (7)\,\nu-12525\,\zeta (5)^2\,\nu 
        +2970\,\zeta (3)^3\,\nu^2\Big)\,\widehat{\lambda}^{\,6} + \cdots\bigg]~,\notag
\end{align}
\begin{align}
\label{datag7}
        \gamma_{7} &= 1-21\,\zeta(3)\,\nu\,\widehat{\lambda}^{\,2}+140\,\zeta(5)\,\nu\,\widehat{\lambda}^{\,3}-\nu\,\Big(\frac{3185}{4}\,\zeta(7)-315\,\zeta (3)^2\,\nu\Big)\,\widehat{\lambda}^{\,4}\notag \\
        &\qquad +\Big(\frac{18081}{4}\,\zeta (9)\,\nu-4515\,\zeta (3) \,\zeta (5) \,\nu ^2\Big)\,\widehat{\lambda}^{\,5} \\
        &\qquad-\Big(\frac{211827}{8}\,\zeta (11)\,\nu
        -\frac{110985}{4}\,\zeta (3) \,\zeta (7)\,\nu ^2-\frac{30025}{2}\,\zeta (5)^2\,\nu ^2 
        +4158\,\zeta (3)^3\,\nu ^3 \Big)\,\widehat{\lambda}^{\,6} + \cdots~,\notag
\end{align}
\begin{align}
\label{datag8}
        \gamma_{8} &= 1-\nu\,\bigg[24\,\zeta(3)\,\widehat{\lambda}^{\,2}-160\,\zeta(5)
        \widehat{\lambda}^{\,3}+\Big(910\,\zeta(7)-396\,\zeta (3)^2\,\nu\Big)\,\widehat{\lambda}^{\,4}\notag \\
        &\qquad 
        -\Big(5166\,\zeta (9)-5640\,\zeta (3) \,\zeta (5) \,\nu\Big)\,\widehat{\lambda}^{\,5} \\
        &\qquad-\Big(30261\,\zeta (11)
        -34440\,\zeta (3) \,\zeta (7)\,\nu-19900\,\zeta (5)^2\,\nu 
        +5616\,\zeta (3)^3\,\nu^2\Big)\,\widehat{\lambda}^{\,6} + \cdots\bigg]~,\notag
\end{align}
\begin{align}
\label{datag9}
        \gamma_{9} &= 1-27\,\zeta(3)\,\nu\,\widehat{\lambda}^{\,2}+180\,\zeta(5)\,\nu\,\widehat{\lambda}^{\,3}-\nu\,\Big(\frac{4095}{4}\,\zeta(7)-486\,\zeta (3)^2\,\nu\Big)\,\widehat{\lambda}^{\,4}\notag \\
        &\qquad +\Big(\frac{23247}{4}\,\zeta (9)\,\nu-6885\,\zeta (3) \,\zeta (5) \,\nu ^2\Big)\,\widehat{\lambda}^{\,5} \\
        &\qquad-\Big(\frac{272349}{8}\,\zeta (11)\,\nu
        -\frac{167265}{4}\,\zeta (3) \,\zeta (7)\,\nu ^2-\frac{48375}{2}\,\zeta (5)^2\,\nu ^2 
        +7371\,\zeta (3)^3\,\nu ^3 \Big)\,\widehat{\lambda}^{\,6} + \cdots~,\notag
\end{align}
\begin{align}
\label{datag10}
        \gamma_{10} &= 1-\nu\,\bigg[30\,\zeta(3)\,\widehat{\lambda}^{\,2}-200\,\zeta(5)
        \widehat{\lambda}^{\,3}+\Big(\frac{2275}{2}\,\zeta(7)-585\,\zeta (3)^2\,\nu\Big)\,\widehat{\lambda}^{\,4}\notag \\
        &\qquad 
        -\Big(\frac{12915}{2}\,\zeta (9)-8250\,\zeta (3) \,\zeta (5) \,\nu\Big)
        \,\widehat{\lambda}^{\,5} \\
        &\qquad-\Big(\frac{151305}{4}\,\zeta (11)
        -49875\,\zeta (3) \,\zeta (7)\,\nu-28875\,\zeta (5)^2\,\nu 
        +5616\,\zeta (3)^3\,\nu^2\Big)\,\widehat{\lambda}^{\,6} + \cdots\bigg]~,\notag
\end{align}
\begin{align}
\label{s4g11}
        \gamma_{11} &= 1-33\,\zeta(3)\,\nu\,\widehat{\lambda}^{\,2}+220\,\zeta(5)\,\nu\,\widehat{\lambda}^{\,3}-\nu\,\Big(\frac{5005}{4}\,\zeta(7)-693\,\zeta (3)^2\,\nu\Big)\,\widehat{\lambda}^{\,4}\notag \\
        &\qquad +\Big(\frac{28413}{4}\,\zeta (9)\,\nu-9735\,\zeta (3) \,\zeta (5) \,\nu ^2\Big)\,\widehat{\lambda}^{\,5} \\
        &\qquad-\Big(\frac{332871}{8}\,\zeta (11)\,\nu
        -\frac{234465}{4}\,\zeta (3) \,\zeta (7)\,\nu ^2-\frac{67925}{2}\,\zeta (5)^2\,\nu ^2 
        +11880\,\zeta (3)^3\,\nu ^3 \Big)\,\widehat{\lambda}^{\,6} + \cdots~,\notag
\end{align}

Additional explicit expressions for the one-point function shift
$\Delta w_{n}$, extending those in (\ref{DW2is}) and (\ref{DW3is}) are:
\begin{align}
     \Delta w_{4} &=- \big(\pi \sqrt{N}\big)^2\,\nu\, \bigg[12\,\zeta (3)\,\mathbf{I}_3\,\widehat{\lambda}^{\,3}-80\,\zeta (5)\,\mathbf{I}_3\,\widehat{\lambda}^{\,4} +
     \Big( 35\,\zeta (7)\,(13 \,\mathbf{I}_3+2 \,\mathbf{I}_4) \notag  \\
     &\qquad\qquad\qquad
     -9\,\zeta (3)^2\,\nu\,(\mathbf{I}_2+8 \,\mathbf{I}_3)\Big)\,\widehat{\lambda}^{\,5}
      +\cdots\bigg]~,
\end{align}
\begin{align}
     \Delta w_{5} &= -\big(\pi  \sqrt{N}\big)^3\, \bigg[30\,\zeta (3)\,\nu \,\mathbf{I}_4\,\widehat{\lambda}^{\,\frac{7}{2}}
     -200\,\zeta (5)\,\nu\,\mathbf{I}_4\,\widehat{\lambda}^{\,\frac{9}{2}}
     +\Big(\frac{2275}{2}\,\zeta (7)\,\nu\,\mathbf{I}_4 \notag \\
&\qquad\qquad\qquad  -\frac{45}{2}\,\zeta (3)^2\,\nu ^2\,(\mathbf{I}_3+8 \,\mathbf{I}_4)\Big)
\,\widehat{\lambda}^{\,\frac{11}{2}}\Big)+\cdots\bigg]~, 
\label{DW5is}
\end{align}
\begin{align}
     \Delta w_{6}&= -\big(\pi  \sqrt{N}\big)^4\,\nu\, \bigg[72 \,\zeta (3)\,\mathbf{I}_5\,\widehat{\lambda}^{\,4}-480\,\zeta (5)\,\mathbf{I}_5\,\widehat{\lambda}^{\,5}+
     \Big(2730\,\zeta (7) \,\mathbf{I}_5 \notag\\
     &\qquad\qquad\qquad-54\,\zeta (3)^2\,\nu\,(\mathbf{I}_4+8 \,\mathbf{I}_5)\Big)\,\widehat{\lambda}^{\,6}+\cdots\bigg] ~,
     \label{DW6is}
\end{align}
\begin{align}     
     \Delta w_{7} &= -\big(\pi  \sqrt{N}\big)^5\, \bigg[168\,\zeta (3)\,\nu \,\mathbf{I}_6\,\widehat{\lambda}^{\,\frac{9}{2}}-1120\,\zeta (5)\,\nu\,\mathbf{I}_6\,\widehat{\lambda}^{\,\frac{11}{2}}+\Big(6370\,\zeta (7)\,\nu \,\mathbf{I}_6\notag \\
&\qquad\qquad\qquad -126\,\zeta (3)^2\,\nu ^2\,(\,\mathbf{I}_5+8 \,\mathbf{I}_6) 
\Big)\,\widehat{\lambda}^{\,\frac{13}{2}} +\cdots\bigg]
\label{DW7is}
\end{align}
where $\bI_n$ are the rescaled Bessel functions (\ref{bI}).

\end{appendix}

\providecommand{\href}[2]{#2}\begingroup\raggedright\endgroup

\end{document}